\documentclass[11pt]{article}

\usepackage[dvips]{graphicx}
\usepackage[english]{babel}

\usepackage{amsmath}
\usepackage{slashed}

\allowdisplaybreaks[4]

\newcommand{\re}{\operatorname{Re}}
\newcommand{\im}{\operatorname{Im}}

\newcommand{\gev}{\operatorname{GeV}}
\newcommand{\mev}{\operatorname{MeV}}

\newcommand{\ms}{\mskip 1.5mu}

\newcommand{\msp}{\phantom{-}}

\newcommand{\jpsi}{J\mskip -2mu/\mskip -0.5mu\Psi}

\newcommand{\lsim}{\raisebox{-4pt}{%
    $\,\stackrel{\textstyle <}{\sim}\,$}}

\newcommand{\dd}{\mathrm{d}}

\newcommand{\sgn}{\operatorname{sgn}}

\newcommand{\Li}{\operatorname{Li}}

\newcommand{\yb}{\bar{y}}
\newcommand{\zb}{\bar{z}}

\newlength{\plotwidth}
\begin{document}

\begin{flushright}
DESY 07-117
\end{flushright}

\begin{center}
\vskip 7.0\baselineskip
{\LARGE \bf Next-to-leading order corrections in \\[0.3em]
  exclusive meson production}
\vskip 4.0\baselineskip
M.~Diehl and W.~Kugler \\[0.5\baselineskip]
\textit{Deutsches Elektronen-Synchroton DESY, 22603 Hamburg, Germany}
\vskip 5.0\baselineskip
\textbf{Abstract}\\[0.8\baselineskip]
\parbox{0.9\textwidth}{We analyze in detail the size of
  next-to-leading order corrections to hard exclusive meson production
  within the collinear factorization approach.  Corrections to the
  cross section are found to be huge at small $x_B$ and substantial in
  typical fixed-target kinematics.  With the models we take for
  nucleon helicity-flip distributions, the transverse target
  polarization asymmetry in vector meson production is strongly
  affected by radiative corrections, except at large $x_B$.  Its
  overall size is very small for $\rho$ production but can be large in
  the $\omega$ channel.}
\end{center}
\vskip 3.0\baselineskip

\newpage

\tableofcontents

\newpage


\section{Introduction}
\label{sec:intro}

Generalized parton distributions (GPDs) have developed into a
versatile tool to quantify important aspects of hadron structure in
QCD.  In particular they contain unique information on the transverse
spatial distribution of partons \cite{Burkardt:2002hr} and on
spin-orbit effects and orbital angular momentum inside the nucleon
\cite{Ji:1996ek,Burkardt:2005km}.  Deeply virtual Compton scattering
is widely recognized as the process providing the theoretically
cleanest access to GPDs, with a wealth of observables calculable in
the large $Q^2$ limit \cite{Belitsky:2001ns} and with the calculation
of the hard-scattering subprocess now pushed to
next-to-next-to-leading order (NNLO) accuracy in $\alpha_s$
\cite{Kumericki:2007sa}.  A quantitative theoretical description of
exclusive meson production remains a challenge.  It would offer the
possibility to obtain important complementary information, difficult
to obtain from Compton scattering alone.  Perhaps most importantly,
vector meson production is directly sensitive to gluon distributions,
which in the Compton process are $\alpha_s$ suppressed relative to
quark distributions and only accessible through scaling violation
(just as in the well-known case of inclusive deep inelastic
scattering).  Given in addition the large number of channels that can
be studied and the wealth of high-quality data in a wide range of
kinematics from collider to fixed-target energies
\cite{Aharon:2007bq,Hadjidakis:2004jc}, it should be worthwhile to try
and push the theory description of exclusive meson production as far
as possible.

In this work we study the exclusive production of light mesons at
large photon virtuality $Q^2$ within the framework of collinear
factorization \cite{Collins:1996fb}.  In Bjorken kinematics, the
process amplitude can be approximated by the convolution of
hard-scattering kernels with generalized parton distributions and the
quark-antiquark distribution amplitude of the produced meson.  The
hard-scattering kernels have been calculated to $O(\alpha_s^2)$, i.e.\
to next-to-leading order (NLO) accuracy
\cite{Melic:1998qr,Belitsky:2001nq,Ivanov:2004zv}.  The aim of the
present paper is to investigate in some detail the size of the NLO
corrections compared with the leading-order (LO) results, on which
phenomenological studies have so far relied.

The collinear factorization approach provides an approximation of the
leading helicity amplitudes for meson production in the Bjorken limit,
up to relative corrections of order $1/Q^2$.  These power corrections
cannot be calculated systematically (and in fact the derivation
\cite{Collins:1996fb} of the factorization theorem suggests that these
corrections do not all factorize into hard-scattering kernels and
nonperturbative quantities pertaining to either the nucleon or the
produced meson).  One particular source of power corrections can
however readily be identified, namely the effect of the transverse
momentum of partons entering the hard-scattering subprocess, which in
the collinear approximation is neglected in the calculation of the
hard-scattering kernel.  A number of approaches include these $k_T$
effects, in particular the studies in
\cite{Vanderhaeghen:1999xj,Goloskokov:2006hr} based on the modified
hard-scattering formalism of Sterman et al.~\cite{Botts:1989kf}, and
calculations like \cite{Frankfurt:1995jw} which are based on the color
dipole formulation.  In the work by Martin, Ryskin and Teubner
\cite{Martin:1996bp}, parton-hadron duality is used to model the meson
formation and thus the transverse momentum of the hadronizing quarks
is included in the calculation, whereas the transverse momentum of
gluons in the proton is treated within high-energy $k_T$
factorization.  The studies just quoted agree in that transverse
momentum effects result in substantial power corrections to the
collinear approximation for $Q^2$ up to several $\gev^2$.
Unfortunately, the calculation of full NLO corrections in $\alpha_s$
remains not only a practical but also a conceptual challenge in all of
these approaches, so that the perturbative stability of their results
cannot be investigated at present.  (The approach of Sterman et al.\
takes partial account of radiative corrections, resumming a certain
class of them into Sudakov form factors.)

A consistent simultaneous treatment of radiative and power corrections
being out of reach at this time, a possible strategy is to study the
NLO corrections in the collinear approximation and in particular to
identify kinematical regions where these corrections are moderate or
small.  There one can then use with greater confidence formulations
incorporating power corrections.  In this spirit the present
investigation should be understood.  We will study both the cross
section for meson production from an unpolarized target and the
transverse target polarization asymmetry.  This asymmetry is one of
the few observables sensitive to the nucleon helicity-flip
distributions (in particular for gluons) and hence to the spin-orbit
and orbital angular momentum effects mentioned above.  We will in
particular see whether corrections tend to cancel in this polarization
asymmetry, as is often assumed.

In the bulk of this paper we concentrate on the production of vector
mesons.  In Sect.~\ref{sec:kernels} we set up our notation and recall
important properties of the hard-scattering kernels at NLO, as well as
giving a one-variable representation of these kernels after Gegenbauer
expansion of the meson distribution amplitude.  In
Sect.~\ref{sec:models} we specify the model of the generalized parton
distributions $H$ and $E$ we use for our numerical studies.  The size
of radiative corrections involving convolutions with distributions
$H$ is then studied in Sects.~\ref{sec:small-x} and \ref{sec:large-x}
for small and large $x_B$, respectively, and the convolutions
involving distributions $E$ are quantified in
Sect.~\ref{sec:E-convolutions}.  In Sect.~\ref{sec:cross} we then look
at the NLO corrections at the level of the observable cross section
and polarization asymmetry.  A brief study of exclusive pion
production in Sect.~\ref{sec:pseudo} complements our work, and in
Sect.~\ref{sec:sum} we summarize our main findings.  A number of more
lengthy formulae is collected in appendices.


\section{Hard-scattering kernels}
\label{sec:kernels}

\begin{figure}
\begin{center}
\includegraphics[width=\textwidth]{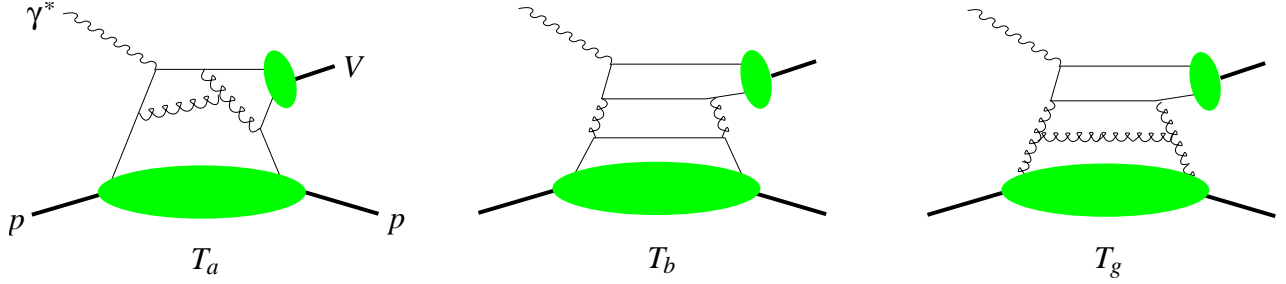}
\end{center}
\caption{\label{fig:graphs} Example graphs for the hard-scattering
  kernels $T_a$, $T_b$ and $T_g$ at order $\alpha_s^2$.}
\end{figure}

In the main part of this paper we are concerned with exclusive
production of a vector meson
\begin{equation}
  \label{vector-prod}
\gamma^*(q) + p(p) \to V(q') + p(p')
\end{equation}
in the limit of large $Q^2 = -q^2$ at fixed Bjorken variable $x_B =
Q^2 /(2 p\cdot q)$ and fixed $t = (p-p')^2$.  To leading order in
$1/Q$, the amplitude for longitudinal polarization of photon and meson
can be written as
\begin{align} 
  \label{meson_amp_NLO}
\mathcal{M} &=
\frac{2\pi\sqrt{4\pi\alpha}}{\xi\ms Q N_c}\, Q_V f_V
\int_0^1 dz\, \phi_V(z) 
\int_{-1}^1 dx\, \biggl\{ T_g(z,x,\xi) \ms F^g(x,\xi,t)
\nonumber \\
&\qquad
+ \frac{1}{n_f} \Bigl[ T_a(\zb,x,\xi) - T_a(z,-x,\xi) \Bigr] \ms
  F^{S}(x,\xi,t)
+ T_b(z,x,\xi) \ms F^{S}(x,\xi,t)
\phantom{\biggl[ \biggr]}
\nonumber \\
&\qquad
+ e_{V}^{(3)}\, \Bigl[ T_a(\zb,x,\xi) - T_a(z,-x,\xi) \Bigr] \,
  \Bigl[ F^{u(+)}(x,\xi,t) - F^{d(+)}(x,\xi,t) \Bigr]
\phantom{\biggl[ \biggr]}
\nonumber \\
&\qquad
+ e_{V}^{(8)}\, \Bigl[ T_a(\zb,x,\xi) - T_a(z,-x,\xi) \Bigr] \,
  \Bigl[ F^{u(+)}(x,\xi,t) + F^{d(+)}(x,\xi,t)
     - 2 F^{s(+)}(x,\xi,t) \Bigr]
\biggr\}
\end{align}
with $\zb = 1-z$, $N_c=3$, and the electromagnetic fine structure
constant $\alpha$.  Throughout this paper we work with $n_f=3$ active
quark flavors.  The proton matrix elements $F$ are parameterized by
generalized parton distributions,
\begin{equation}
  \label{GPD-def}
F^{q,g}(x,\xi,t) =
    \frac{1}{(p+p')\cdot n} \left[
    H^{q,g}(x,\xi,t)\, \bar u(p') \ms\slashed{n}\ms u(p)
  + E^{q,g}(x,\xi,t)\, \bar u(p')\,
    \frac{i\sigma^{\alpha\beta}n_{\alpha} (p'-p)_\beta}{2 m_p}\, u(p)
    \right]
\end{equation}
for quarks and gluons, where we use the conventions of
\cite{Diehl:2003ny}.  Here $n$ is a light-like auxiliary vector, $\xi
= x_B /(2-x_B)$ is the skewness variable, and $m_p$ denotes the
nucleon mass.  We have further introduced the combination
\begin{equation}
  \label{def-Fplus}
F^{q(+)}(x,\xi,t) = F^q(x,\xi,t) - F^q(-x,\xi,t)
\end{equation}
with positive charge conjugation parity.  In \eqref{meson_amp_NLO} we
have arranged the terms containing quark distributions into the flavor
singlet
\begin{align}
  \label{def-FS}
F^{S} &= F^{u(+)} + F^{d(+)} + F^{s(+)}
\end{align}
and the flavor triplet and octet combinations, $F^{u(+)} - F^{d(+)}$
and $F^{u(+)} + F^{d(+)} - 2F^{s(+)}$.  The factors
\begin{align}
  \label{charge_factors}
Q_\rho   &= \tfrac{1}{\sqrt{2}} \,,
&
Q_\omega &= \tfrac{1}{3\sqrt{2}} \,,
&
Q_\phi   &= -\tfrac{1}{3}
\end{align}
and
\begin{align}
  \label{quark_charge_factors}
e_{\rho}^{(3)} &= e_{\rho}^{(8)} = e_{\omega}^{(8)} = \tfrac{1}{6} \,,
&
e_{\omega}^{(3)} &= \tfrac{3}{2} \,,
&
e_{\phi}^{(3)} &= 0 \,,
&
e_{\phi}^{(8)} &= -\tfrac{1}{3}
\end{align}
correspond to a respective flavor content
\begin{equation}
\tfrac{1}{\sqrt{2}} \bigl( |u\bar{u}\rangle-|d\bar{d}\rangle \bigr) \,,
\qquad
\tfrac{1}{\sqrt{2}} \bigl( |u\bar{u}\rangle+|d\bar{d}\rangle \bigr) \,,
\qquad
|s\bar{s}\rangle
\end{equation}
of the $\rho$, $\omega$ and $\phi$.  The meson distribution amplitudes
$\phi_V(z)$ are normalized as $\int_0^1 dz\, \phi_V(z) = 1$, and the
decay constants have the values $f_\rho = 209 \mev$, $f_\omega = 187
\mev$, $f_\phi = 221 \mev$ \cite{Beneke:2003zv}.
We finally have hard-scattering kernels in \eqref{meson_amp_NLO},
where $T_g$ goes with gluon and $T_a$, $T_b$ go with quark
distributions in the proton.  In the graphs for $T_a$ quark lines
connect the proton and meson side, whereas in the graphs for $T_b$ the
proton and meson side are only connected by gluon lines.  $T_b$ thus
starts at order $\alpha_s^2$ and only goes with the quark singlet
distribution $F^S$.  Example graphs for the three kernels at NLO are
shown in Fig.~\ref{fig:graphs}.  We will refer to $T_g$, $T_a$, $T_b$
as the gluon, the quark non-singlet, and the pure quark singlet
kernel, respectively.

For better legibility we have not displayed the dependence on the
renormalization and factorization scales in \eqref{meson_amp_NLO}.
The renormalization scale $\mu_R$ appears as argument of $\alpha_s$
and through explicit logarithms in the hard-scattering kernels $T$.
The kernels further contain logarithms of the respective factorization
scales $\mu_{DA}$ and $\mu_{GPD}$ for the meson distribution amplitude
and the generalized parton distributions.  The NLO kernels in
\cite{Belitsky:2001nq,Ivanov:2004zv} are given for a common
factorization scale $\mu_F = \mu_{DA} = \mu_{GPD}$.  We can restore
the individual logarithms of $\mu_{DA}$ and $\mu_{GPD}$ from the
requirement that within the calculated precision the process amplitude
\eqref{meson_amp_NLO} must be independent of these scales.  As an
example consider the term
\begin{align}
  \label{scale-derivative}
& \frac{\dd}{\dd\ln\mu_{DA}^{2} \rule{0pt}{0.9em}}\, \int_0^1 dz\,
  \phi_V(z; \mu_{DA})\,
  T_a(\zb,x,\xi;\ms \alpha_s(\mu_R),\mu_R, \mu_{GPD},\mu_{DA},Q)
\nonumber \\[0.3em]
&\qquad
= \int_0^1 dz\,
  \biggl[ \frac{\dd}{\dd\ln\mu_{DA}^{2} \rule{0pt}{0.9em}}\,
          \phi_V(z; \mu_{DA}) \biggr]\,
  T_a(\zb,x,\xi;\ms \alpha_s(\mu_R),\mu_R, \mu_{GPD},\mu_{DA},Q)
\nonumber \\[0.2em]
&\qquad
+ \int_0^1 dz\, \phi_V(z; \mu_{DA})\,
  \biggl[ \frac{\dd}{\dd\ln\mu_{DA}^{2} \rule{0pt}{0.9em}}\,
  T_a(\zb,x,\xi;\ms \alpha_s(\mu_R),\mu_R, \mu_{GPD},\mu_{DA},Q)
\biggr] \,,
\end{align}
where the scale dependence of $\phi_V(z; \mu_{DA})$ is given by the
ERBL evolution equation \cite{Efremov:1980qk}.  At leading order this
gives a term $\dd/\dd (\ln\mu_{DA}^{2}\ms )\, \phi_V(z; \mu_{DA})$ of
order $\alpha_s$, whose convolution with the $O(\alpha_s)$ part of
$T_a$ must cancel against the contribution from explicit logarithms of
$\mu_{DA}$ in the $O(\alpha_s^2)$ part of $T_a$.  An analogous
argument holds for the dependence on $\mu_{GPD}$, with the
complication that the gluon and quark singlet distributions mix under
evolution.  More precisely, the convolution of $\dd/\dd
(\ln\mu_{GPD}^{2}\ms )\, F^S(x,\xi,t; \mu_{GPD})$ with the
$O(\alpha_s)$ part of $T_a$ cancels at $O(\alpha_s^2)$ against the
contributions from logarithms of $\mu_{GPD}$ in $T_a$ and in $T_g$.
Likewise, the convolution of $\dd/\dd (\ln\mu_{GPD}^{2}\ms )\,
F^g(x,\xi,t; \mu_{GPD})$ with the Born term of $T_g$ cancels at
$O(\alpha_s^2)$ against the contributions from logarithms of
$\mu_{GPD}$ in $T_g$ and in the pure singlet kernel $T_b$.
We have explicitly checked that the scale dependence of the
hard-scattering kernels given in \cite{Ivanov:2004zv} cancels in the
process amplitude \eqref{meson_amp_NLO} as just described, using the
LO evolution equations for GPDs given in App.~\ref{app:evolution}.

Separating the $\mu_{DA}$ and $\mu_{GPD}$ dependence, we can write the
kernels as
\begin{align}
  \label{orig_kernels}
T_g(z,x,\xi) &=
- \alpha_s\,
\frac{\xi}{(\xi-x-i\epsilon) (\xi+x-i\epsilon)}\, \frac{1}{z\zb}
\biggl[1 + \frac{\alpha_s}{4\pi}\;
           \mathcal{I}_g\Bigl( z,\frac{\xi-x}{2\xi} \Bigr)
\biggr] \,,
\nonumber \\
T_b(z,x,\xi) &=
\phantom{-}C_F\, \frac{\alpha_s^2}{8\pi}\,
\frac{1}{z\zb}\; \mathcal{I}_b\Bigl( z,\frac{\xi-x}{2\xi} \Bigr) \,,
\nonumber \\
T_a(\zb,x,\xi) &=
- C_F\ms \alpha_s\, \frac{\xi}{\xi-x-i\epsilon}\, \frac{1}{\zb}\,
\biggl[ 1 + \frac{\alpha_s}{4\pi}\;
            \mathcal{I}_a\Bigl( \zb,\frac{\xi-x}{2\xi} \Bigr)
\biggr]
\end{align}
with
\begin{align}
  \label{I_gluon}
\mathcal{I}_g(z,y) &= \left[\ms
2 C_A \left(\frac{\yb}{y}+\frac{y}{\yb}\right)
       \bigl( y\ln y + \yb \ln\yb \bigr)
- C_F \left( \frac{y}{\yb} \ln y + \frac{\yb}{y} \ln\yb \right)
\right] \ln\frac{Q^2}{\mu^2_{GPD}}
\nonumber \\
&\quad
+ \beta_0 \ln\frac{\mu^2_{R}}{\mu_{GPD}^2}
+ C_F\ms \bigl( 3 + 2 z \ln\zb + 2 \zb \ln z \bigr)
         \ln\frac{Q^2}{\mu^2_{DA}}
+ \mathcal{K}_g(z,y) \,,
\nonumber \\
\mathcal{I}_b(z,y) &=
2 (\yb-y) \left( \frac{\ln y}{\bar{y}}+\frac{\ln\bar{y}}{y} \right)
  \ln\frac{Q^2}{\mu_{GPD}^2}
+ \mathcal{K}_b(z,y)
\intertext{and}
  \label{I_a}
\mathcal{I}_a(v,u) &=
\beta_0\ms \biggl( \frac{5}{3} - \ln(vu)
- \ln\frac{Q^2}{\mu^2_R} \biggr)
+ C_F\, \bigl( 3 + 2\ln u \bigr) \ln\frac{Q^2}{\mu^2_{GPD}}
+ C_F\, \bigl( 3 + 2\ln v \bigr) \ln\frac{Q^2}{\mu^2_{DA}}
\nonumber \\
&\quad
+ \mathcal{K}_a(v,u) \,,
\phantom{\biggl[ \biggr]}
\end{align}
where $\yb = 1-y$ and we use the standard notation
\begin{align}
  \label{color_factors}
C_F &= \frac{N_c^2-1}{2N_c} \,,
&
C_A &= N_c \,,
&
\beta_0 &= \frac{11}{3}\ms N_c - \frac{2}{3}\ms n_f \,.
\end{align}
The functions $\mathcal{K}_g$, $\mathcal{K}_b$ and $\mathcal{K}_a$ are
independent of $Q^2$ and the renormalization and factorization scales.
They contain factors $C_F$ or $C_A$ but not $\beta_0$.
Their expressions can be found in \cite{Ivanov:2004zv}, taking into
account that the kernels $T_g$ and $T_b$ here are denoted by $T_g$ and
$T_{(+)}$ there, and that
\begin{align}
  \label{dima-trans}
T(v,u) \,\bigg|_{\text{\protect\cite{Ivanov:2004zv}}}
&= \frac{C_F\ms \alpha_s}{4 v u}
\biggl[ 1 + \frac{\alpha_s}{4\pi}\;
            \mathcal{I}_a(v,u) \biggr]_{\text{here}} \,,
&
y \,\big|_{\text{\protect\cite{Ivanov:2004zv}}}
&= - y \,\big|_{\text{here}} \,.
\end{align}
Note that the pure singlet kernel $T_b$ does not contain logarithms of
$\mu_{DA}$ and $\mu_{R}$ at $O(\alpha_s^2)$, since there is no Born
level contribution against which they could cancel in the scale
dependence of the process amplitude.  There is however a logarithm of
$\mu_{GPD}$, since the corresponding derivative of the Born level
convolution of $T_g$ with $F^g$ contains a term going with the quark
singlet distribution $F^S$, as already mentioned after
\eqref{scale-derivative}.

The kernels in \eqref{orig_kernels} have singularities for real-valued
arguments.  One readily finds that $x/\xi = (\hat{s} - \hat{u}) /Q^2$,
where $\hat{s}$ and $\hat{u}$ are the Mandelstam variables for the
parton-level subprocess $\gamma^* q\to (q\bar{q})\ms q$ or $\gamma^*
g\to (q\bar{q})\ms g$.  The prescriptions $\hat{s}+i\epsilon$ for the
$\hat{s}$-channel and $\hat{u}+i\epsilon$ for the $\hat{u}$-channel
singularities thus instruct us to take $x+i\epsilon$ for $x>0$ and
$x-i\epsilon$ for $x<0$.  Correspondingly, the second argument $(\xi -
x)/(2\xi)$ of $\mathcal{I}_g$, $\mathcal{I}_b$ and $\mathcal{I}_a$
must be taken with $-i\epsilon$ for $x>0$ and $+i\epsilon$ for $x<0$.
In $T_a(z,-x,\xi)$ the second argument of $\mathcal{I}_a$ is $(\xi +
x)/(2\xi)$, which has to be taken with $-i\epsilon$ for $x<0$ and
$+i\epsilon$ for $x>0$.
We remark that, as it is written, the $i\epsilon$ prescription in
\cite{Ivanov:2004zv} for the gluon and the pure singlet kernel is
correct for $x>0$ but incorrect for $x<0$.  Likewise, the prescription
given in \cite{Belitsky:2001nq,Ivanov:2004zv} for the quark
non-singlet kernel is correct for $x>0$ if the corresponding argument
is $(\xi - x)/(2\xi)$ and for $x<0$ if the argument is $(\xi +
x)/(2\xi)$, but incorrect in the other cases.\footnote{%
  We thank Dima~Ivanov for discussions on this point.  The numerical
  results in \protect\cite{Ivanov:2004zv} were obtained with the
  correct prescription.}


\subsection{Gegenbauer expansion}
\label{sec:gegenbauer}

Let us expand the meson distribution amplitude on Gegenbauer
polynomials, 
\begin{equation}
  \label{gegen-phi}
\phi_V(z;\mu) = 6 z(1-z) \sum_{n=0}^\infty 
  a_n(\mu)\, C_n^{3/2}(2z-1) \,,
\end{equation}
where $a_0 = 1$ according to the normalization condition $\int_0^1
dz\, \phi_V(z) = 1$.  To leading order, the Gegenbauer coefficients
evolve as
\begin{equation}
  \label{DA-evol}
a_n(\mu) = a_n(\mu_0)\, 
  \left( \frac{\alpha_s(\mu)}{\alpha_s(\mu_0)}
  \right)^{\gamma_n/\beta_0}
\end{equation}
with anomalous dimensions
\begin{align}
  \label{anom-dim}
\gamma_0 &= 0 \,,  & \gamma_2 &= \tfrac{25}{6}\, C_F \,, &
\gamma_4 &= \tfrac{91}{15}\, C_F \,,
\end{align}
where $\alpha_s(\mu)$ is the running coupling at one-loop accuracy.
One has $\gamma_n \approx 4 C_F \ln(n+1)$ within at most $6\%$ for all
$n$.
For $V=\rho, \omega, \phi$ only coefficients $a_n$ with even $n$ are
nonzero due to charge conjugation invariance, and in all subsequent
expressions of this paper we consider $n$ to be even.  Calculations of
the distribution amplitudes in models or on the lattice typically give
values for the first or the first two nonvanishing moments, see e.g.\
\cite{Ball:1996tb,Bakulev:2005cp,Braun:2006dg}, so that a truncated
version of the expansion \eqref{gegen-phi} is very often used in
phenomenological studies.  Convolution with individual terms in
\eqref{gegen-phi} also allows us to reduce the hard-scattering kernels
for meson production to functions of a single longitudinal variable.
More precisely, we can rewrite the process amplitude
\eqref{meson_amp_NLO} as
\begin{align}
  \label{gegen-amp}
\mathcal{M} =
\frac{2\pi\sqrt{4\pi\alpha}}{\xi\ms Q N_c}\, Q_V f_V
\sum_{n=0}^\infty a_n^{}\ms
\biggl[ \mathcal{F}^g_n + \mathcal{F}^{S(a)}_n + \mathcal{F}^{S(b)}_n
      + e_{V}^{(3)}\, \mathcal{F}^{(3)}_{n\phantom{V}}
      + e_{V}^{(8)}\, \mathcal{F}^{(8)}_{n\phantom{V}} \biggr]
\end{align}
with convolutions in $x$
\begin{align}
  \label{F-def}
\mathcal{F}^g_n
 &= \int_{-1}^1 dx\, T_{g,n}(x,\xi)\ms F^g(x,\xi,t) \,,
\qquad\qquad\qquad
\mathcal{F}^{S(b)}_n
  = \int_{-1}^1 dx\, T_{b,n}(x,\xi)\ms F^{S}(x,\xi,t) \,,
\nonumber\\
\mathcal{F}^{S(a)}_n
 &= \int_{-1}^1 dx\, \Bigl[ T_{a,n}(x,\xi) - T_{a,n}(-x,\xi) \Bigr] \ms
    \frac{1}{n_f}\, F^{S}(x,\xi,t) \,,
\nonumber \\
\mathcal{F}^{(3)}_n
 &= \int_{-1}^1 dx\, \Bigl[ T_{a,n}(x,\xi) - T_{a,n}(-x,\xi) \Bigr] \,
    \Bigl[ F^{u(+)}(x,\xi,t) - F^{d(+)}(x,\xi,t) \Bigr] \,,
\displaybreak
\nonumber \\
\mathcal{F}^{(8)}_n
 &= \int_{-1}^1 dx\, \Bigl[ T_{a,n}(x,\xi) - T_{a,n}(-x,\xi) \Bigr] \,
    \Bigl[ F^{u(+)}(x,\xi,t) + F^{d(+)}(x,\xi,t)
       - 2 F^{s(+)}(x,\xi,t) \Bigr] \,,
\end{align}
which depend on $\xi$ and $t$, and logarithmically on $Q^2$ and on the
factorization and renormalization scales.  At order $\alpha_s^2$ the
dependence on $\mu_R$ and on $\mu_{DA}$ cancels in each separate
convolution, while the dependence on $\mu_{GPD}$ cancels in
$\mathcal{F}^{\smash{(3)}}_{n\phantom{i}}$ and
$\mathcal{F}^{\smash{(8)}}_{n\phantom{i}}$ and in the sum
$\mathcal{F}^{\smash[b]{g}}_n + \mathcal{F}^{\smash{S(a)}}_n +
\mathcal{F}^{\smash{S(b)}}_n$ as discussed after
\eqref{scale-derivative}.
In analogy to \eqref{F-def} we define convolutions $\mathcal{H}$ and
$\mathcal{E}$ for the individual distributions $H$ and $E$ in
\eqref{GPD-def}.
The kernels $T_{g,n}$, $T_{a,n}$, $T_{b,n}$ are obtained from $T_g$,
$T_a$, $T_b$ by multiplying with $6 z(1-z)\, C_{\smash{n}}^{3/2}(2z -
1)$ and integrating over $z$.  For $n=0$ we find
\begin{align}
  \label{gegen-kernels}
T_{g,n}(x,\xi) &= - 3 \alpha_s\,
  \frac{2\xi}{(\xi-x-i\epsilon) (\xi+x-i\epsilon)}
  \biggl[ 1 + \frac{\alpha_s}{4\pi}\,
          t_{g,n}\left( \frac{\xi-x}{2\xi} \right) \biggr] \,,
\nonumber \\
T_{b,n}(x,\xi) &= \phantom{-}3  C_F\, \frac{\alpha_s^2}{4\pi}\,
                  t_{b,n}\left( \frac{\xi-x}{2\xi} \right) \,,
\nonumber \\
T_{a,n}(x,\xi) &= - 3  C_F\, \alpha_s\,
  \frac{\xi}{\xi-x-i\epsilon}\,
  \biggl[ 1 + \frac{\alpha_s}{4\pi}\,
          t_{a,n}\left( \frac{\xi-x}{2\xi} \right) \biggr]
\end{align}
with
\begin{align}
  \label{kernels_asy}
t_{g,0}(y) &=
\biggl[ 2C_A\ms (y^2+\yb^2) - C_F\ms y \biggr] \frac{\ln y}{\yb}\,
  \ln\frac{Q^2}{\mu^2_{GPD}}
+ \frac{\beta_0}{2}\ms \ln\frac{\mu_R^2}{\mu^2_{GPD}}
\nonumber \\
&\quad + C_F \biggl[
- \frac{5}{2}
+ \left(\frac{1}{\yb}+1-4y\right) \ln y
- \frac{y}{2}\, \frac{\ln^2 y}{\yb}
\nonumber \\
& \qquad\qquad
- 2 (\yb-y) \Li_2\yb
- 4y\yb \biggl(3\Li_3\yb - \ln y\, \Li_2 y
             - \frac{\pi^2}{6} \ln y \biggr)
\biggr]
\nonumber \\
&\quad + C_A \left[
- \left( \frac{6}{\yb} - 8y \right) \ln y
+ \left( \frac{1}{\yb} - 2y \right) \ln^2 y + 2(\yb-y)\Li_2\yb
\right]
+ \{y\to\yb\}\,, \phantom{\biggl[ \biggr]}
\nonumber \\
t_{b,0}(y) &=
2 (\yb-y)\, \frac{\ln y}{\yb}\,
\biggl[ \ln\frac{Q^2}{\mu_{GPD}^2} - 3 \biggr]
+ (\yb-y)\, \frac{\ln^2 y}{\yb} + 4 \Li_2\yb
- \{y\to\yb\} \,,
\nonumber \\
t_{a,0}(y) &=
\beta_0 \left[ \frac{19}{6} - \ln y
  - \ln\frac{Q^2}{\mu^2_R} \right]
+ C_F \biggl[
  \left( 3 + 2 \ln y \right)\ms \ln\frac{Q^2}{\mu^2_{GPD}}
- \frac{77}{6} - \left( \frac{1}{\yb} - 3 \right) \ln y + \ln^2 y
\biggr]
\nonumber \\
&\quad
+ \left( 2 C_F - C_A \right)
  \biggl\{ - \frac{1}{3} - 4(2-3y) \ln\yb
+ 2(1-6y) \ln y + 4(1-3y)\ms \bigl( \Li_2 y - \Li_2\yb \bigr)
\nonumber \\
&\qquad\qquad
+ 2(1-6y\yb) \left[ 3 \bigl( \Li_3\yb + \Li_3 y \bigr)
  - \ln y\, \Li_2 y - \ln\yb\, \Li_2\yb
  - \frac{\pi^2}{6}\, \bigl( \ln y + \ln\yb \bigr)
\right]
\biggr\} \,.
\end{align}
The corresponding kernels for $n=2$ and $n=4$ are given in
App.~\ref{app:kernels}. 
The $i\epsilon$ prescription to be used in \eqref{gegen-kernels} is
the same as specified at the end of the previous subsection.  This
implies that in $t_{g,n}(y)$, $t_{b,n}(y)$, $t_{a,n}(y)$ and
$t_{a,n}(\yb)$ one has to take $\ln(y -i\epsilon)$, $\Li_2(\yb +
i\epsilon)$ and $\Li_3(\yb + i\epsilon)$ for $y<0$.
For the gluon and pure singlet kernel, which dominate in process
amplitudes at small $\xi$, we have in particular
\begin{align}
  \label{im_gluon}
\frac{1}{\pi} \im t_{g,0}(y) &=
- \biggl[ 2C_A\ms (y^2+\yb^2) - C_F\ms y \ms\biggr] \frac{1}{\yb}\,
  \ln\frac{Q^2}{\mu_{GPD}^2}
\nonumber \\
&\quad
- C_F \biggl[ 1-4y + \frac{1 - y \ln(-y)}{\yb}
  + 2(\yb-y) \ln\yb + 2y\yb \left(
    \ln^2\yb + 2 \Li_2 y + \frac{\pi^2}{3} \right) \biggr]
\nonumber \\
&\quad
+ 2C_A \biggl[ \frac{3}{\yb} - 4y
  - \left( \frac{1}{\yb} - 2y \right) \ln(-y)
  + (\yb-y) \ln\yb \biggr] \,,
\nonumber \\
\frac{1}{\pi} \im t_{b,0}(y) &=
2\, \frac{\yb-y}{\yb}\, \biggl[ 3 - \ln(-y) 
    - \ln\frac{Q^2}{\mu_{GPD}^2} \biggr] + 4 \ln\yb
\end{align}
in the region $y<0$.  In the limit $y\to 0$ all three expressions in
\eqref{kernels_asy} contain singular terms proportional to $\ln y$ and
$\ln^2 y$.  For the convolution \eqref{F-def} we should however
consider $(y \yb)^{-1}\ms t_{g,n}(y)$, $y^{-1}\ms t_{a,n}(y)$ and
$\yb^{-1}\ms t_{a,n}(\yb)$ according to \eqref{gegen-kernels}.  With
the appropriate $i\epsilon$ prescription, these kernels contain terms
which for $y\to 0$ go like $(y-i\epsilon)^{-1} \ln^m(y-i\epsilon)$,
where $m=0,1,2$.


\section{Model for the unpolarized GPDs}
\label{sec:models}

It is difficult to study the impact of NLO corrections at the level of
the hard-scattering kernels given in the previous subsection,
especially since they are not smooth functions but distributions with
singularities at $y=0$.  We will therefore use model GPDs to
investigate the radiative corrections at the level of the convolution
integrals \eqref{F-def}.  The aim of this work is not a systematic
improvement of existing models, nor a detailed exploration of model
uncertainties on observables in exclusive meson production.  We do
however require that the models we use are consistent with known
theoretical requirements and basic phenomenological constraints.

For $H^q$ and $H^g$ we adopt the widely used ansatz of
\cite{Musatov:1999xp,Goeke:2001tz} based on double distributions,
where a $\xi$ dependence is generated according to
\begin{align}
  \label{dd-models}
H^{q(+)}(x,\xi,t) &=
\int_{-1}^1 d\beta \int_{-1+|\beta|}^{1-|\beta|} d\alpha\;
  \delta(x-\beta-\xi\alpha)\, h^{(2)}(\beta,\alpha)\,
  H^{q(+)}(\beta,0,t) \,,
\nonumber \\
H^g(x,\xi,t) &=
\int_{-1}^1 d\beta \int_{-1+|\beta|}^{1-|\beta|} d\alpha\;
  \delta(x-\beta-\xi\alpha)\, h^{(2)}(\beta,\alpha)\,
  H^g(\beta,0,t)
\end{align}
with
\begin{equation}
  \label{profile}
h^{(b)}(\beta,\alpha) = \frac{\Gamma(2b+2)}{2^{2b+1}\Gamma^2(b+1)}\,
\frac{[ (1-|\beta|)^2- \alpha^2 ]^b}{(1-|\beta|)^{2b+1}} \,.
\end{equation}
The distributions at zero skewness are taken as
\begin{align}
  \label{t-dep-h}
H^{q(+)}(x,0,t) &=
   q_v(x) \exp\bigl[ t f_{q_v}(x) \bigr]
 + 2\ms \bar{q}(x) \exp\bigl[ t f_{\bar{q}}(x) \bigr] \,,
\nonumber \\[0.3em]
H^g(x,0,t)      &= x g(x) \exp\bigl[ t f_{g}(x) \bigr]
\end{align}
for $x>0$, with the values for $x<0$ following from the symmetry
properties of the distributions.  Here $q_v(x) = q(x) - \bar{q}(x)$,
$\bar{q}(x)$ and $g(x)$ are the usual unpolarized densities for
valence quarks, antiquarks and gluons, for which we take the CTEQ6M
parameterization \cite{Pumplin:2002vw}.  This parameterization has an
identical strange and antistrange sea, so that $s_v(x)=0$.
The ansatz \eqref{dd-models} is taken at a starting scale $\mu_0$ and
then evolved with the LO evolution equations given in
App.~\ref{app:evolution}.  For the studies in Sects.~\ref{sec:small-x}
and \ref{sec:large-x} we take $\mu_0 = 1.3\gev$, which is the starting
scale of evolution for the CTEQ6M densities.  In
Sects.~\ref{sec:E-convolutions} and \ref{sec:cross} we will instead
take $\mu_0 = 2\gev$, since this will allow us to use the results for
the $t$ dependence of valence distributions obtained in
\cite{Diehl:2004cx}.

For the $t$ dependence in the ansatz \eqref{dd-models} we follow the
modeling strategy of \cite{Goeke:2001tz} and take an exponential
behavior in $t$ with an $x$ dependent slope.  For valence quarks we
take the slope functions
\begin{equation}
  \label{DFJK4-f}
f_{q_v}(x) = \alpha'_v (1-x)^3 \ln\frac{1}{x} + B_{q_v} (1-x)^3
  + A_{q_v} x (1-x)^2
\end{equation}
with parameters $\alpha'_v = 0.9 \gev^{-2}$ and
\pagebreak[3]
\begin{align}
  \label{DFJK-params}
A_{u_v} &= 1.26 \gev^{-2} \,,  & B_{u_v} &= 0.59 \gev^{-2} \,,
\nonumber \\
A_{d_v} &= 3.82 \gev^{-2} \,,  & B_{d_v} &= 0.32 \gev^{-2} \,,
\end{align}
from \cite{Diehl:2004cx}.  We recall the sum rule
\begin{align}
  \label{F1-sum-rule}
F_1^q(t) &= \int_{-1}^1 dx\, H^q(x,0,t) 
          = \int_0^1 dx\, q_v(x) \exp\bigl[ t f_{q_v}(x) \bigr] \,,
\end{align}
from which one obtains the electromagnetic Dirac form factors of
proton and neutron by appropriate quark flavor combinations.  Together
with the CTEQ6M distributions at $\mu_0 = 2\gev$, the ansatz in
\eqref{DFJK4-f} and \eqref{DFJK-params} gives a good description of
the data for these form factors.
For gluons we take a slightly simpler form than \eqref{DFJK4-f} and
set
\begin{equation}
  \label{gluon-profile}
f_g(x) = \alpha'_g (1-x)^2 \ln\frac{1}{x} + B_g (1-x)^2 \,.
\end{equation}
For the parameters we take
\begin{align}
  \label{g-prof-param}
  \alpha'_g &= 0.164 \gev^{-2} \,, &
  B_g &= 1.2 \gev^{-2}
\end{align}
so as to match recent H1 data on $\jpsi$ photoproduction, whose $t$
dependence is well fitted by \cite{Aktas:2005xu}
\begin{equation}
  \label{h1-jpsi}
\frac{d\sigma}{dt} \propto \exp\biggl[ 
  \left( b_0 + 4\alpha'_g \ln\frac{W_{\gamma p}}{W_0} \right) t 
\,\biggr]
\end{equation}
with central values $b_0 = 4.63 \gev^{-2}$ and $\alpha'_g = 0.164
\gev^{-2}$ for $W_{0} = 90 \gev$.  To connect \eqref{h1-jpsi} with
\eqref{gluon-profile} we have used the approximate relation
$d\sigma/dt \propto |H^g(\xi,\xi,t)|^2$, which is obtained when only
keeping the imaginary part of the tree-level amplitude, where $2 \xi =
(M_{\jpsi} /W_{\gamma p})^2$ in terms of the $\gamma p$ c.m.\ energy.
With the ansatz \eqref{dd-models} one approximately has
$H^g(\xi,\xi,t) \propto \exp\bigl[ t f_g(2\xi) \bigr]$ for the $t$
dependence of the GPD \cite{Goloskokov:2006hr}.

Whereas information on valence quark GPDs can be obtained from the sum
rules \eqref{F1-sum-rule} and information on gluon GPDs from $\jpsi$
production, almost nothing is so far known about the $t$ dependence of
GPDs for antiquarks.  As a simple ansatz we shall take their slope
functions equal to those in the valence sector,
\begin{align}
f_{\bar{u}} &= f_{u_v} \,, &
f_{\smash{\bar{d}}} &= f_{d_v} \,, &
f_{\bar{s}} &= f_{d_v} \,,
\end{align}
bearing in mind that it remains an outstanding task to develop more
realistic models.


\subsection{Nucleon helicity-flip distributions}
\label{sec:e-model}

The nucleon helicity-flip distributions $E^q$ and $E^g$ are less-well
known than their counterparts $H^q$ and $H^g$, because their values at
$\xi=0$ and $t=0$ cannot be measured in inclusive processes and are
thus subject to considerable uncertainty.

The model described in this subsection refers to a scale of $\mu_0
=2\gev$.  We make a double distribution based ansatz
\begin{align}
  \label{dd-models-e}
E^{q(+)}(x,\xi,t) &=
\int_{-1}^1 d\beta \int_{-1+|\beta|}^{1-|\beta|} d\alpha\;
  \delta(x-\beta-\xi\alpha)\, h^{(2)}(\beta,\alpha)\,
  E^{q(+)}(\beta,0,t) \,,
\nonumber \\
E^g(x,\xi,t) &=
\int_{-1}^1 d\beta \int_{-1+|\beta|}^{1-|\beta|} d\alpha\;
  \delta(x-\beta-\xi\alpha)\, h^{(2)}(\beta,\alpha)\,
  E^g(\beta,0,t)
\end{align}
as in \eqref{dd-models}, and for $x>0$ set
\begin{align}
  \label{t-dep-e}
E^{q(+)}(x,0,t) &=
   e_{q_v}(x) \exp\bigl[ t\ms g_{q_v}(x) \bigr]
 + 2 e_{\bar{q}}(x) \exp\bigl[ t\ms g_{\bar{q}}(x) \bigr] \,,
\nonumber \\[0.3em]
E^g(x,0,t)      &= x e_g(x) \exp\bigl[ t\ms g_{g}(x) \bigr] \,,
\end{align}
with the corresponding values for $x<0$ determined by the symmetry
properties of the distributions.
For the forward limit of the valence distribution we take
\begin{equation}
  \label{e-val-model}
e_{q_v}(x) = \kappa_q \, N(\alpha_v,\beta_{q_v})\,
  x^{-\alpha_v}\ms (1-x)^{\beta_{q_v}} \,,
\end{equation}
whose normalization factor
\begin{equation}
\label{norm-fact}
N(\alpha,\beta) =
  \frac{\Gamma(2-\alpha+\beta)}{\Gamma(1-\alpha)\, \Gamma(1+\beta)}
\end{equation}
ensures the sum rules
\begin{align}
  \label{E-kappa-sum}
\kappa_q &= \int_{-1}^1 dx\, E^q(x,0,0)
          = \int_0^1 dx\, e_{q_v}(x) \,,
\end{align}
where $\kappa_u \approx 1.67$ and $\kappa_d \approx -2.03$ are the
contributions of $u$ and $d$ quarks to the anomalous magnetic moment
of the proton.  For the functions controlling the $t$ dependence we
take the same form as in~\eqref{DFJK4-f},
\begin{equation}
  \label{DFJK4-g}
g_{q_v}(x) = \alpha'_v (1-x)^3 \ln\frac{1}{x} + D_{q_v} (1-x)^3
  + C_{q_v} x (1-x)^2 \,.
\end{equation}
With the parameters $\alpha_v^{} = 0.55$, $\alpha'_v = 0.9 \gev^{-2}$
and
\begin{align}
  \label{DFJK4-E-par}
\beta_u &= 3.99 \,, &
C_{u_v \,} &= 1.22 \gev^{-2} \,,  & D_{u_v} &= 0.38 \gev^{-2} \,,
\nonumber \\
\beta_d &= 5.59 \,, &
C_{d_v} &= 2.59 \gev^{-2} \,,  & D_{d_v} &= -0.75 \gev^{-2} \,,
\end{align}
from \cite{Diehl:2004cx} one obtains a good fit to the electromagnetic
Pauli form factors of proton and neutron via the generalization of
the sum rule \eqref{E-kappa-sum} to finite $t$.

For the forward limit of the distributions of antiquarks and gluons we
make the same simple ansatz as in \eqref{e-val-model},
\begin{align}
  \label{eg-model}
e_{\bar{q}}(x) &= k_{\bar{q}} \,
  x^{-\alpha_{\bar{q}}}\ms (1-x)^{\beta_{\bar{q}}} \,,
&
e_g(x) &= k_g \, x^{-\alpha_g}\ms (1-x)^{\beta_g} \,,
\end{align}
and for the $t$ dependence in the gluon sector we set
\begin{equation}
  \label{e-g-profile}
g_g(x) = \alpha'_g (1-x)^2 \ln\frac{1}{x} + D_g (1-x)^2 \,,
\end{equation}
in analogy to the form \eqref{gluon-profile} we used for $H^g$.  We
presently have not no phenomenological information on these
distributions, but two theoretical constraints.  There is a condition
that ensures positive semidefinite densities of partons in the
transverse plane \cite{Burkardt:2003ck}, which with our ansatz for the
GPDs reads~\cite{Diehl:2004cx}
\begin{align}
  \label{pos-cond}
\left[ \frac{e_{\bar{q}}(x)}{\bar{q}(x)} \right]^2 & \le
8\ms \mathrm{e}\ms m_p^2  \,
\left[ \frac{g_{\bar{q}}(x)}{f_{\bar{q}}(x)} \right]^3
  \bigl[ f_{\bar{q}}(x) - g_{\bar{q}}(x) \bigr] \,,
\nonumber \\
\left[ \frac{e_g(x)}{g(x)} \right]^2 & \le
8\ms \mathrm{e}\ms m_p^2 \,
\left[ \frac{g_g(x)}{f_g(x)} \right]^3
  \bigl[ f_g(x) - g_g(x) \bigr]
\end{align}
if we neglect for simplicity the polarized antiquark and gluon
distributions compared with the unpolarized ones.  On the other hand
we have the sum rule
\begin{align}
  \label{zero-sum-rule}
0 &= \int_{0}^1 dx\, E^g(x,0,0) + \sum_q \int_{-1}^1 dx\, x E^q(x,0,0)
\nonumber \\
&= \int_{0}^1 dx\, x e^g(x)
 + \sum_q \int_{0}^1 dx\, x \bigl[ e_{q_v}(x) + 2 e_{\bar{q}}(x)
   \bigr]
\end{align}
following from the conservation of the energy-momentum tensor.  For
the parameters in \eqref{e-g-profile} we take
\begin{align}
  \alpha'_g &= 0.164 \gev^{-2} \,, &
  D_g &= 1.08 \gev^{-2} \,,
\end{align}
with $\alpha'_g$ as in \eqref{g-prof-param} and $D_g$ slightly smaller
than its counterpart $B_g$ for $H^g$, so that the positivity condition
\eqref{pos-cond} can be fulfilled.
Assuming a similar small-$x$ behavior of the distributions for proton
helicity-flip and non-flip, we take in \eqref{eg-model} the values
$\alpha_{\bar{q}} = 1.25$ and $\alpha_g = 1.10$, which we obtain when
fitting the CTEQ6M distributions to a power law in the $x$ range from
$10^{-4}$ to $10^{-3}$.

\begin{figure}[t]
\begin{center}
\includegraphics[width=\plotwidth]
{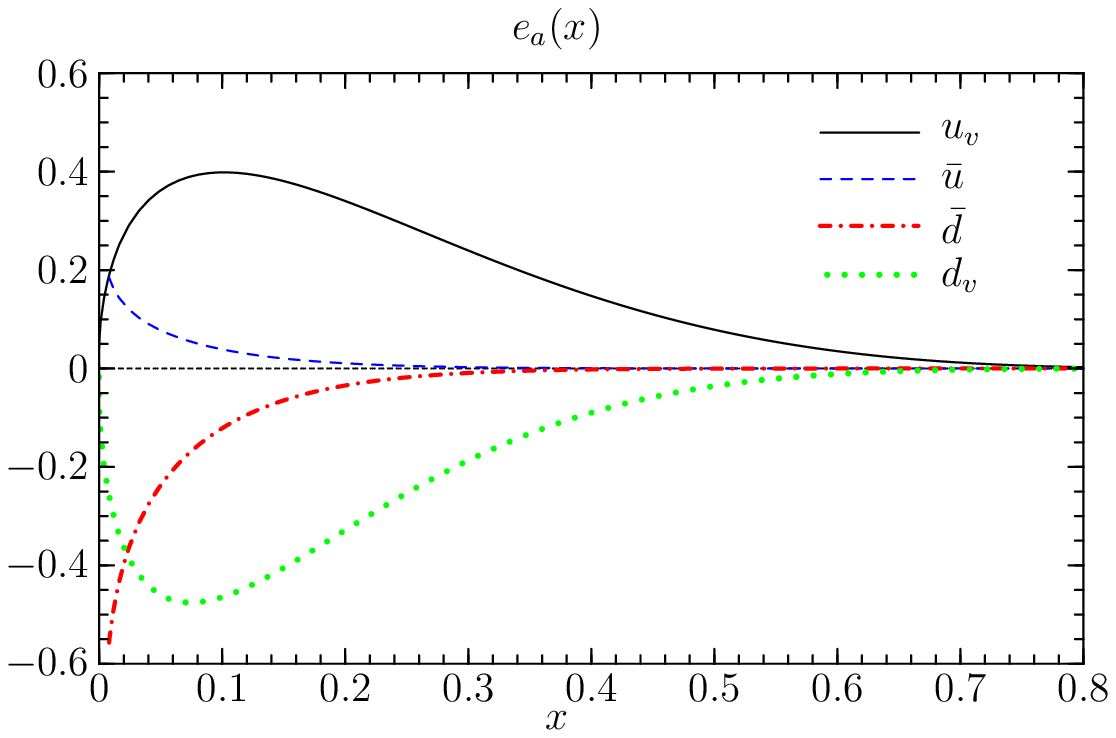}\hspace{1ex}
\includegraphics[width=\plotwidth]
{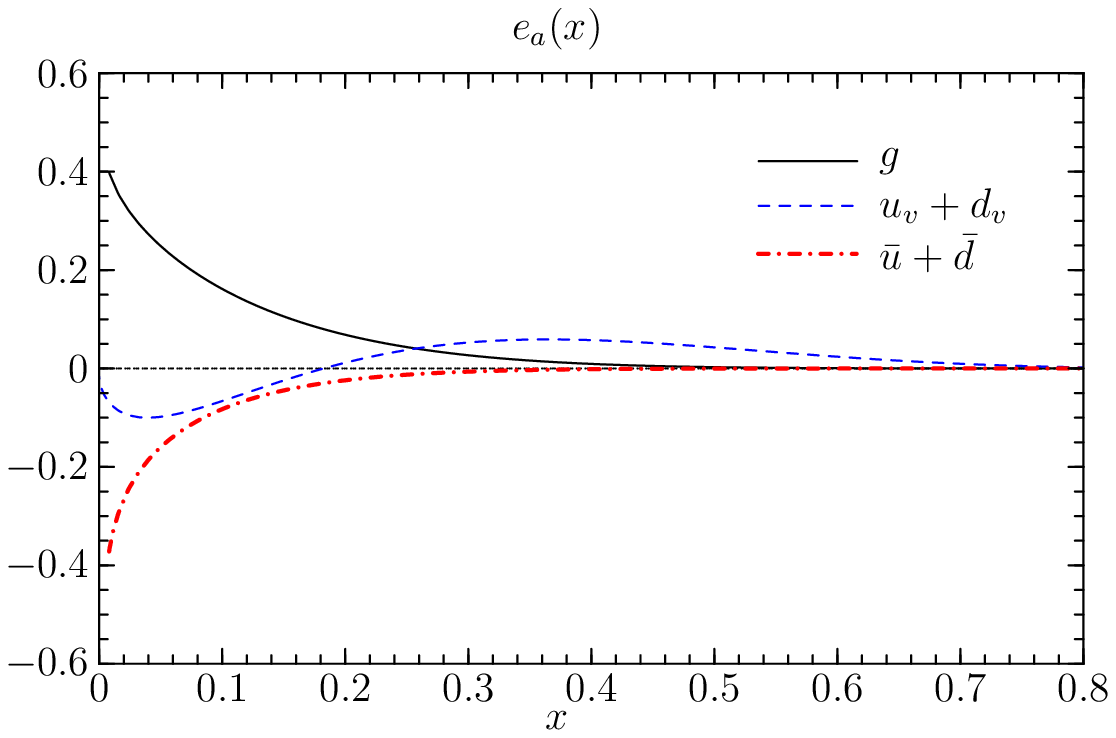}
\end{center}
\caption{\label{E-model} The forward limits $e_a(x)$ of the nucleon
  helicity-flip distributions at $\mu= 2\gev$ for different parton
  species $a$ in model 1.}
\end{figure}

Since it turns out that the transverse target polarization asymmetry
in $\rho$ production is very sensitive to the details of the
helicity-flip distributions, we will explore two model scenarios in
our numerical studies:
\begin{enumerate}
\item a scenario where the sea quark distributions $e_{\bar{q}}$
behave similarly to the valence distributions $e_{q_v}$.
For the $t$ dependence we then take $g_{\bar{u}}(x) = g_{u_v}(x)$ and
$g_{\smash{\bar{d}}}(x) = g_{d_v}(x)$.  The parameters $k_{\bar{q}}$
in \eqref{eg-model} are taken such that second moments at $t=0$ fulfill
\begin{equation}
  \label{val-scen}
\frac{\int_0^1 dx\, x e_{\bar{q}}(x)}{\int_0^1 dx\, x e_{q_v}(x)}
  = \frac{\int_0^1 dx\, x \bar{q}(x)}{\int_0^1 dx\, x q_v(x)}
\end{equation}
for $q = u,d$, where the ratio on the r.h.s.\ is taken from the CTEQ6M
parameterization at $\mu=2 \gev$.  Its value is $0.095$ for $u$ and
$0.30$ for $d$ quarks.  This fixes the values of $k_{\bar{q}}\,
N^{-1}(\alpha_{\bar{q}} -1, \beta_{\bar{q}})$ with $N$ given in
\eqref{norm-fact}.
For the strange distribution we set $e_s = e_{\bar{s}} =0$, and $k_g\,
N^{-1}(\alpha_g -1, \beta_g)$ is then fixed by the sum rule
\eqref{zero-sum-rule}.

The powers $\beta_{\bar{q}}$ and $\beta_g$ controlling the large-$x$
behavior are finally taken to have the smallest values for which the
positivity condition \eqref{pos-cond} holds in the range $x<0.9$ (for
higher $x$ even the unpolarized densities are so uncertain that we do
not insist on the positivity conditions to be fulfilled).
\item a scenario where  $e_{\bar{q}}$ behaves similarly to the gluon
distribution $e_g$ .
The $t$ dependence is now modeled by taking $g_{\bar{q}}(x) = g_g(x)$
for $q = u,d,s$.  For the second moments we impose
\begin{equation}
  \label{glu-scen}
\frac{\int_0^1 dx\, x e_{\bar{q}}(x)}{\int_0^1 dx\, x e_g(x)}
  = \frac{\int_0^1 dx\, x \bar{q}(x)}{\int_0^1 dx\, x g(x)}
\end{equation}
for the three light quark flavors, where with the CTEQ6M distributions
the r.h.s.\ is equal to $0.064$, $0.083$, $0.036$ for $u$, $d$, $s$,
respectively.  We now have a nonzero $e_{s} = e_{\bar{s}}$.  The
values of $k_{\bar{q}}\, N^{-1}(\alpha_{\bar{q}} -1, \beta_{\bar{q}})$
and $k_g\, N^{-1}(\alpha_g -1, \beta_g)$ are taken to fulfill both
\eqref{glu-scen} and \eqref{zero-sum-rule}, and the powers
$\beta_{\bar{q}}$, $\beta_g$ are set to the minimal values for which
positivity holds in the range $x<0.9$.
\end{enumerate}
The parameters resulting from this modeling procedure are collected in
Table~\ref{tab:params}, and the distributions at $\xi=0$ and $t=0$ for
model 1 are shown in Fig.~\ref{E-model}.

\begin{table}[t]
  \caption{\label{tab:params} Parameters in the ansatz
    \protect\eqref{eg-model} for different parton species $a$ in
    the two models described in the text.  The values for valence
    quarks apply to both models, with normalization parameters 
    given by $k_{q_v} = \kappa_q\ms N(\alpha_{v}, \beta_{q_v})$
    according to \protect\eqref{e-val-model}.  The last line gives the
    second Mellin moment at $\mu= 2\gev$ in the forward limit.}
\begin{center}
\renewcommand{\arraystretch}{1.3}
\renewcommand{\tabcolsep}{4pt}
\begin{tabular}{cllllllllll} \hline
 & & & \multicolumn{3}{c}{model 1} &
     & \multicolumn{4}{c}{model 2} 
\\ \cline{4-6} \cline{8-11}
 & ~~~~$u_v$ & ~~~~$d_v$ & ~~~~$\bar{u}$ & ~~~~$\bar{d}$ & ~~~~$g$ & 
 & ~~~~$\bar{u}$ & ~~~~$\bar{d}$ & ~~~~$\bar{s}$ & ~~~~$g$ \\ \hline
$\alpha_a$ & $\msp 0.55$ & $\msp 0.55$
           & $\msp 1.25$ & $\msp 1.25$ & $\msp 1.10$ &
           & $\msp 1.25$ & $\msp 1.25$ & $\msp 1.25$ & $\msp 1.10$ \\
$\beta_a$ 
  & $\msp 3.99$ & $\msp 5.59$ & $\msp 9.6$ & $\msp 9.2$ & $\msp 6.7$ &
  & $\msp 7.6$ & $\msp 6.5$ & $\msp 5.5$ & $\msp 2.5$ \\
$k_a$ & $\msp 1.71$ & $-2.36$ & $\msp 0.06$ & $-0.18$ & $\msp 0.26$ &
      & $-0.0016$ & $-0.0018$ & $-0.0007$ & $-0.017$
\\ \hline
$\int_0^1 dx\, x e_a(x)$ & $\msp 0.138$ & $-0.130$
            & $\msp 0.013$ & $-0.039$ & $\msp 0.044$ &
            & $-0.0004$ & $-0.0005$ & $-0.0002$ & $-0.0059$ \\ \hline
\end{tabular}
\end{center}
\end{table}

We find that in model 2, both sea quark and gluon distributions are
nearly zero (so that we do not attach importance to the
unrealistically small value of $\beta_g$ obtained with our above
procedure).  Their smallness can be traced back to the small value of
the flavor singlet integral
\begin{equation}
\label{e-singlet-int}
\int_0^1 dx\, x \bigl[ e_{u_v}(x) + e_{d_v}(x) \bigr] = 0.008
\end{equation}
in the valence sector of our ansatz.  In model 2, the distributions
$e_{\bar{q}}$ and $e_g$ have the same sign as a consequence of
\eqref{glu-scen} and due to the sum rule \eqref{zero-sum-rule} can
only be tiny.  Somewhat larger distributions for sea quarks and gluons
are obtained in model 1, where they have opposite sign because of
\eqref{val-scen}.

We note that the parameters \eqref{DFJK4-E-par} we have taken for the
valence part of $E^q$ are by no means precisely determined by a fit to
the Pauli form factors: alternative fits in \cite{Diehl:2004cx} gave a
similarly good description of the form factor data, with some
variation of the resulting value of the integral in
\eqref{e-singlet-int}.  Nevertheless, any model where $e_{u_v}$ and
$e_{d_v}$ have similar shapes and no zeroes in $x$ will yield rather
small values of this integral, given the strong cancellation between
$u$ and $d$ quark contributions in the moment $\int_0^1 dx\, \bigl[
e_{u_v}(x) + e_{d_v}(x) \bigr] = \kappa_u + \kappa_d \approx -0.36$.
It would be interesting to explore how much the integral
\eqref{e-singlet-int} and as a consequence the sea quark and gluon
distributions can vary in realistic models, but such an investigation
is beyond the scope of this work.

We end this section by quoting the values for the total angular
momentum carried by quarks and antiquarks of a given flavor in our
model, given by
\begin{equation}
J_{q} = \frac{1}{2} \int_{-1}^1 dx\, x 
  \bigl[ H^q(x,0,0) + E^q(x,0,0) \bigr]
\end{equation}
according to Ji's sum rule \cite{Ji:1996ek}.  With the parameters in
Table~\ref{tab:params} and the CTEQ6M distributions we find
\begin{align}
 J_u &= 0.25 \,, & J_d &= -0.01 \,,
& & \text{(model 1)}
\nonumber \\
 J_u &= 0.24 \,, & J_d &= \msp 0.03
& & \text{(model 2)}
\end{align}
at the scale $\mu= 2\gev$ of our model.  We note that this is in
rather good agreement with the results of recent lattice calculations,
with $J_u = 0.214(16)$ and $J_d = -0.001(16)$ reported in
\cite{Hagler:2007xi}, and $J_u = 0.33(2)$ and $J_d = -0.02(2)$ in
\cite{Schierholz:2007}.  Let us reiterate that with just two sets of
model parameters we cannot exhaust the range of possible scenarios but
only provide two representatives that are consistent with presently
known constraints.  As just discussed, the relative smallness of sea
quark and gluon distributions compared with the nucleon helicity
conserving case should however be typical of a rather wide class of
models.


\section{Vector meson production at small $x_B$}
\label{sec:small-x}

We now study numerically the importance of NLO corrections in vector
meson production.  Here and in the following sections we use the
two-loop strong coupling for $n_f=3$ flavors with a QCD scale
parameter $\Lambda^{(3)} = 226 \mev$.  This value corresponds to
$\Lambda^{(4)} = 326 \mev$, $\Lambda^{(5)} = 372 \mev$ and to
$\alpha_{\smash{s}}^{(5)}(M_Z) = 0.118$ when matching at $m_c=1.3\gev$
and $m_b=4.5\gev$, which are the values used in the CTEQ6M parton
analysis \cite{Pumplin:2002vw}.
We also take $n_f=3$ fixed in the evolution and the hard-scattering
kernels.  Taking $n_f=4$ with massless charm or $n_f=5$ with massless
charm and bottom would not be a good approximation for the rather
moderate values of $Q^2$ we will discuss for fixed-target kinematics.
On the other hand, taking $n_f=3$ and neglecting charm altogether is
admittedly not a good approximation for the larger $Q^2$ relevant in
collider kinematics.  However, with $\alpha_{\smash{s}}^{(3)} = 0.164$
compared to $\alpha_{\smash{s}}^{(5)} = 0.178$ at $\mu=10\gev$ we
expect that this inaccuracy will not affect the conclusions at high
$Q^2$ we shall draw from our studies.

We have performed the evolution of the GPDs at LO using the
momentum-space evolution code of \cite{Vinnikov:2006xw}.  As explained
in Sect.~\ref{sec:kernels}, taking LO evolution together with the NLO
hard-scattering kernels is sufficient to obtain scale independence of
the process amplitude up to uncalculated corrections of order
$\alpha_s^3$.  With the input scale of evolution not taken too small,
NLO evolution effects should be rather moderate at the $Q^2$ values
relevant in fixed-target kinematics, whereas our general conclusions
for high $Q^2$ and small $x_B$ will again not depend on this level of
detail.  We note that the NLO kernels in momentum space are available
in the literature \cite{Belitsky:1999hf}, but their considerable
length makes it difficult to implement them in a fast numerical
evaluation.  For including NLO effects in the evolution it should be
more efficient to use the Mellin space approach recently followed for
deeply virtual Compton scattering in \cite{Kumericki:2007sa}.

Here and in the following section we consider the convolutions of
hard-scattering kernels with GPDs at $t=0$.  For nonzero $\xi = x_B
/(2-x_B)$ this should be understood in the sense of an analytic
continuation, since the physical region for meson production is $-t
\ge 4 m_p^2\ms \xi^2/(1-\xi^2)$ in Bjorken kinematics.  To explore the
importance of NLO corrections we do not see this as a shortcoming.

\begin{figure}
\begin{center}
\includegraphics[width=\plotwidth]
{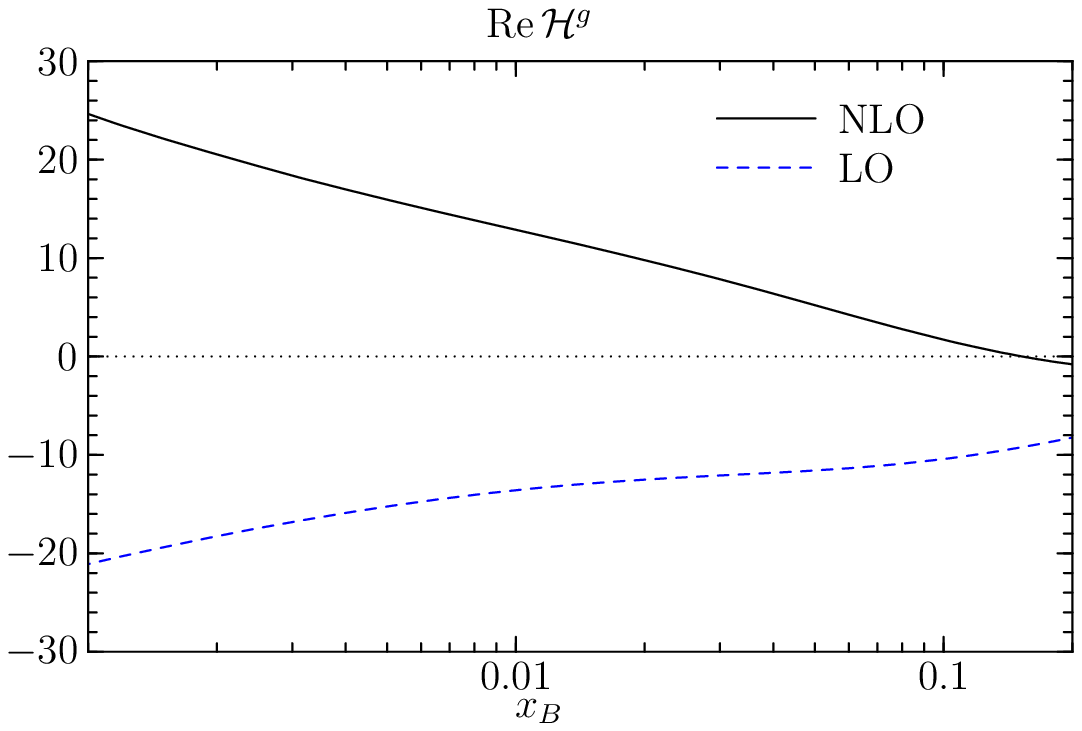}\hspace{1ex}
\includegraphics[width=\plotwidth]
{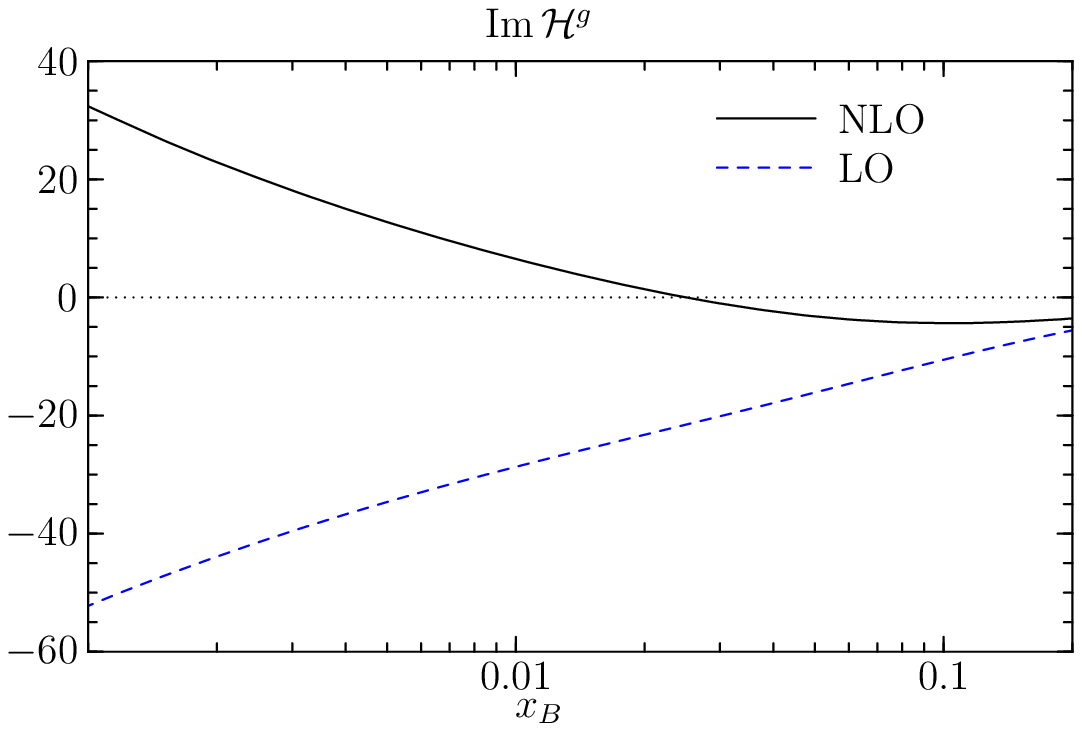}\\[1.05ex]
\includegraphics[width=\plotwidth]
{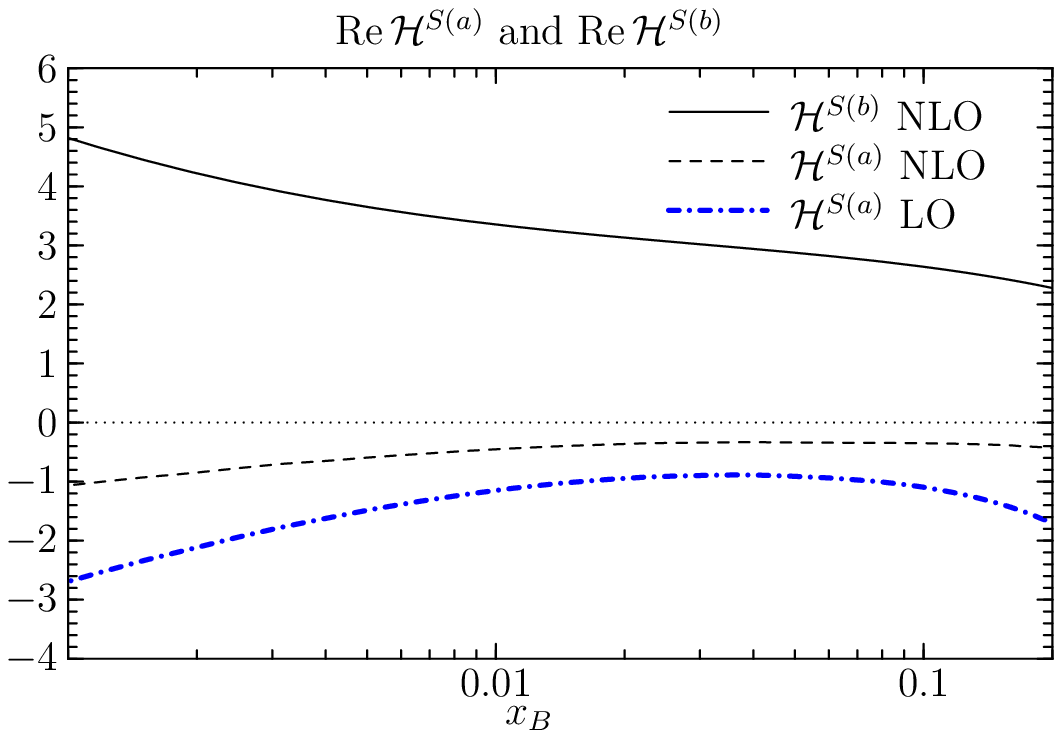}\hspace{1ex}
\includegraphics[width=\plotwidth]
{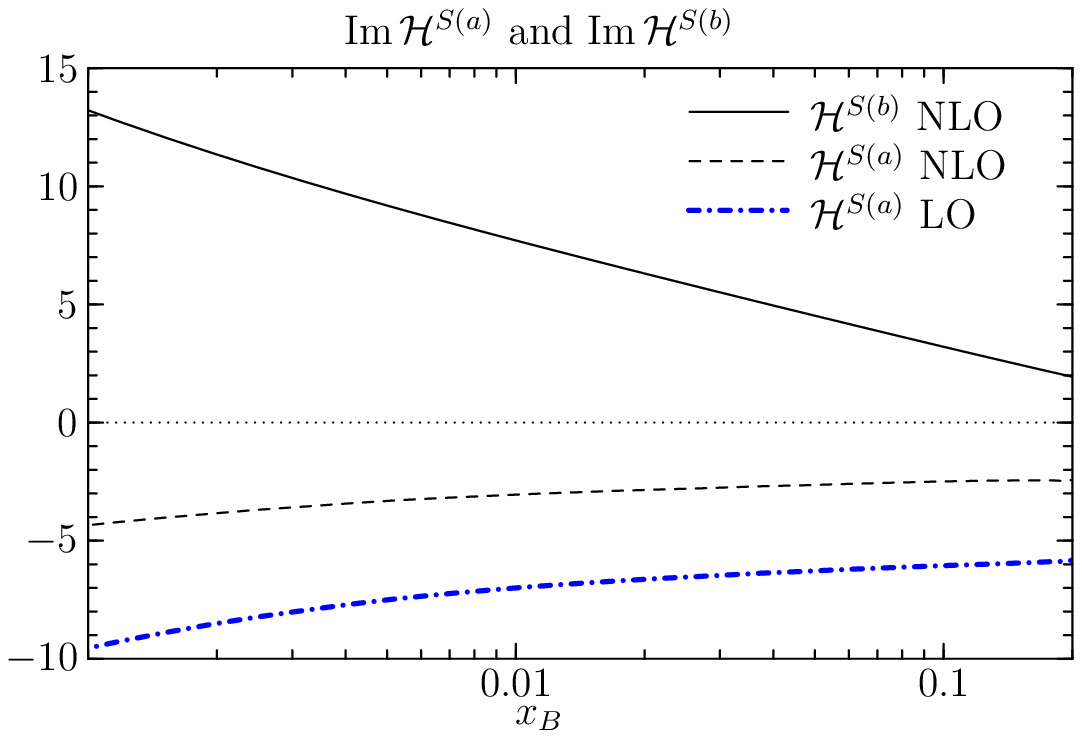}\\[1.05ex]
\includegraphics[width=\plotwidth]
{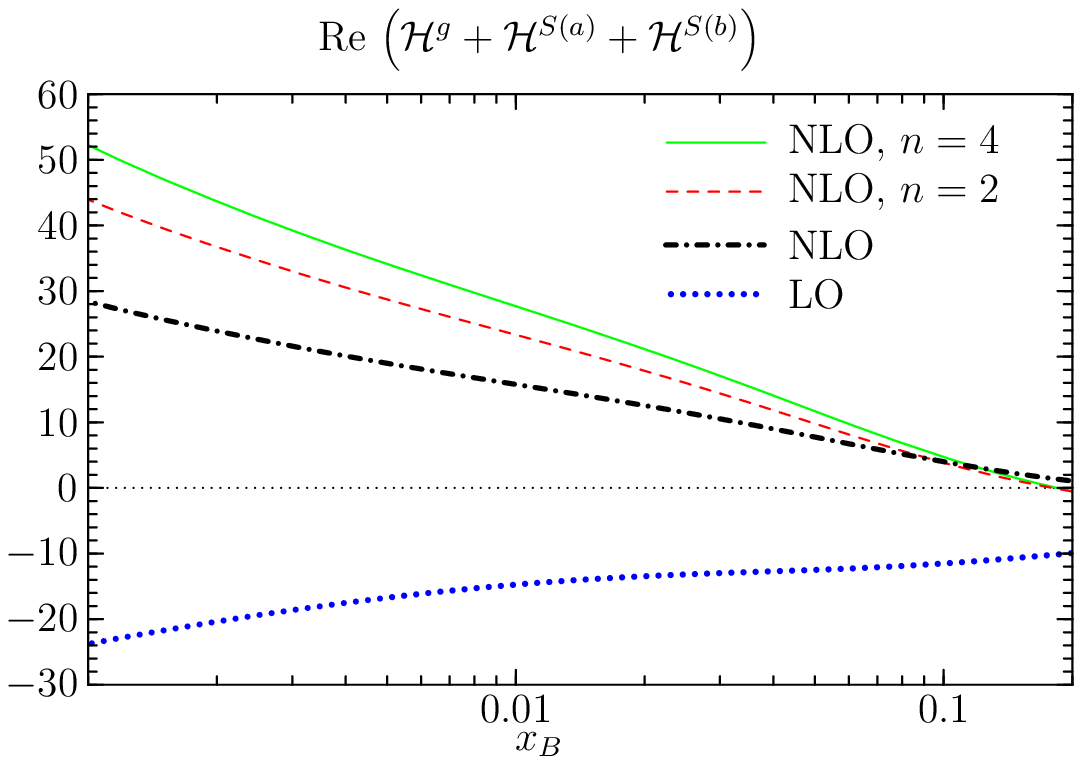}\hspace{1ex}
\includegraphics[width=\plotwidth]
{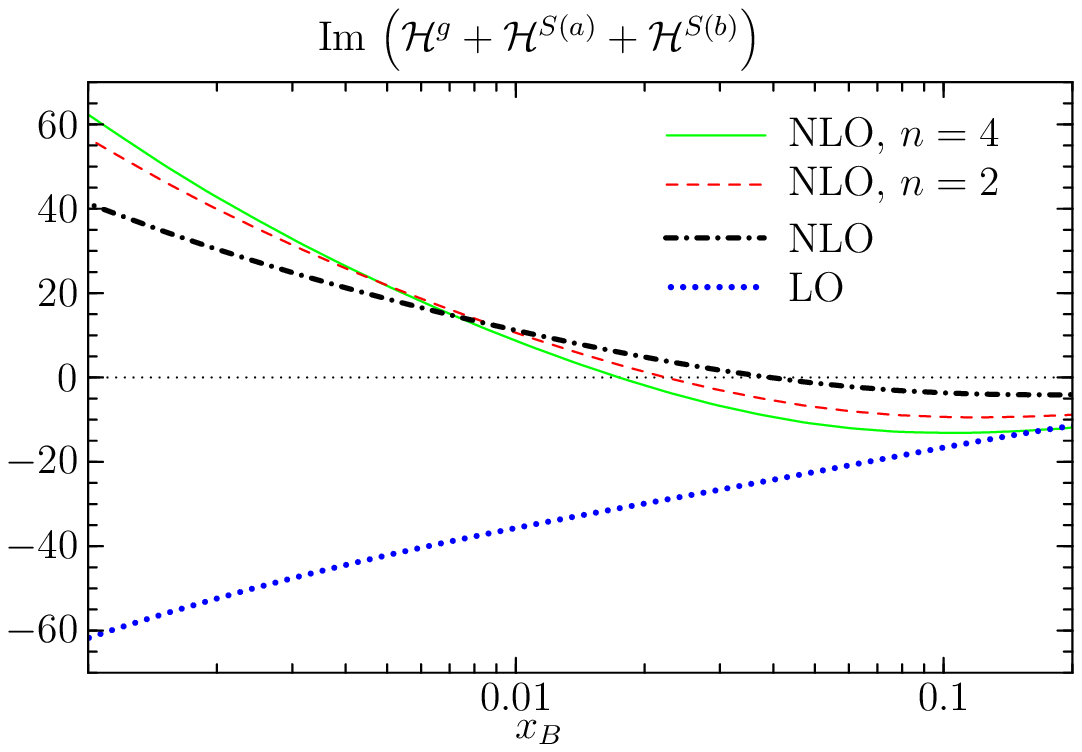}
\end{center}
\caption{\label{LO_NLO_comparison_small_x} LO and NLO terms of the
  convolutions in the gluon and quark singlet sector at $Q = 4 \gev$.
  The scales are set to $\mu_R = \mu_{GPD} = \mu_{DA} = Q$.  The NLO
  terms are for Gegenbauer index $n=0$ unless specified explicitly.
  Here and in the following plots the label ``NLO'' denotes the
  $O(\alpha_s^2)$ part of the convolutions, whereas the sum of
  $O(\alpha_s)$ and $O(\alpha_s^2)$ terms is labeled by ``LO+NLO''.}
\end{figure}

\begin{figure}
\begin{center}
\includegraphics[width=\plotwidth]
{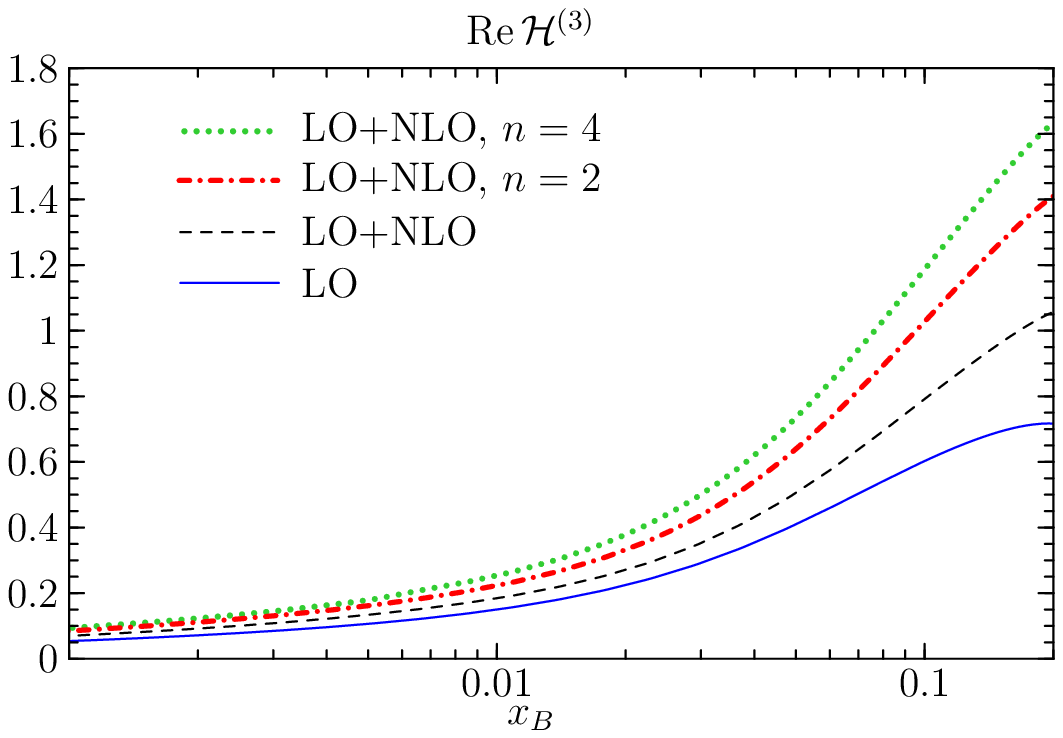}\hspace{1ex}
\includegraphics[width=\plotwidth]
{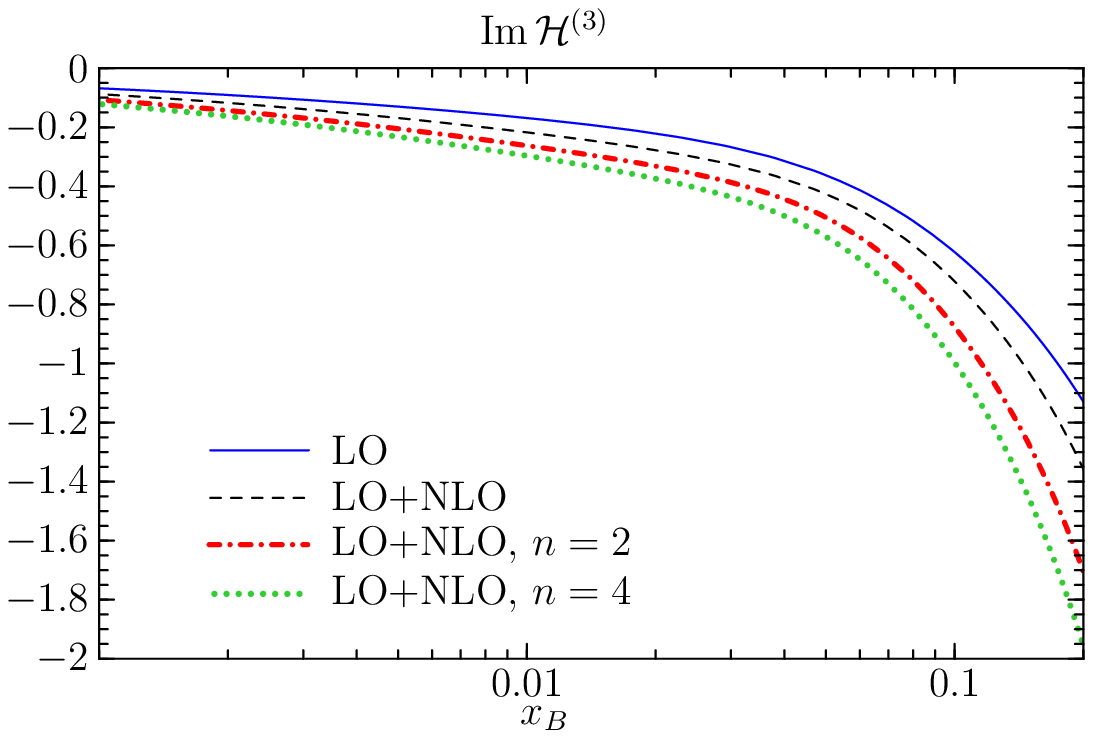}\\[0.8ex]
\includegraphics[width=\plotwidth]
{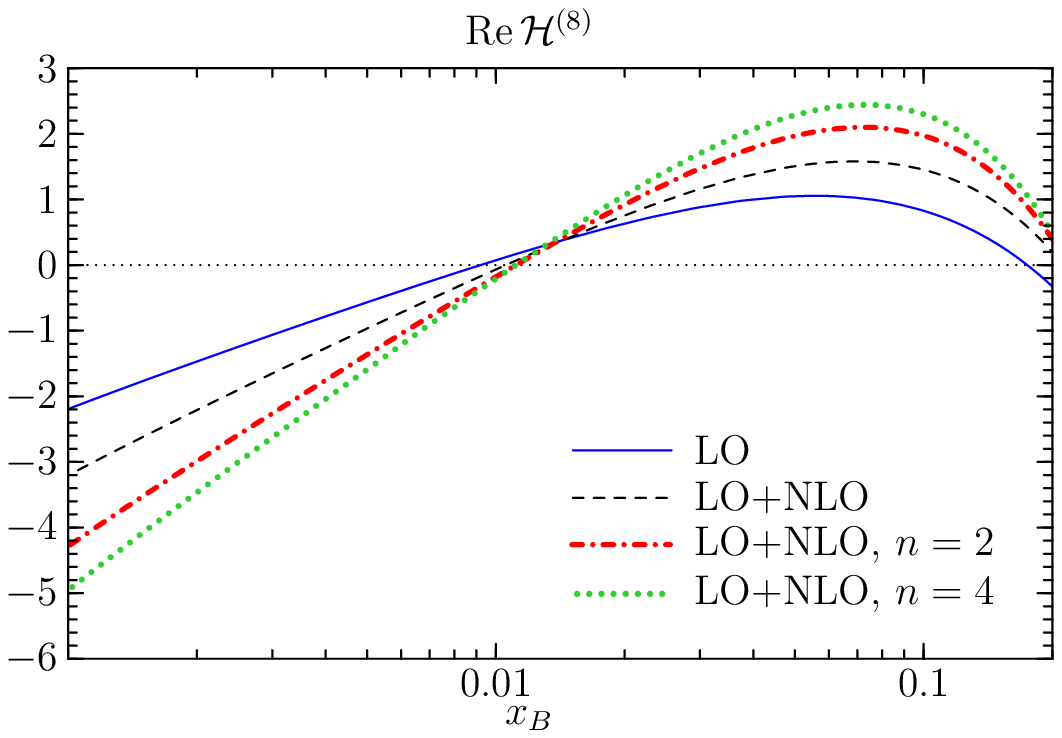}\hspace{1ex}
\includegraphics[width=\plotwidth]
{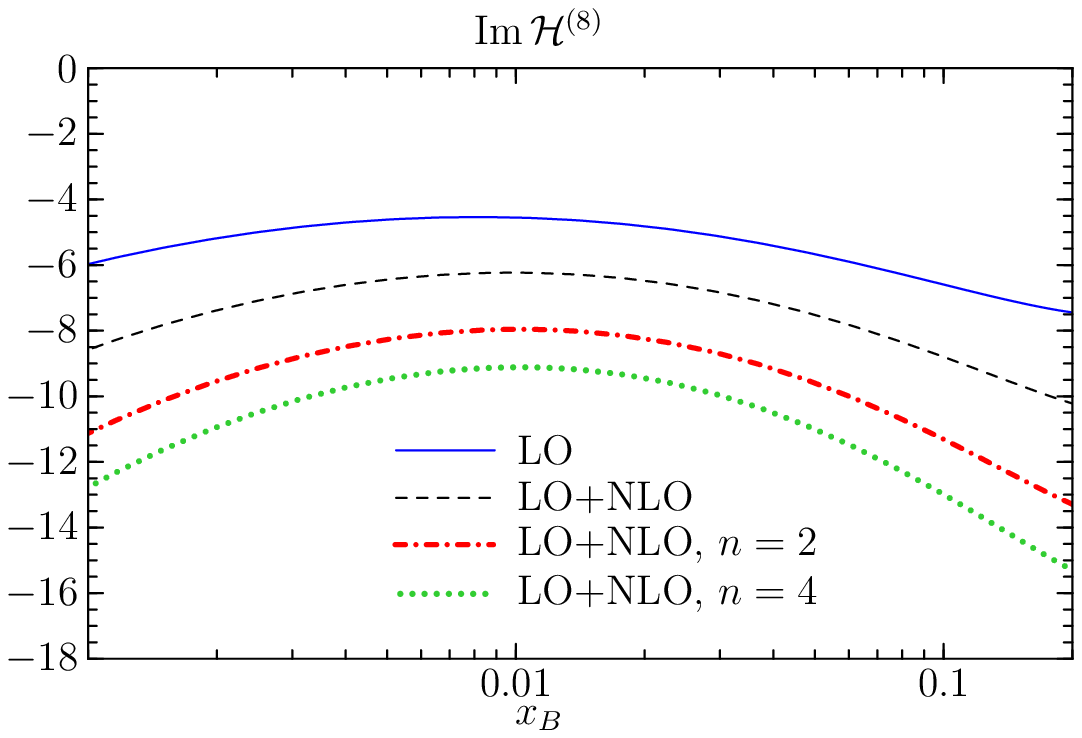}
\end{center}
\caption{\label{small_x_nonsinglet_LO_NLO} LO terms and the sum of LO
  and NLO terms of the convolutions in the quark non-singlet sector at
  $Q = 4 \gev$, with $\mu_R = \mu_{GPD} = \mu_{DA} = Q$.}
\end{figure}

\begin{figure}
\begin{center}
\includegraphics[width=\plotwidth]
{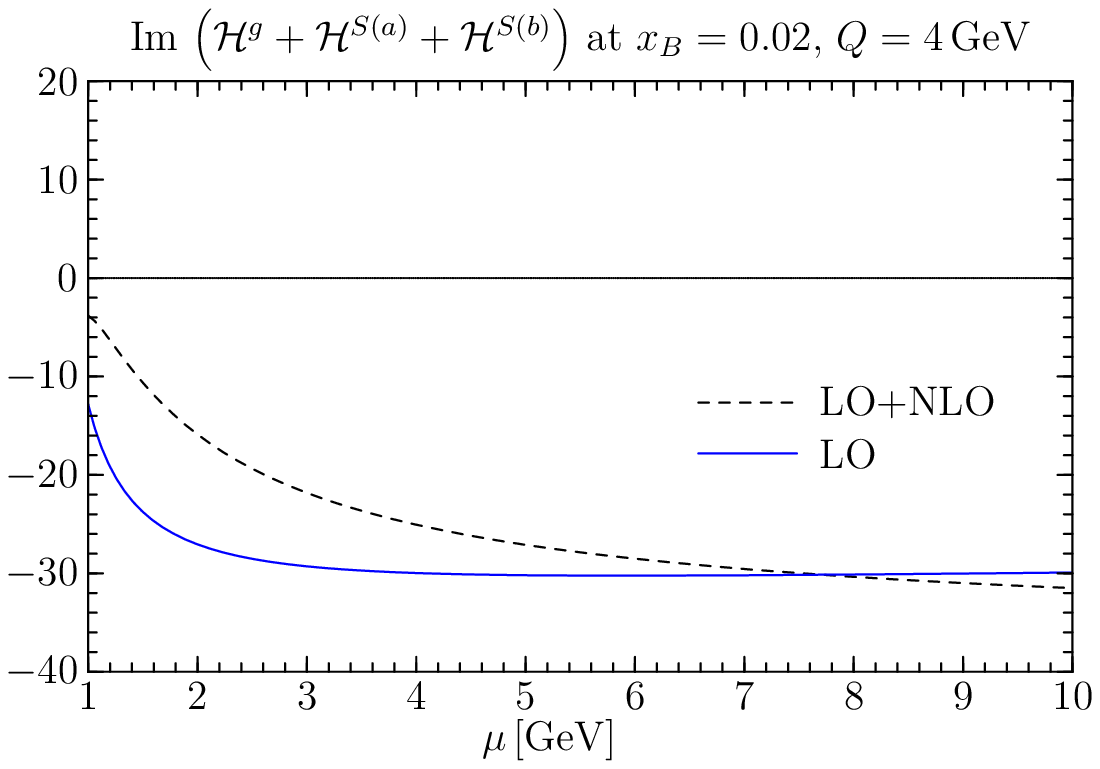}\hspace{1ex}
\includegraphics[width=\plotwidth]
{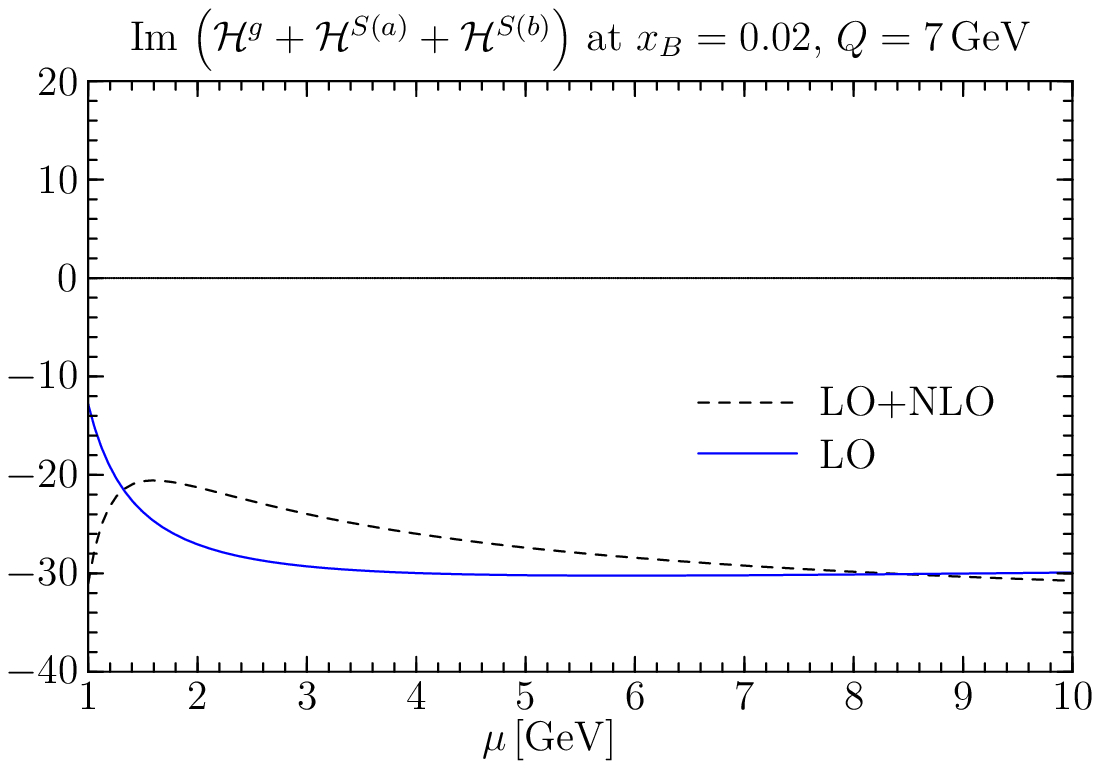}\\[1.5ex]
\includegraphics[width=\plotwidth]
{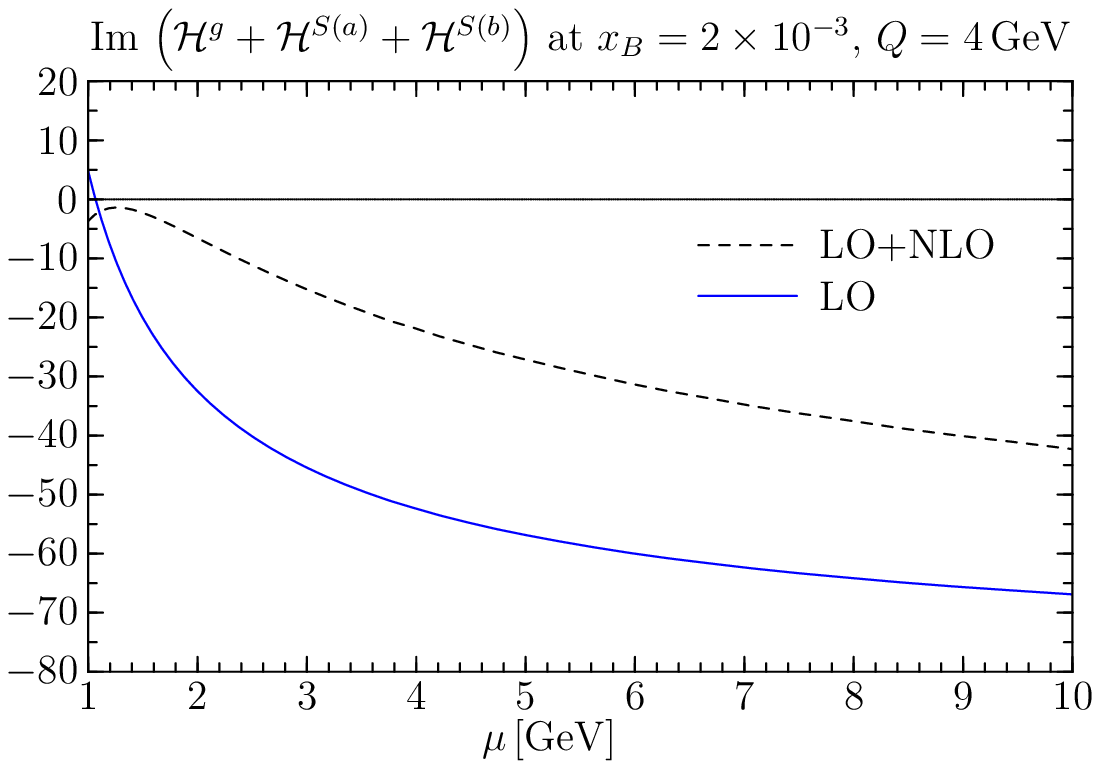}\hspace{1ex}
\includegraphics[width=\plotwidth]
{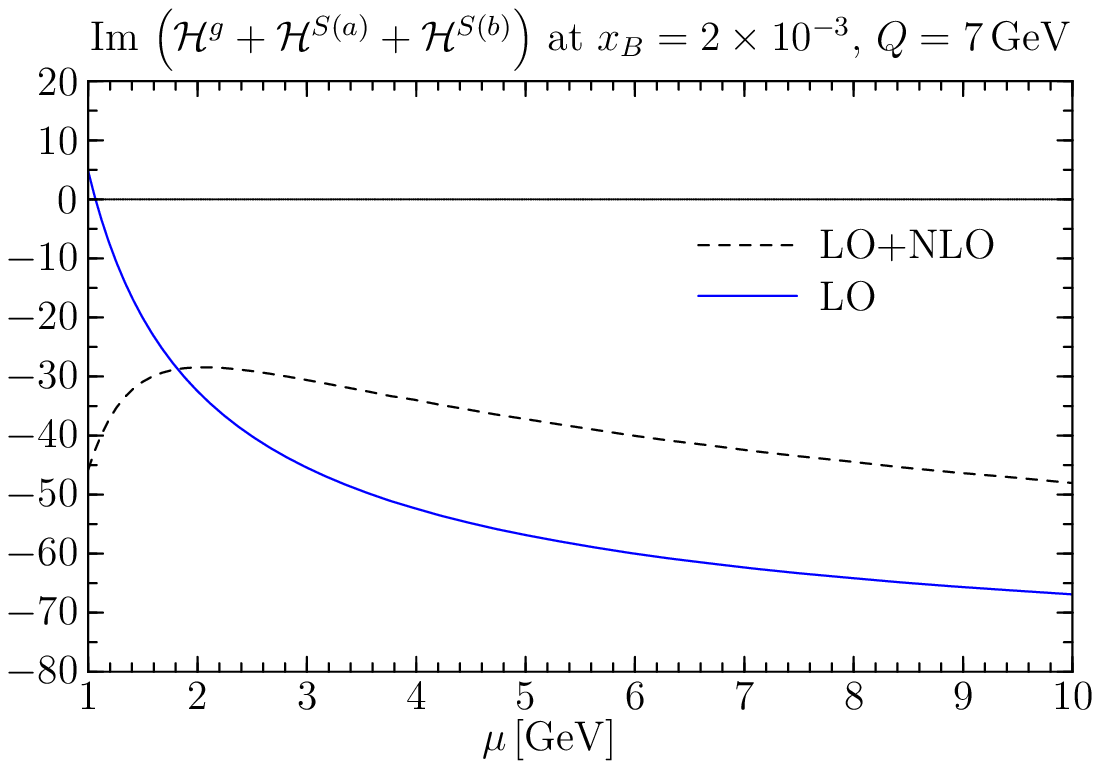}\\[1.5ex]
\includegraphics[width=\plotwidth] 
{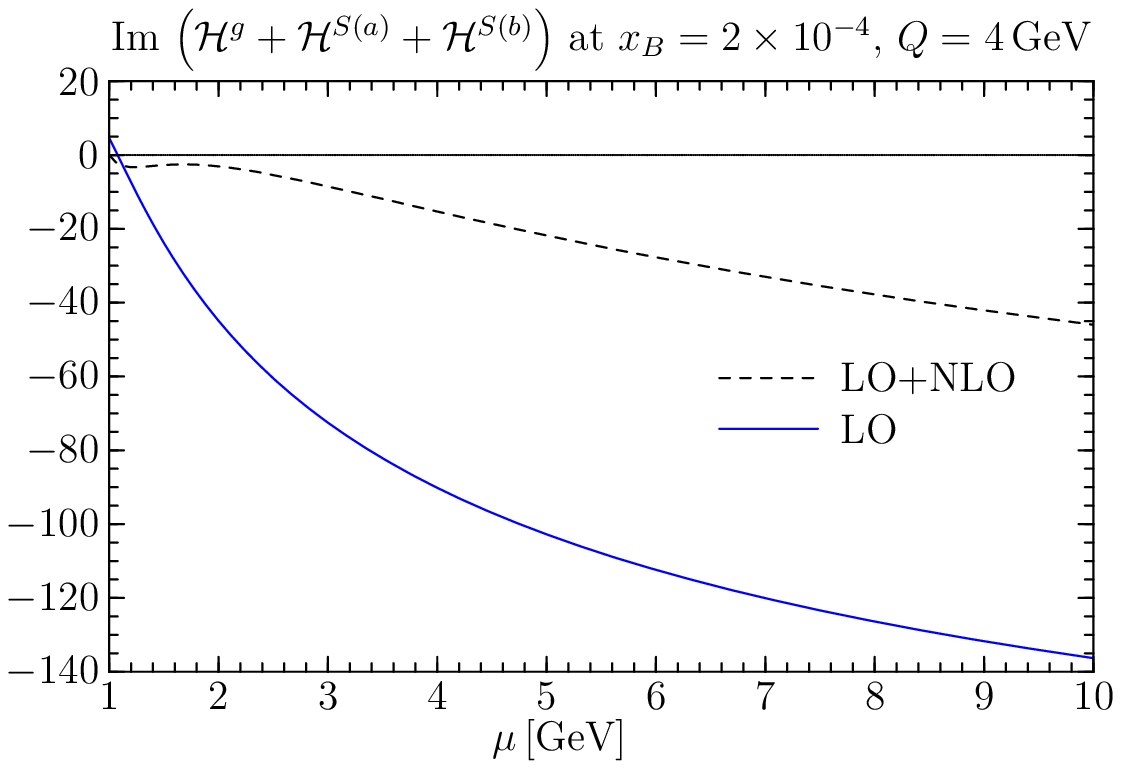}\hspace{1ex}
\includegraphics[width=\plotwidth]
{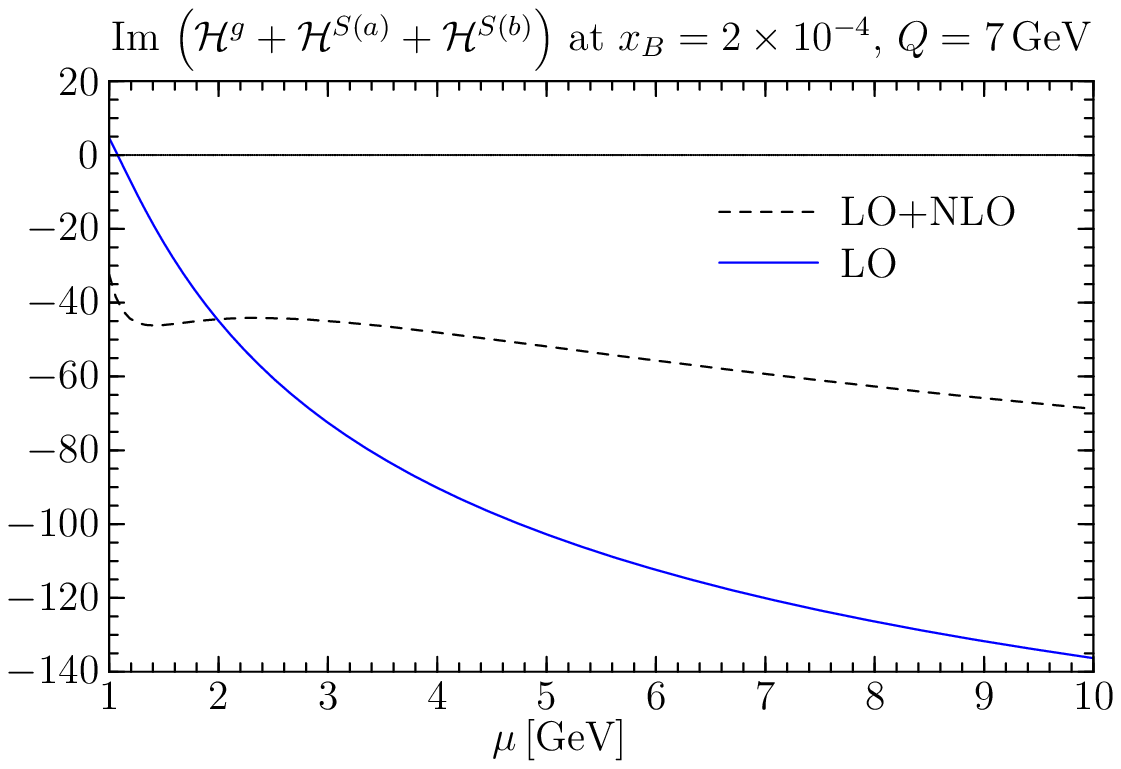}
\end{center}
\caption{\label{imag_diff_Q2} Dependence on the common scale $\mu =
  \mu_R = \mu_{GPD}$ for the sum of convolutions in the gluon and
  quark singlet sector.}
\end{figure}

Let us start our discussion with the gluon and quark singlet sector.
Here and in following we shall always present the convolutions
\eqref{F-def} for Gegenbauer index $n=0$ unless indicated otherwise.
In Fig.~\ref{LO_NLO_comparison_small_x} we show the LO and NLO
pieces of the convolutions for the scale choice $\mu_R = \mu_{GPD} =
\mu_{DA} = Q$.  The size of corrections at small $x_B$ is dramatic: we
have large NLO corrections with opposite sign compared to the LO term
for $\mathcal{H}^g$, and a similarly large NLO contribution from
$\mathcal{H}^{S(b)}$ with sign opposite to the LO result for
$\mathcal{H}^{S(a)}$.  In the sum of gluon and quark singlet terms,
the NLO corrections drastically reduce the LO result or even lead to a
change of sign between LO and the sum of LO and NLO results.  We also
observe that for higher Gegenbauer index the NLO corrections tend to
be even more important.  Note that the LO term of the convolutions is
the same for all $n$ as can be seen from \eqref{F-def} and
\eqref{gegen-kernels}.
The size of NLO corrections in $\mathcal{H}^{S(a)}$ is comparatively
moderate, at least for lower Gegenbauer moments.  The same is seen for
the quark non-singlet convolutions in
Fig.~\ref{small_x_nonsinglet_LO_NLO}.  Of course, the gluon and quark
singlet terms will dominate meson production at small $x_B$ in those
channels where it is allowed by the meson quantum numbers.

In Fig.~\ref{imag_diff_Q2} we explore the influence of the scale
choice by varying $\mu_R = \mu_{GPD}$ simultaneously.  For $x_B=
2\times 10^{-3}$ we find an indication for the onset of perturbative
stability at $Q= 7\gev$ but not yet at $Q= 4\gev$.  For $x_B= 2\times
10^{-2}$ the situation is less severe, with moderate corrections in a
wide $\mu$ range already at $Q= 4\gev$.  In contrast, when going down
to $x_B= 2\times 10^{-4}$ we find very large corrections even at $Q=
7\gev$.  We have checked that the conclusions in the respective
kinematics do not change when we vary $\mu_{GPD}$ while keeping $\mu_R
= Q$ fixed.

\begin{figure}
\begin{center}
  \includegraphics[width=\plotwidth]{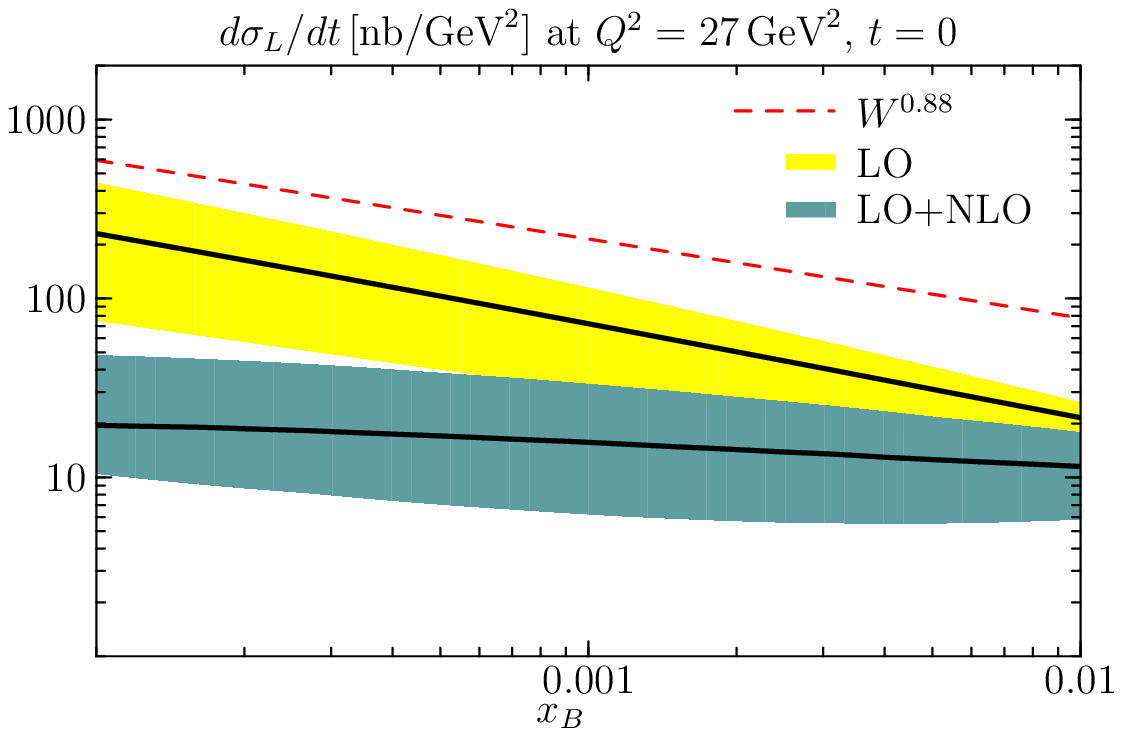}
\end{center} 
\caption{\label{cs_rho_zeus} Cross section for $\gamma^* p\to \rho\ms
  p$ with longitudinal photon polarization.  Bands correspond to the
  range $Q/2 < \mu < 2Q$ and solid lines to $\mu=Q$.  We also show the
  power-law behavior $\sigma \propto W^{0.88}$ (with arbitrary
  normalization) obtained from a fit to data in the range $0.001
  \protect\lsim x_B \protect\lsim 0.005$ \protect\cite{ZEUS:2001}.}
\end{figure}

Figure~\ref{cs_rho_zeus} shows how the perturbative instability we
observed in the convolutions affects the longitudinal cross section
for $\rho$ production.  Here we have taken the asymptotic form of the
meson distribution amplitude, i.e.\ set $a_n=0$ for $n\ge 2$.  In the
NLO result for the cross section we have squared the coherent sum of
LO and NLO terms in the process amplitude,\footnote{%
  We thus keep terms of $O(\alpha_s^3)$ in the cross section, although
  the accuracy of the NLO calculation is only up to $O(\alpha_s^2)$.
  This should not be seen as a problem, as it will not make a
  considerable difference in situations where perturbative corrections
  are moderate, whereas in situations where NLO corrections are huge
  we would neither trust the cross section with or without the
  partially included $O(\alpha_s^3)$ terms.}
i.e.\ we have taken $|\mathcal{M}_{\mathrm{LO}} +
\mathcal{M}_{\mathrm{NLO}}|^2$.  We see that the NLO corrections
severely decrease the LO result.  As a consequence of the
cancellations between LO and NLO contributions, the scale dependence
of the cross section does not decrease.  We also show in the figure
the power-law behavior $\sigma \propto W^{0.88}$ obtained from a fit
to data in the range $0.001 \protect\lsim x_B \protect\lsim 0.005$
\protect\cite{ZEUS:2001}.  As observed in \cite{Goloskokov:2006hr}, a
double distribution model with the CTEQ6M distributions as input lead
to a rather good description of this experimentally observed energy
dependence if the cross section is evaluated at LO.  With the strong
cancellations from the $O(\alpha_s^2)$ corrections, one obtains an NLO
result whose energy behavior is much too weak.

\begin{figure}
\begin{center}
\includegraphics[width=\plotwidth]
{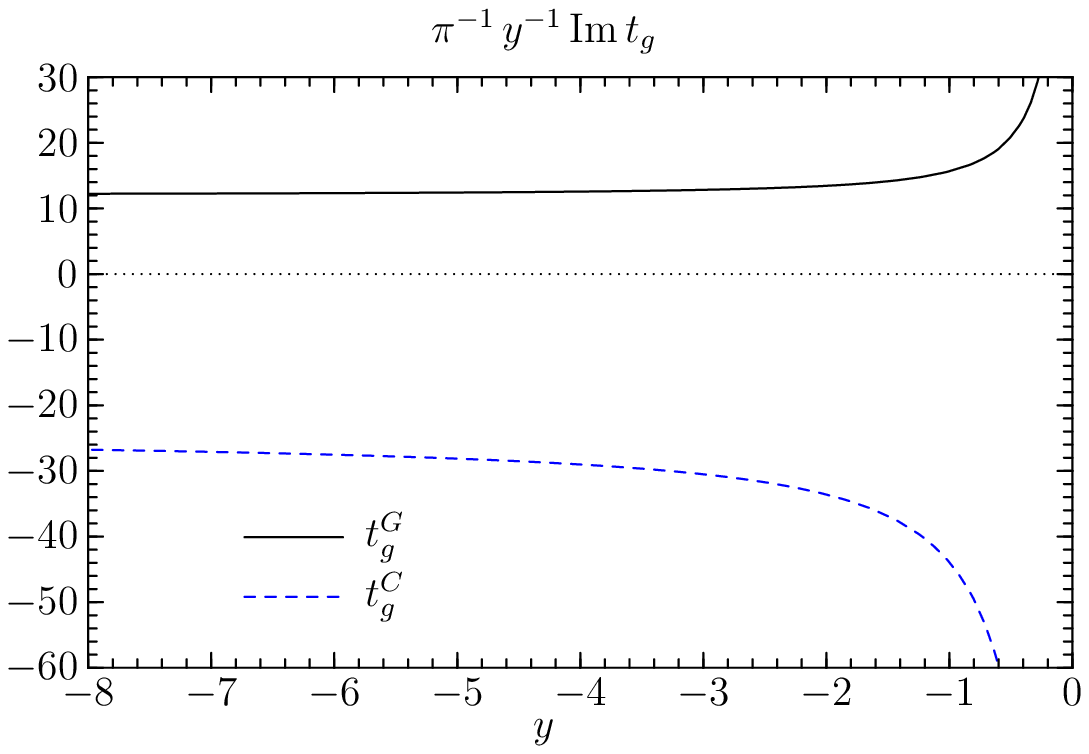}\hspace{1ex}
\includegraphics[width=\plotwidth]
{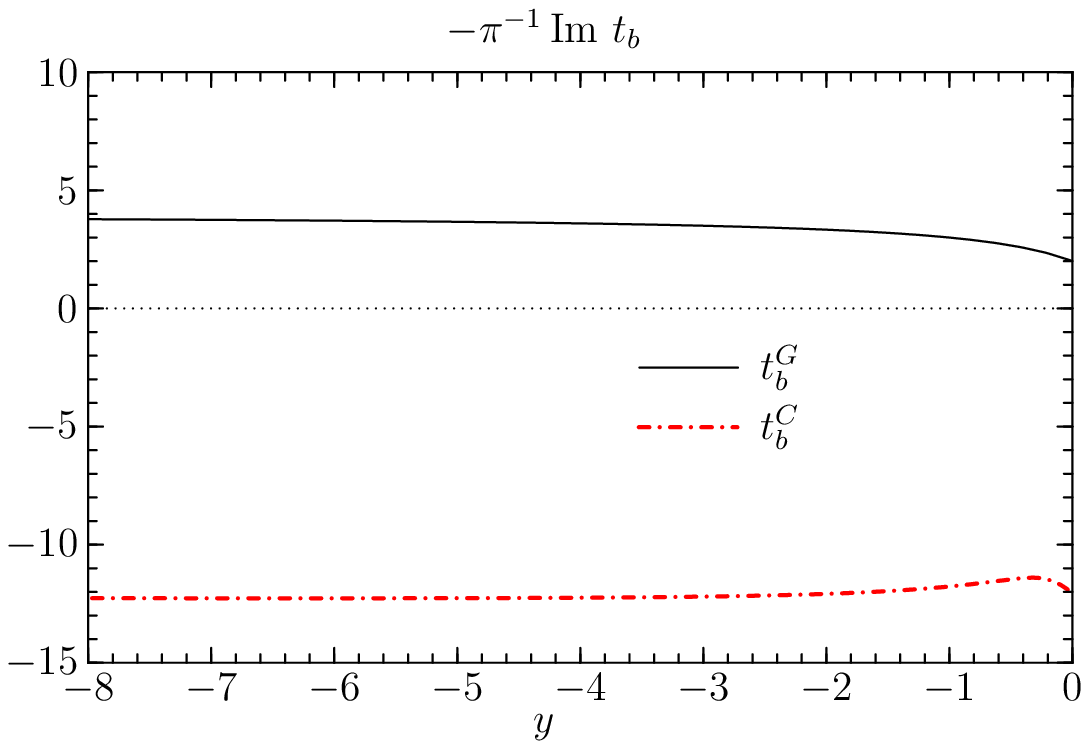}
\end{center}
\caption{\label{fig:kernels} The factorization scale dependent and
  independent terms of $\im t_{g,n}$ and $\im t_{b,n}$ as specified in
  \protect\eqref{kernels-CG}, shown for $n=0$ and $y<0$.}
\end{figure}

Let us discuss how the huge size of corrections can be understood at
an analytical level, following the line of argument given in
\cite{Ivanov:2004zv,Ivanov:2004vd}.
Using \eqref{poly-asy1} and \eqref{poly-asy2} we can approximate the
hard-scattering kernels for large negative $y$ as
\begin{align}
  \label{asy-g}
\frac{1}{\pi}\, \im \mathcal{I}_g(z,y)
&= 4 C_A \biggl[ \ln(z\zb) + \ln \frac{Q^2}{\mu_{GPD}^2} \biggr] \ms y
+ O(1) \,,
&
\re \mathcal{I}_g(z,y)
&= O(1) \,,
\nonumber \\[0.2em]
\frac{1}{\pi} \im \mathcal{I}_b(z,y) &= 4 \biggl[
  1 - \ln(z\zb) - \ln \frac{Q^2}{\mu_{GPD}^2} \biggr]
+ O\bigl( y^{-1} \bigr) \,,
&
\re \mathcal{I}_b(z,y) &= O\bigl( y^{-1} \bigr) \,,
\end{align}
where here and in the following the order of corrections is given up
to powers of $\ln\yb$.  The quark non-singlet kernel is subleading
compared with the pure singlet one,
\begin{align}
  \label{asy-a}
\frac{\mathcal{I}_a(\zb, y)}{y} \sim
\frac{\mathcal{I}_a(z, \yb)}{\yb} \sim
O\bigl( y^{-1} \bigr) \,,
\end{align}
where we have divided $\mathcal{I}_a(\zb, y)$ by $y$ corresponding to
the prefactor in the complete kernel \eqref{orig_kernels}.  {}From
\eqref{asy-g} we readily obtain
\begin{align}
  \label{tgb-approx}
\frac{1}{\pi} \im t_{g,n}(y) &= - 4 C_A
  \biggl[ c_n - \ln\frac{Q^2}{\mu^2_{GPD}} \biggr] \ms y
+ O(1) \,,
\nonumber \\
\frac{1}{\pi} \im t_{b,n}(y) &=
4 \biggl[ c_n+1 - \ln\frac{Q^2}{\mu^2_{GPD}} \biggr]
+ O\bigl( y^{-1} \bigr)
\end{align}
with constants
\begin{align}
c_{0} &= 2 \,,  &
c_{2} &= \frac{11}{3} \approx 3.7 \,, &
c_{4} &= \frac{137}{30} \approx 4.6 \,, &
c_{n} &= - \int_0^1 dz\, \ln(z\zb)\, C_{n}^{3/2}(2z-1)
\end{align}
that increase with the Gegenbauer index $n$.  
In Fig.~\ref{fig:kernels} we show for the case $n=0$ that these
approximations become very good for increasing $|y|$, where we have
decomposed the exact kernels as
\begin{align}
  \label{kernels-CG}
\im t_{g,n}^{}(y) &= \im t_{g,n}^{C}(y) 
  + \im t_{g,n}^{G}(y)\ms \ln\frac{Q^2}{\mu^2_{GPD}} \,,
\nonumber \\
\im t_{b,n}^{}(y) &= \im t_{b,n}^{C}(y) 
  + \im t_{b,n}^{G}(y)\ms \ln\frac{Q^2}{\mu^2_{GPD}} \,.
\end{align}

Let us now rewrite the convolutions of kernels and GPDs in terms of
the variable $\omega = x/\xi$,
\begin{align}
  \label{small-x-con}
& \im\ms \Bigl[ \mathcal{H}^g_n + \mathcal{H}^{S(b)}_n \Bigr]
= - 6 \alpha_s\, 
\biggl[ \pi H^g(\xi,\xi,t)
+ \frac{\alpha_s}{4\pi} \im \int_0^{1/\xi} d\omega
\nonumber \\[0.3em]
& \qquad \times \biggl\{
  \frac{2}{1-\omega-i\epsilon}\;
  t_{g,n}\biggl( \frac{1-\omega-i\epsilon}{2} \biggr)\,
  \frac{H^g(\omega\xi,\xi,t)}{1+\omega}
- C_F\, t_{b,n}\biggl( \frac{1-\omega-i\epsilon}{2} \biggr)\,
  \xi H^S(\omega\xi,\xi,t) \biggr\}
\biggr] \,.
\end{align}
For $\omega \ge \omega_0$ with some $\omega_0 \gg 1$ we can use the
approximation \eqref{tgb-approx} of the hard-scattering kernels, and
further approximate $1+\omega \approx \omega$ in the first term on the
second line.  This gives
\begin{align}
  \label{small-x-app}
&\hspace{-1.4em} - \frac{1}{6 \pi \alpha_s}
  \im\ms \Bigl[ \mathcal{H}^g_n + \mathcal{H}^{S(b)}_n \Bigr]
\,\approx\,
H^g(\xi,\xi,t) + \frac{\alpha_s}{\pi}\,
\int_0^{\omega_0} d\omega\, \ldots
\nonumber \\[0.3em]
& - \frac{\alpha_s}{\pi}\,
\int_{\omega_0}^{1/\xi} d\omega\,
 \biggl\{
  C_A \biggl[ c_n - \ln\frac{Q^2}{\mu^2_{GPD}} \biggr]\,
  \frac{H^g(\omega\xi,\xi,t)}{\omega}
+ C_F \biggl[ c_n+1 - \ln\frac{Q^2}{\mu^2_{GPD}} \biggr]\,
  \xi H^S(\omega\xi,\xi,t) 
\biggr\} \,,
\end{align}
where the integral over $\omega$ on the first line is to be taken with
the unapproximated integrand from \eqref{small-x-con}.  It grows with
$\xi$ like $H^g(\omega\xi,\xi,t)$ or $\xi H^S(\omega\xi,\xi,t)$ but
lacks the enhancement due to the upper limit $1/\xi$ of the integral
on the second line.
Restricting our discussion to $t=0$ for simplicity, we can for
sufficiently large $\omega$ neglect the effect of skewness in the GPDs
and then have
\begin{align}
  H^g(\omega\xi,\xi,0) &\approx \omega\xi\ms g(\omega\xi) \,,
&
  H^S(\omega\xi,\xi,0) &\approx S(\omega\xi) 
  = \sum\nolimits_q\, \bigl[ q(\omega\xi) + \bar{q}(\omega\xi) \bigr] \,,
\end{align}
where $S(x)$ is the usual quark singlet distribution.  In a very rough
approximation one may treat $x g(x)$ and $x S(x)$ as constant at small
$x$.  In \eqref{small-x-app} one then has loop integrals $\int d\omega
/\omega$ for both the gluon and the quark term, which generate large
logarithms $\ln(\omega_0\ms\xi)$ for $1/\xi \gg \omega_0$.  These
logarithms are of BFKL type and correspond to graphs with $t$-channel
gluon exchange in the hard-scattering kernel, such as those for $T_b$
and $T_g$ in Fig.~\ref{fig:graphs}.

In a phenomenologically more realistic approximation one has $x g(x)
\approx a\ms x^{-\lambda}$ at small $x$ and a similar behavior with
different values of $a$ and $\lambda$ for $x S(x)$.  This gives
\begin{equation}
  \label{small-x-pow-approx}
\int_{\omega_0}^{1/\xi} d\omega\, 
  \frac{H^g(\omega\xi,\xi,0)}{\omega} \approx
  a\ms \xi^{-\lambda} \int_{\omega_0}^{1/\xi}
     d\omega\, \omega^{-\lambda-1}
\approx
\frac{a}{\lambda}\, \bigl( \omega_0\ms\xi \bigr)^{-\lambda}
\end{equation}
for $1/\xi \gg \omega_0$, when the bulk of the integral comes from the
region where the small-$x$ approximation of the gluon density is
valid.  With $\lambda$ being rather small for the gluon distribution
in a wide range of the factorization scale, the term
\eqref{small-x-pow-approx} has the same power behavior
$\xi^{-\lambda}$ as the Born term $H^g(\xi,\xi,0)$ in
\eqref{small-x-app} but is numerically enhanced by $1/\lambda$.  A
contribution analogous to \eqref{small-x-pow-approx} is obtained from
the quark singlet term in \eqref{small-x-app} and comes with a similar
enhancement.

Concerning the choice of factorization scale, it is clear that the
size of the corrections in \eqref{small-x-app} is decreased if
$\mu_{GPD}$ is taken smaller than $Q$.  It is also clear that no scale
choice can eliminate both the gluon and quark singlet contribution in
this expression.  To make at least the gluon term for $n=0$ disappear
one needs $\mu_{GPD}^2 = e^{-2}\ms Q^2 \approx 0.14\, Q^2$.  For a
wide range of $Q^2$ this is outside the perturbative region or at
least so low that the quark singlet distribution has a rather small
power $\lambda$ and can thus give important corrections.  We note that
previous analyses of vector meson production at small $x_B$ have
argued for a factorization scale well below $Q^2$, based on different
estimates of the typical virtualities in the leading-order graphs
\cite{Frankfurt:1995jw,Ryskin:1995hz}.  We also note that the $\mu_R$
dependent term $\beta_0 \ln(\mu_R^2 /\mu^2_{GPD})$ in the gluon kernel
\eqref{I_gluon} does not appear in the approximation \eqref{asy-g}
which dominates the convolutions at small $x_B$.  The choice of
$\mu_R$ can thus not cure the huge NLO corrections we have discussed.


\section{Vector meson production at moderate to large $x_B$}
\label{sec:large-x}

Let us now investigate the NLO corrections in typical fixed-target
kinematics, as it is accessible at HERMES, JLab and COMPASS.  We take
again $t=0$ and for definiteness present estimates at $Q^2 =4 \gev^2$.
For larger $Q^2$, which will in particular be accessible with the JLab
energy upgrade to $12\gev$, the corrections are in general smaller.

\begin{figure}
\begin{center}
\includegraphics[width=\plotwidth]
{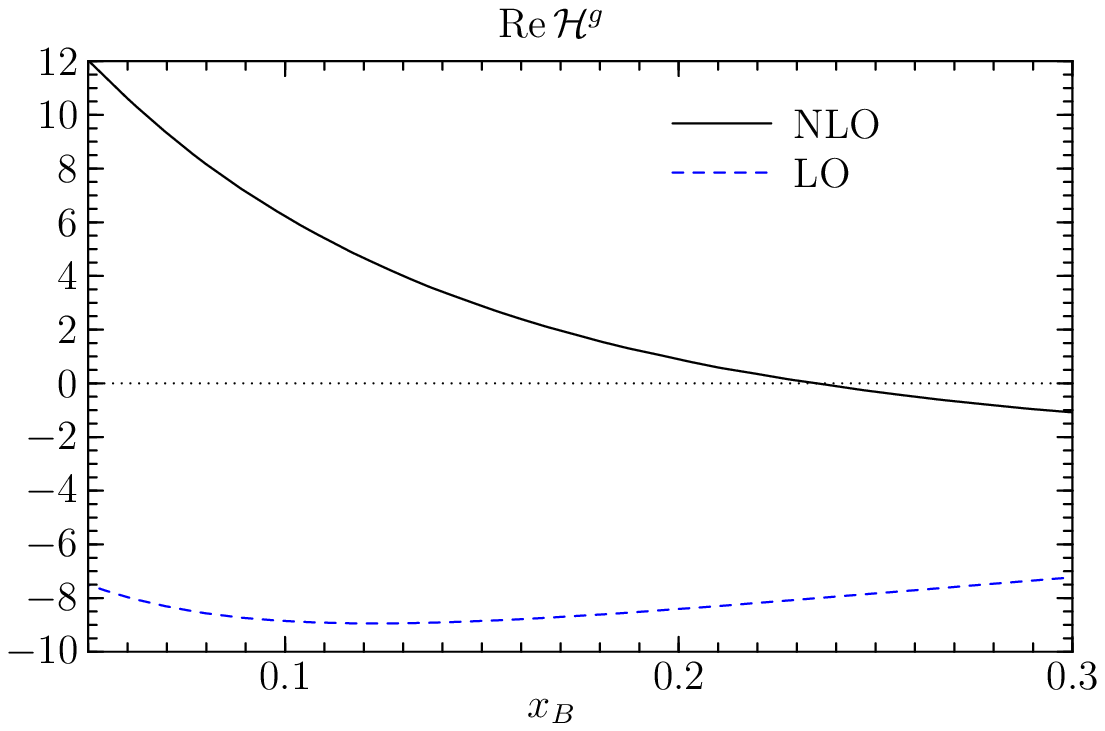}\hspace{1ex}
\includegraphics[width=\plotwidth]
{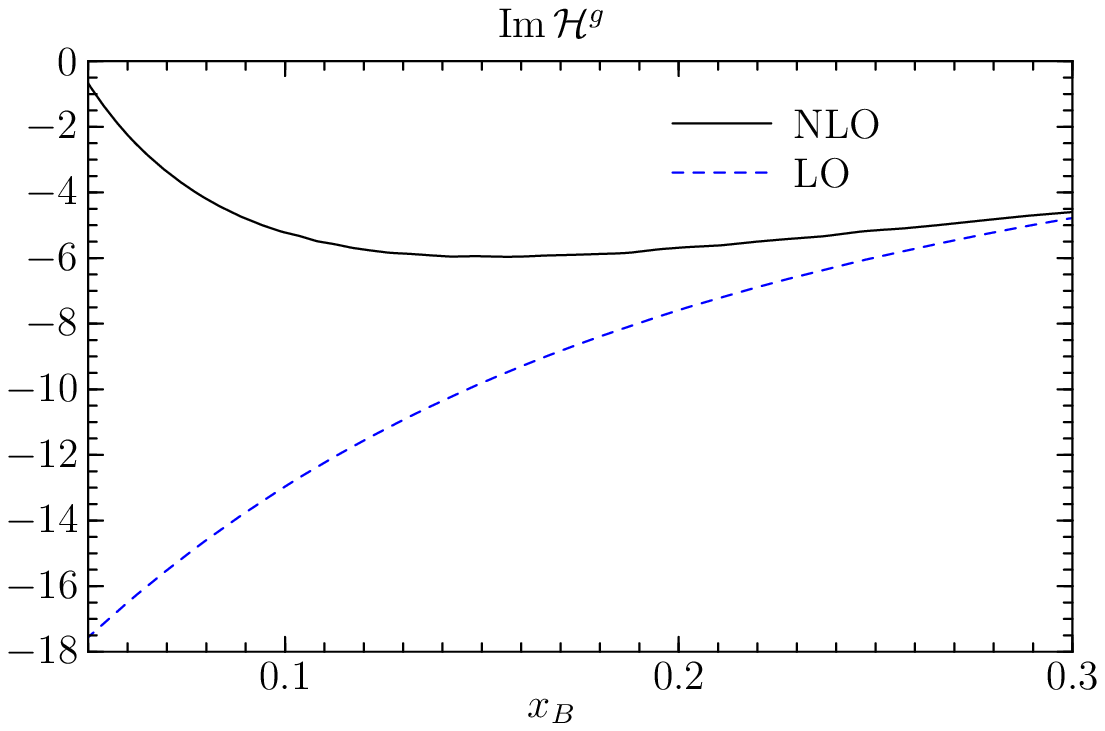}\\[1.5ex]
\includegraphics[width=\plotwidth]
{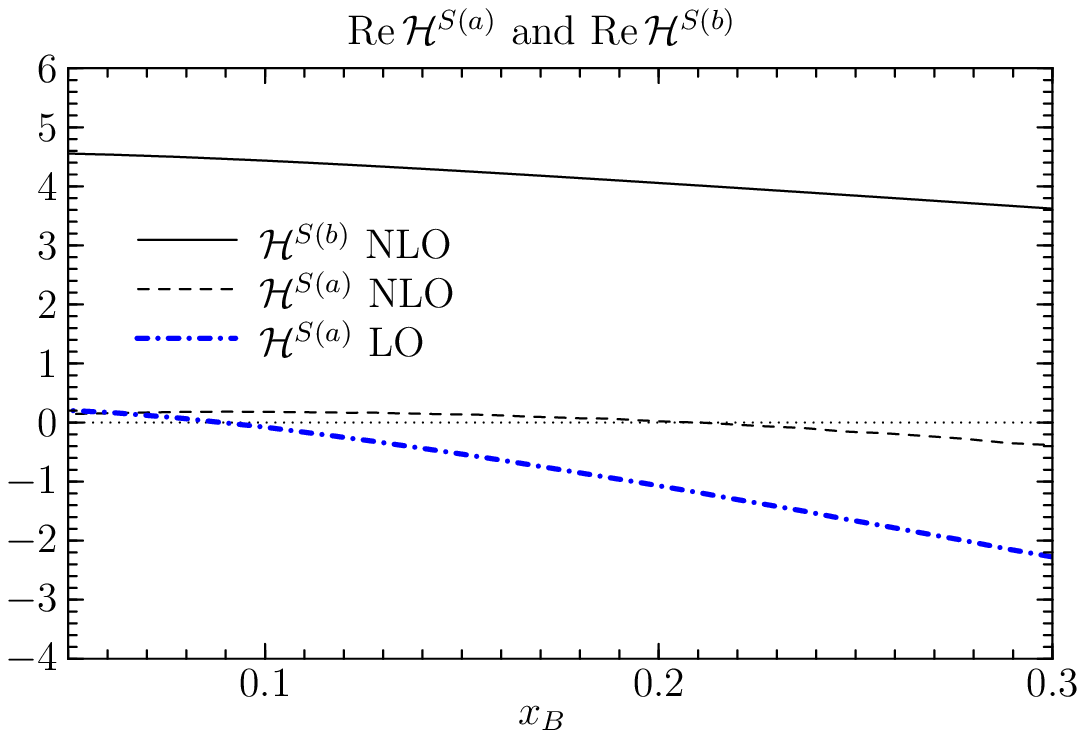}\hspace{1ex}
\includegraphics[width=\plotwidth]
{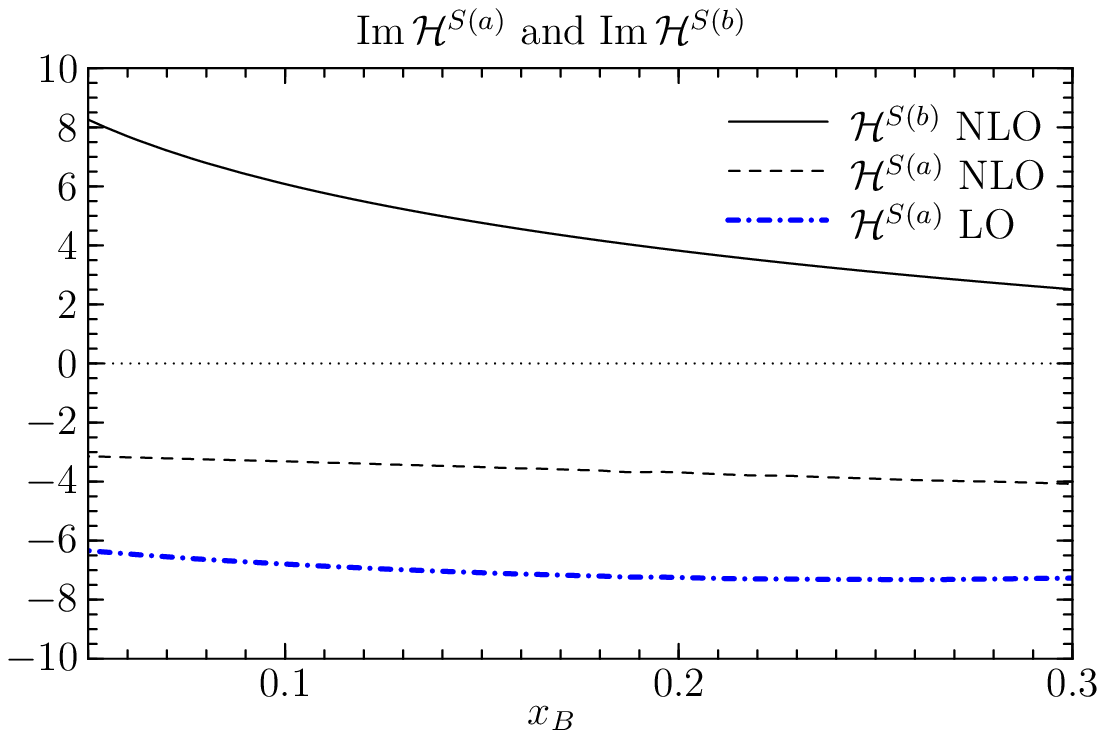}\\[1.5ex]
\includegraphics[width=\plotwidth]
{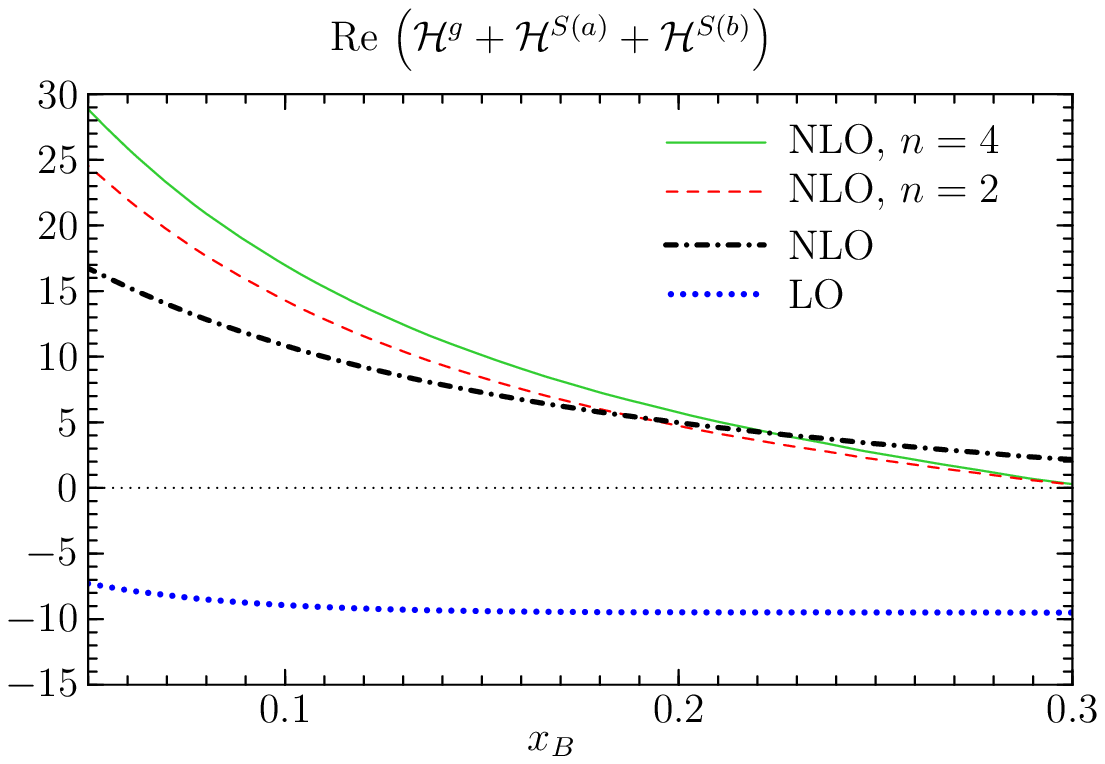}\hspace{1ex}
\includegraphics[width=\plotwidth]
{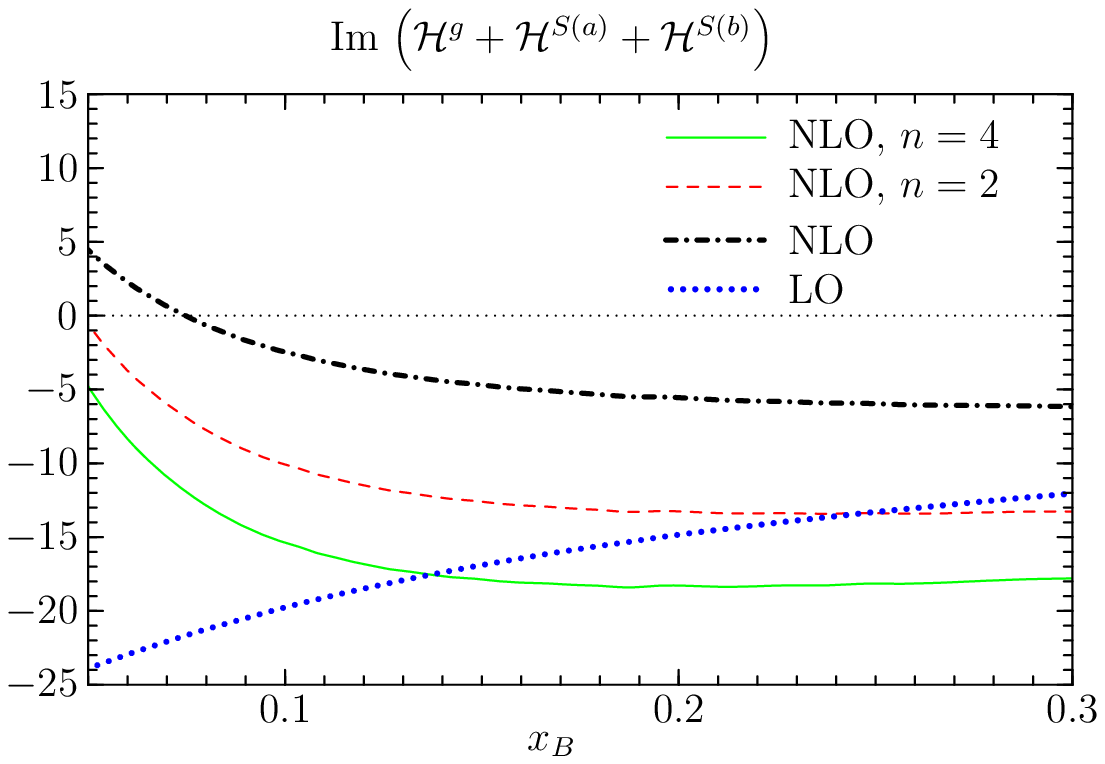}
\end{center}
\caption{\label{LO_NLO_comparison_large_x} LO and NLO terms of the
  convolutions in the gluon and quark singlet sector at $Q = 2 \gev$,
  with $\mu_R = \mu_{GPD} = \mu_{DA} = Q$.}
\end{figure}

\begin{figure}
\begin{center}
\includegraphics[width=\plotwidth]
{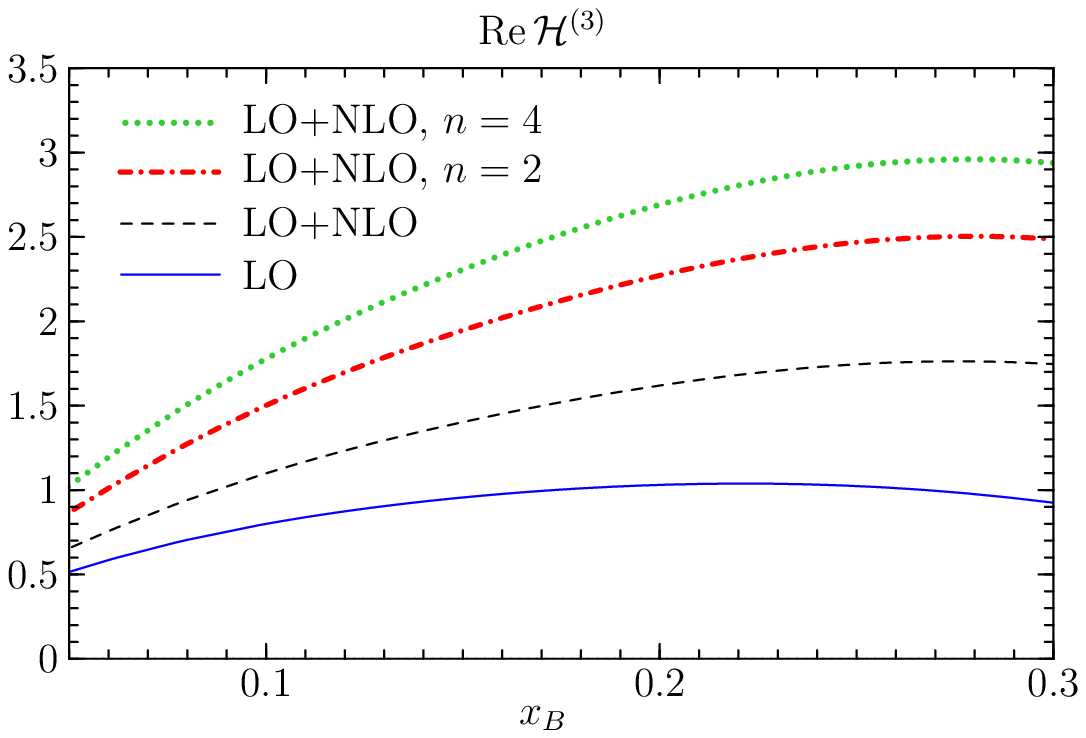}\hspace{1ex}
\includegraphics[width=\plotwidth]
{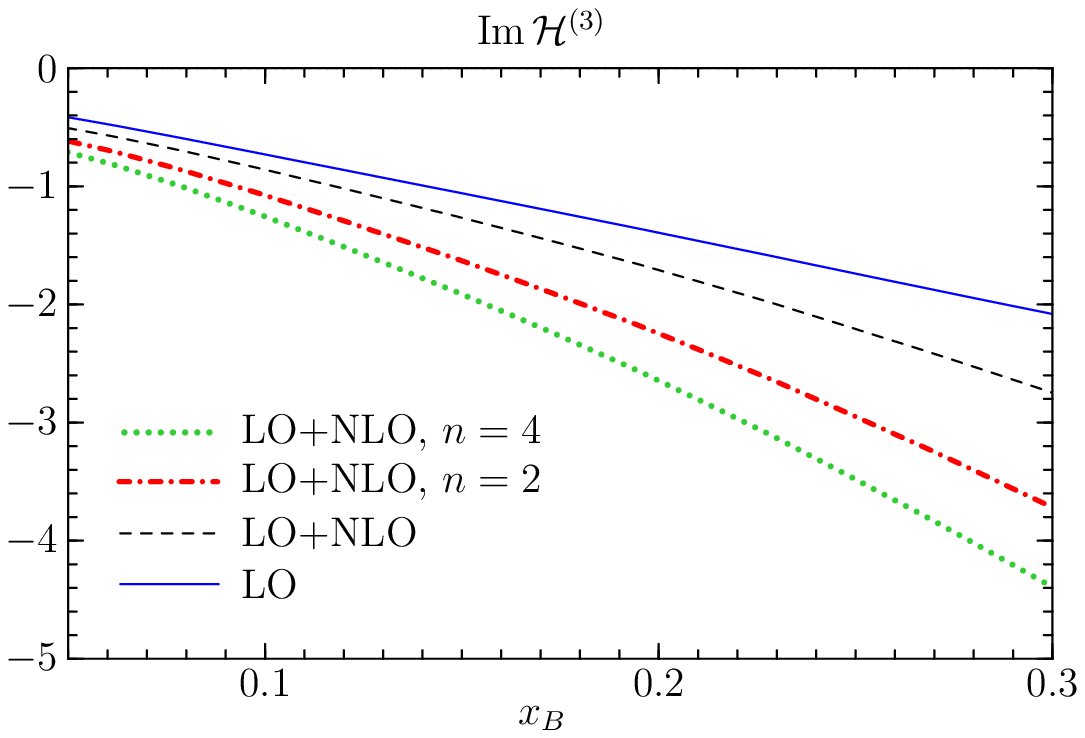}\\[1ex]
\includegraphics[width=\plotwidth]
{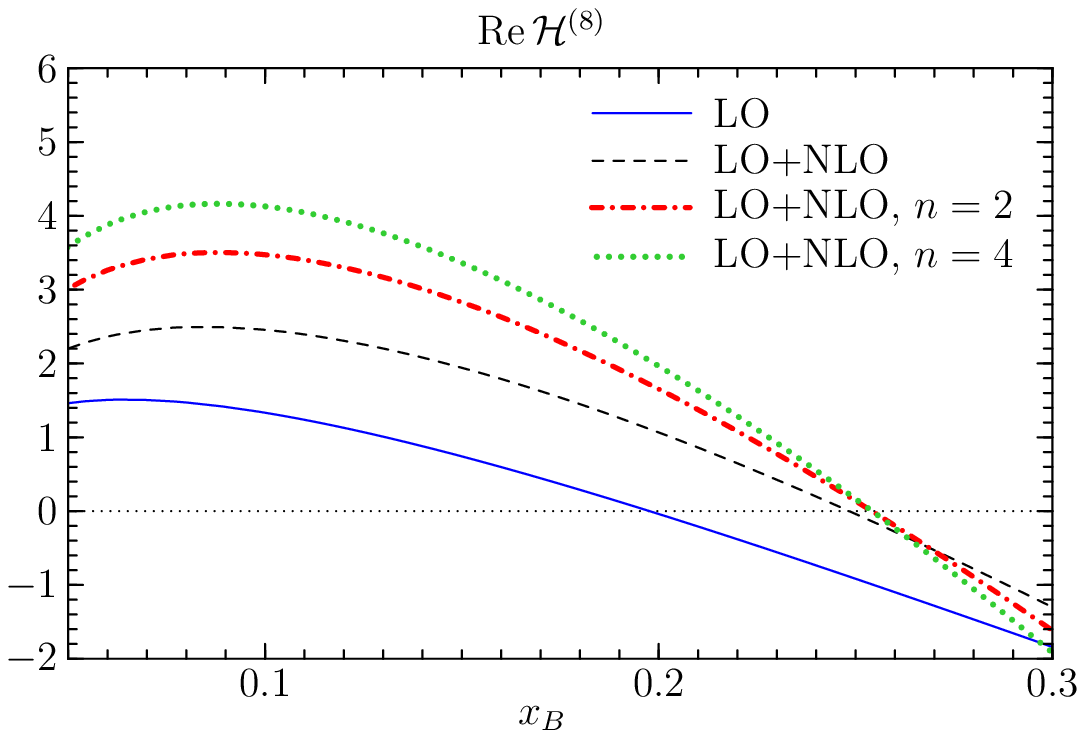}\hspace{1ex}
\includegraphics[width=\plotwidth]
{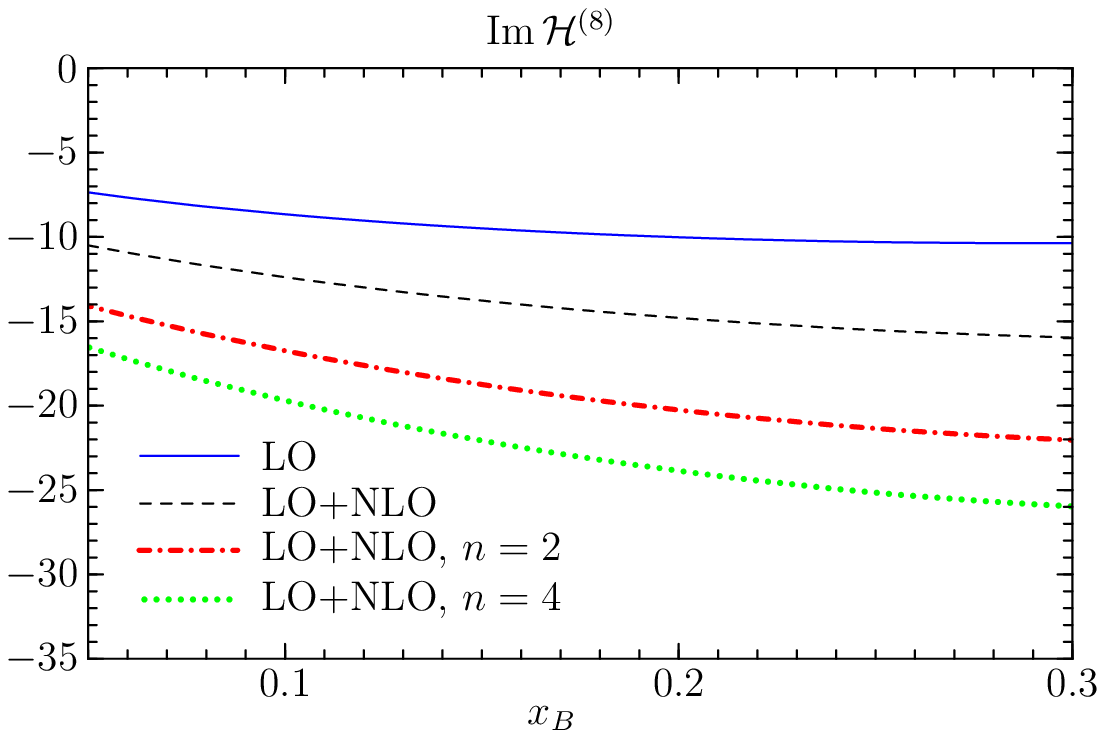}
\end{center}
\caption{\label{ns_LO_NLO_comparison_large_x} LO terms and the sum of
  LO and NLO terms of the convolutions in the quark non-singlet sector
  at $Q = 2 \gev$, with $\mu_R = \mu_{GPD} = \mu_{DA} = Q$.}
\end{figure}

In Fig.~\ref{LO_NLO_comparison_large_x} we compare the LO and NLO
parts of the convolution integrals.  In the gluon sector we find no
simple picture, with relative corrections that are typically moderate
but become large for $\re\mathcal{H}^g$ at smaller $x_B$ and for
$\im\mathcal{H}^g$ at larger $x_B$.  For the quark singlet the
situation is similar to the one in the small-$x_B$ region, i.e.\ we
have rather large NLO corrections from $\mathcal{H}^{S(b)}$ with sign
opposite to the LO part of $\mathcal{H}^{S(a)}$, whereas the NLO
corrections in $\mathcal{H}^{S(a)}$ are smaller.
Adding gluon and quark singlet contributions, we find that for $n=0$
the NLO corrections are of reasonable size for the imaginary part.
For the real part at lower $x_B$, the corrections are however large
and of opposite sign compared to the Born term.  We note that the
convolutions $\mathcal{H}$ satisfy a dispersion relation in $1/x_B$
for fixed $Q^2$ and $t$ \cite{Anikin:2007yh}.  In this representation
their real parts at a given $x_B$ are sensitive to the imaginary part
at smaller values of $x_B$, where the NLO corrections rapidly increase
as we have seen in the previous section.
Turning to the quark non-singlet convolutions, we see in
Fig.~\ref{ns_LO_NLO_comparison_large_x} that for $n=0$ the NLO
corrections are comparatively moderate for the imaginary part and
larger for the real part.

Going from $n=0$ to higher Gegenbauer indices $n=2$ and $n=4$, the NLO
corrections become larger, as we see in
Figs.~\ref{LO_NLO_comparison_large_x} and
\ref{ns_LO_NLO_comparison_large_x} and already observed at small
$x_B$.  Generically this is not unexpected, since the $z$ dependent
kernels \eqref{orig_kernels} contain logarithms $\ln z$ and $\ln\zb$
which enhance the endpoint regions of the $z$ integration, and those
endpoint regions are more prominent for higher Gegenbauer polynomials
in the expansion \eqref{gegen-phi}.  Note that according to
phenomenological estimates or lattice calculations the coefficients
$a_n$ of these polynomials are clearly smaller than $a_0$, so that
increasing corrections to $\mathcal{H}_n$ for higher $n$ do not affect
the sum $\sum_n a_n \mathcal{H}_n$ as much.
We note that in the modified hard-scattering approach of Sterman et
al.~\cite{Botts:1989kf}, which goes beyond the collinear approximation
used in the present work, the endpoint regions in $z$ are suppressed
by radiative corrections that are resummed into Sudakov form factors.
As just discussed, we do not observe such a suppression in the
fixed-order results analyzed here, where various positive and negative
corrections compete with each other---only some of them related to the
Sudakov factor.  How the situation will be at higher orders is an
important question, which goes beyond the scope of the present work.

\begin{figure}
\begin{center}
\includegraphics[width=0.66\textwidth]{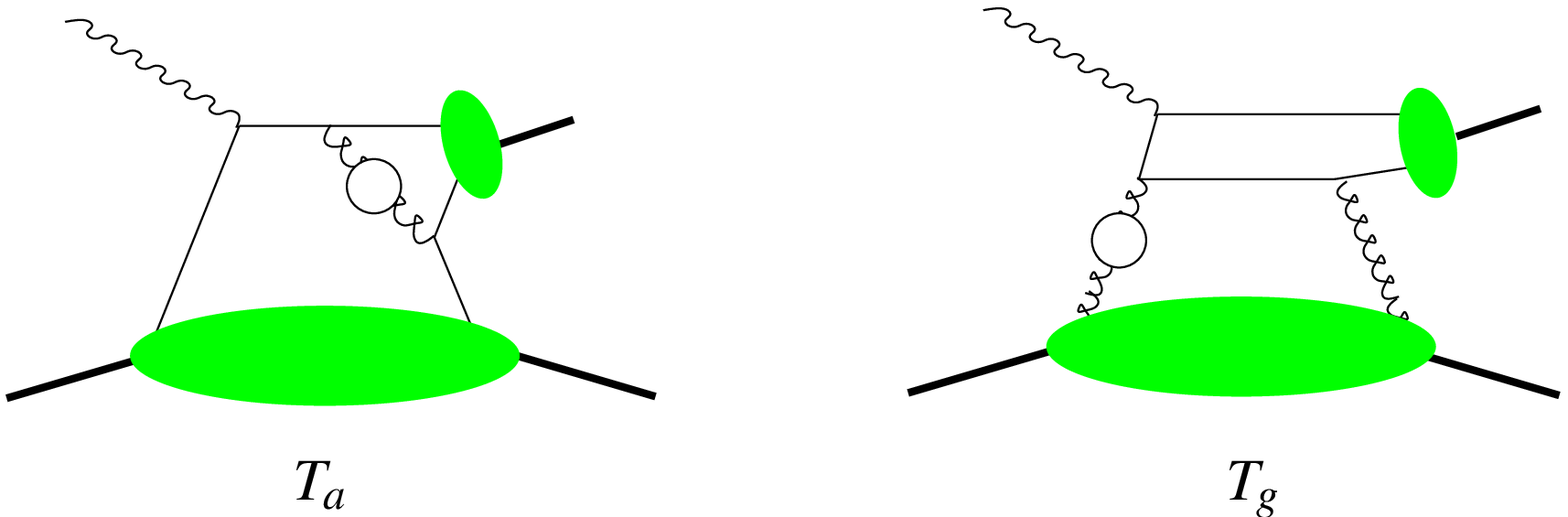}
\end{center}
\caption{\label{fig:beta-graphs} Example graphs giving rise to terms
  proportional to $\beta_0$ in the hard-scattering kernels $T_a$ and
  $T_g$.}
\end{figure}

Let us now take a closer look at the $\mu_R$ dependence of the
corrections.  As we explained in Sect.~\ref{sec:kernels}, the pure
quark singlet kernel $T_b$ is independent of this scale at
$O(\alpha_s^2)$.  According to \eqref{I_gluon} the gluon kernel $T_g$
depends on $\mu_R^2$ only through $\beta_0 \ln(\mu_R^2 /\mu^2_{GPD})$,
which originates from graphs with gluon propagator corrections such as
the one shown in Fig.~\ref{fig:beta-graphs}.  The $\mu_{GPD}$
dependence of this term is connected with the contribution
proportional to $\beta_0$ in the evolution kernel $V^{gg}$ for the
gluon GPD, given in \eqref{evol-kernels}.  As already pointed out in
Sect.~\ref{sec:small-x}, the term $\beta_0 \ln(\mu_R^2 /\mu^2_{GPD})$
does not contribute to the large-$|y|$ behavior of $\im t_{g,n}(y)$
and is hence not relevant for the huge NLO corrections at small $x_B$.

\begin{figure}[t]
\begin{center}
\includegraphics[width=\plotwidth]
{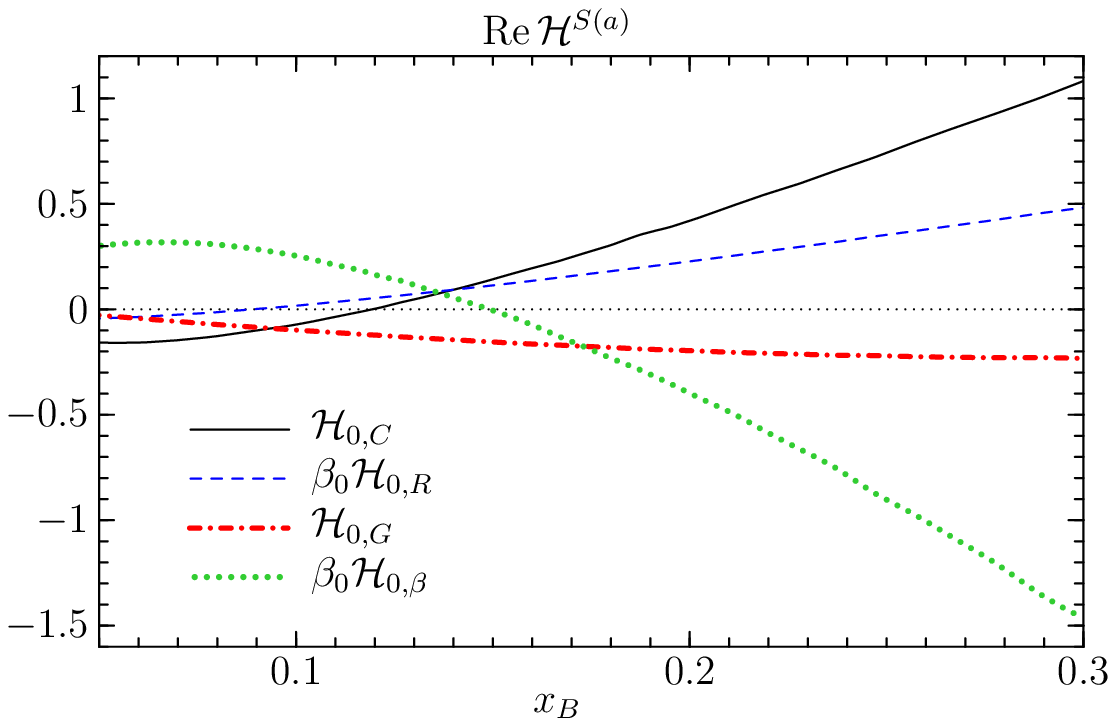}\hspace{1ex}
\includegraphics[width=\plotwidth]
{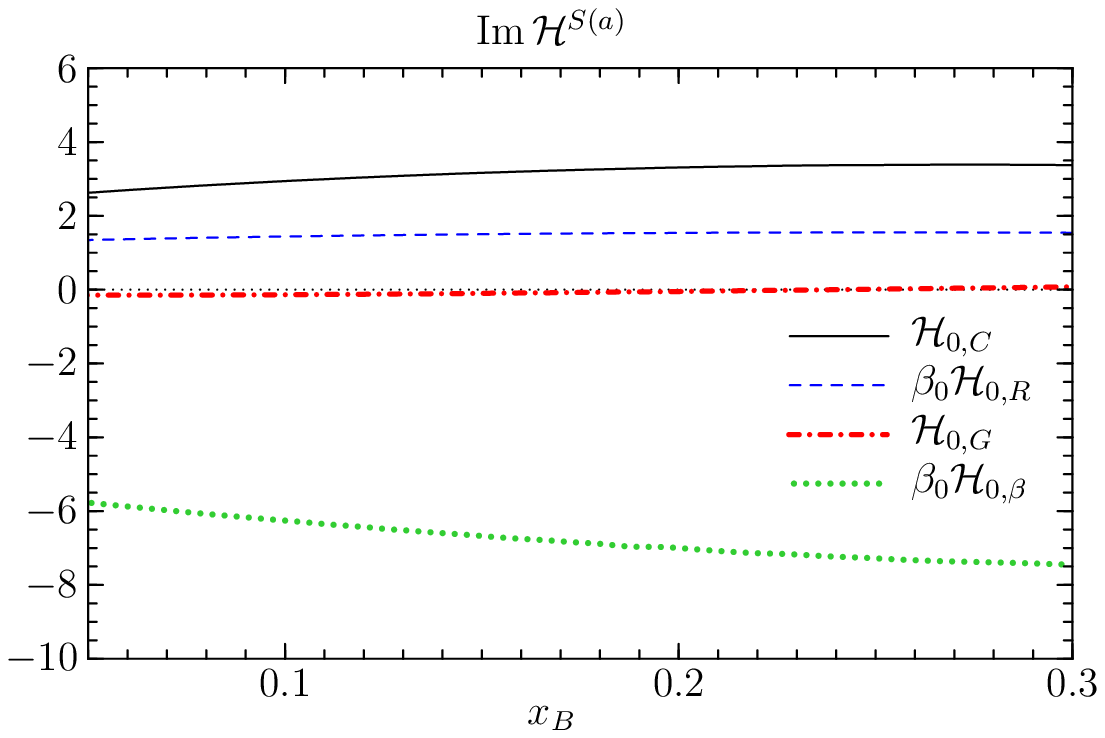}
\includegraphics[width=\plotwidth]
{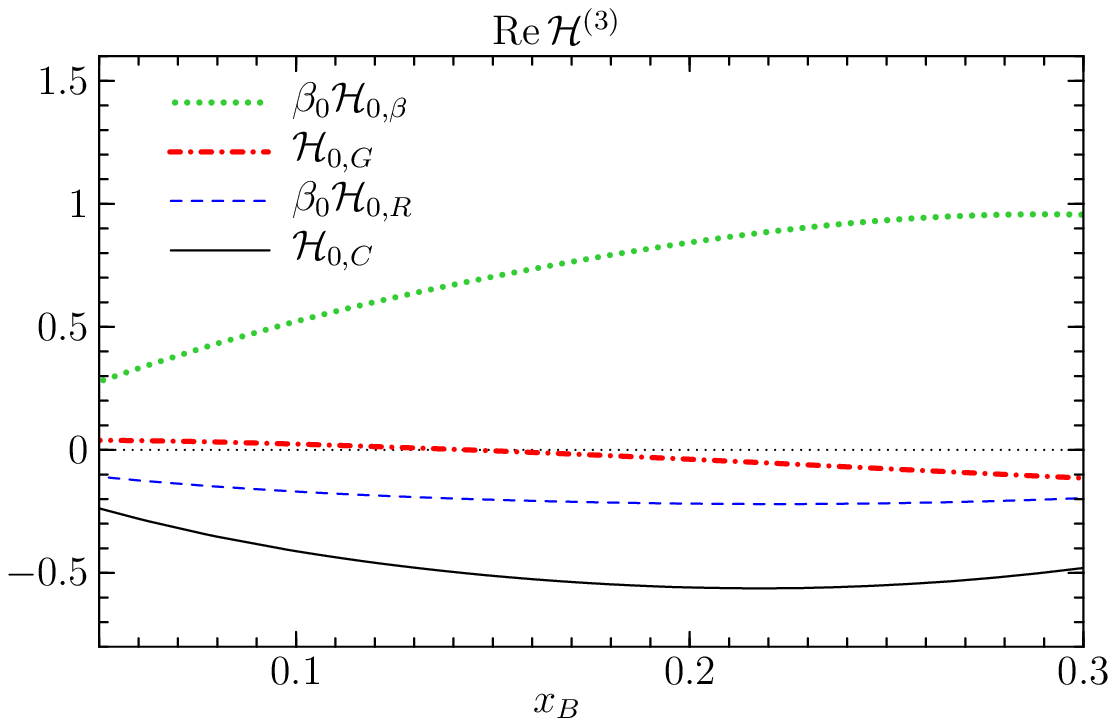}\hspace{1ex}
\includegraphics[width=\plotwidth]
{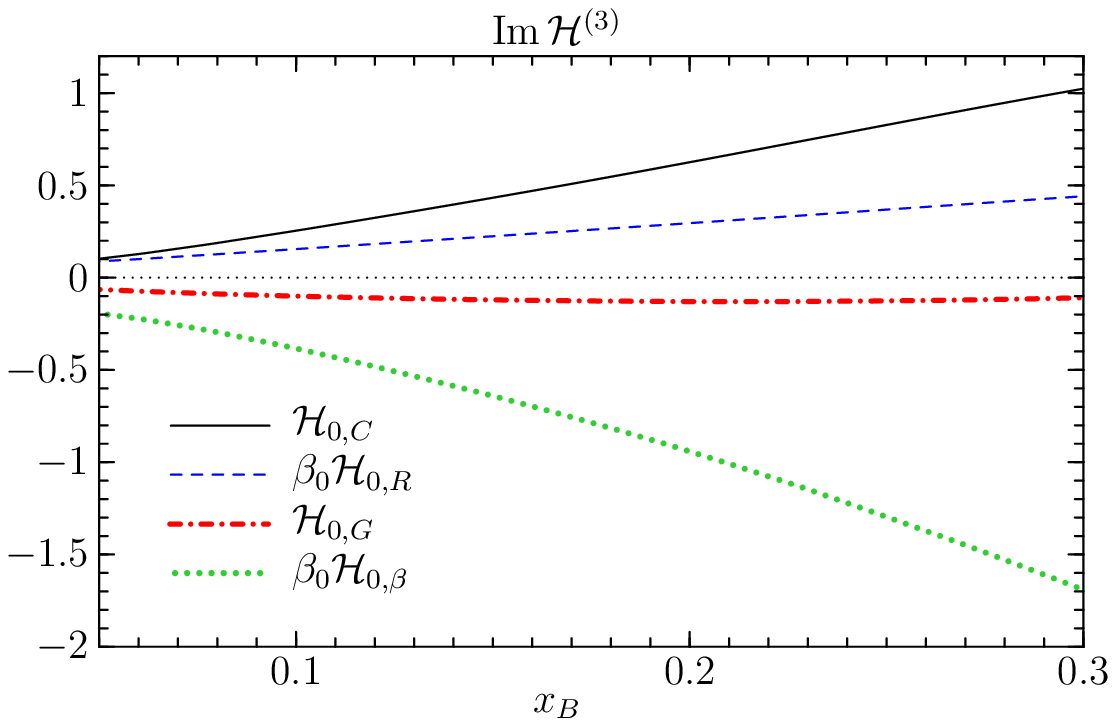}
\end{center}
\caption{\label{NLO_single_terms} Individual terms
  \protect\eqref{F-terms} in the convolutions of the quark non-singlet
  kernel for $n=0$.  The quark distributions are evaluated at
  $\mu_{GPD}= 2 \gev$ and the running coupling in the kernels at
  $\mu_R = 2\gev$.}
\end{figure}

For the kernel $T_{a}$ the situation is more involved.  The general
structure of its convolution with the quark singlet distribution $H^S$
can be written as
\begin{align}
  \label{F-terms}
\mathcal{H}^{S(a)}_{n\phantom{P}} &=
\beta_0^{}\, \biggl( \mathcal{H}^{S(a)}_{n, \beta}
                   + \mathcal{H}^{S(a)}_{n, R}\, \ln\frac{Q^2}{\mu^2_{R}}
\biggr)
+ \mathcal{H}^{S(a)}_{n, C} 
+ \mathcal{H}^{S(a)}_{n, G} \ln\frac{Q^2}{\mu^2_{GPD}}
+ \mathcal{H}^{S(a)}_{n, D} \ln\frac{Q^2}{\mu^2_{DA}}
\end{align}
with an analogous decomposition for the convolutions
$\mathcal{H}^{(3)}_n$ and $\mathcal{H}^{(8)}_n$.  The terms
proportional to $\beta_0$ originate from graphs with gluon propagator
corrections such as in Fig.~\ref{fig:beta-graphs}, whereas the terms
with subscripts $C, G, D$ do not contain $\beta_0$.
In Fig.~\ref{NLO_single_terms} we show the corresponding contributions
for $n=0$.  We see that terms multiplying $\ln(Q^2 /\mu^2_{GPD})$ are
rather small, whereas those going with $\ln(Q^2 /\mu^2_{DA})$ are of
course absent for $n=0$.  The term $\mathcal{H}_{0, R}$ is clearly
smaller than $\mathcal{H}_{0, \beta}$ and has opposite sign.
$\mathcal{H}_{0, C}$ has also the opposite sign compared to
$\mathcal{H}_{0, \beta}$ but is similar in magnitude.  We note that
$|\mathcal{H}_{n, \beta}|$ increases with $n$, as can be seen from
\eqref{app:non-sing} and \eqref{higher-kern}.

Let us briefly comment on the BLM scale setting prescription
\cite{Brodsky:1982gc}, which has been discussed in the context of
exclusive meson production in \cite{Belitsky:2001nq,Anikin:2004jb}.
This prescription aims at including the corrections from graphs like
those of Fig.~\ref{fig:beta-graphs} in the argument of the running
coupling, and for the case at hand takes $\mu_R$ such that the
contribution from $\mathcal{H}_{n, \beta}$ cancels against the one
from $\mathcal{H}_{n, R}\ms \ln(Q^2 /\mu^2_{R})$ in \eqref{F-terms}.
As is evident from Fig.~\ref{NLO_single_terms}, this requires
$\mu_R^2$ to be substantially lower than $Q^2$.  For most of
experimentally accessible kinematics, the resulting $\mu_R$ is in fact
far below the region where perturbation theory can be applied.  In
such a situation, the perturbative running of $\alpha_s$ is often
modified such that the coupling saturates for decreasing $\mu_R$.  We
note that in the context of our NLO analysis, the logarithm $\beta_0
\ln(Q^2 /\mu^2_{R})$ in the hard-scattering kernel is intimately
related with the perturbative running of $\alpha_s(\mu_R)$, so that
keeping one while modifying the other is not obviously consistent.

We also remark that if $\mathcal{H}_{n, \beta}$ and $\mathcal{H}_{n,
R}\ms \ln(Q^2 /\mu^2_{R})$ are made to cancel by the BLM scale choice,
one is left with a relatively large correction from $\mathcal{H}_{n,
C}$.  For scale choices where $\mu_R^2$ is closer to $Q^2$, one
instead has a partial cancellation between $\mathcal{H}_{n, C}$ and
$\mathcal{H}_{n, \beta}$.  A more detailed analysis for the similar
case of the electromagnetic form factor is given in
\cite{Bakulev:2000uh}, which also discusses the issue of Sudakov-type
corrections we raised above.

\begin{figure}
\begin{center}
\includegraphics[width=\plotwidth]
{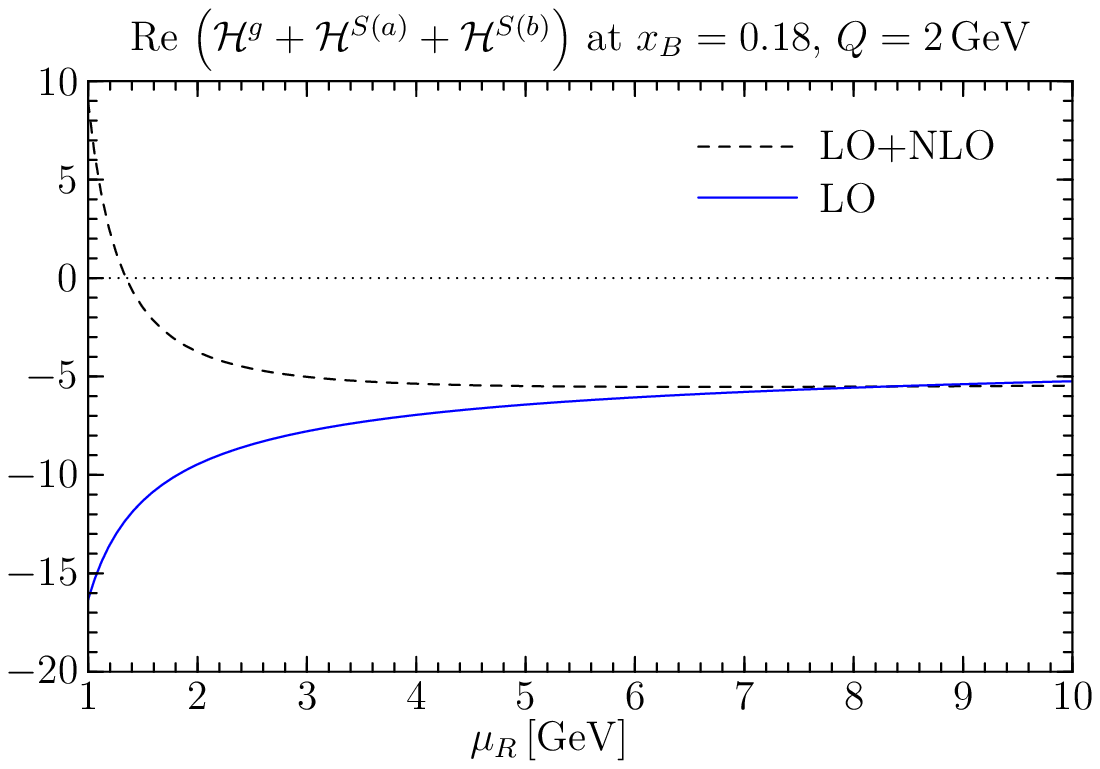}\hspace{1ex}
\includegraphics[width=\plotwidth]
{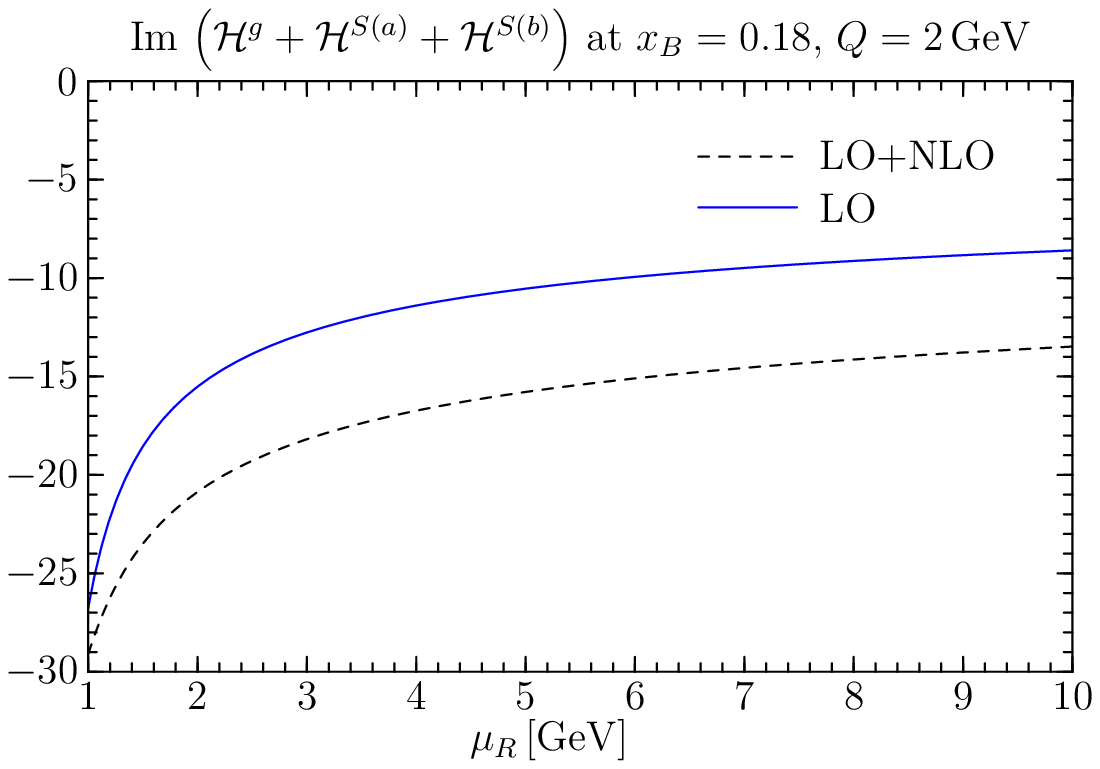}\\[1.5ex]
\includegraphics[width=\plotwidth]
{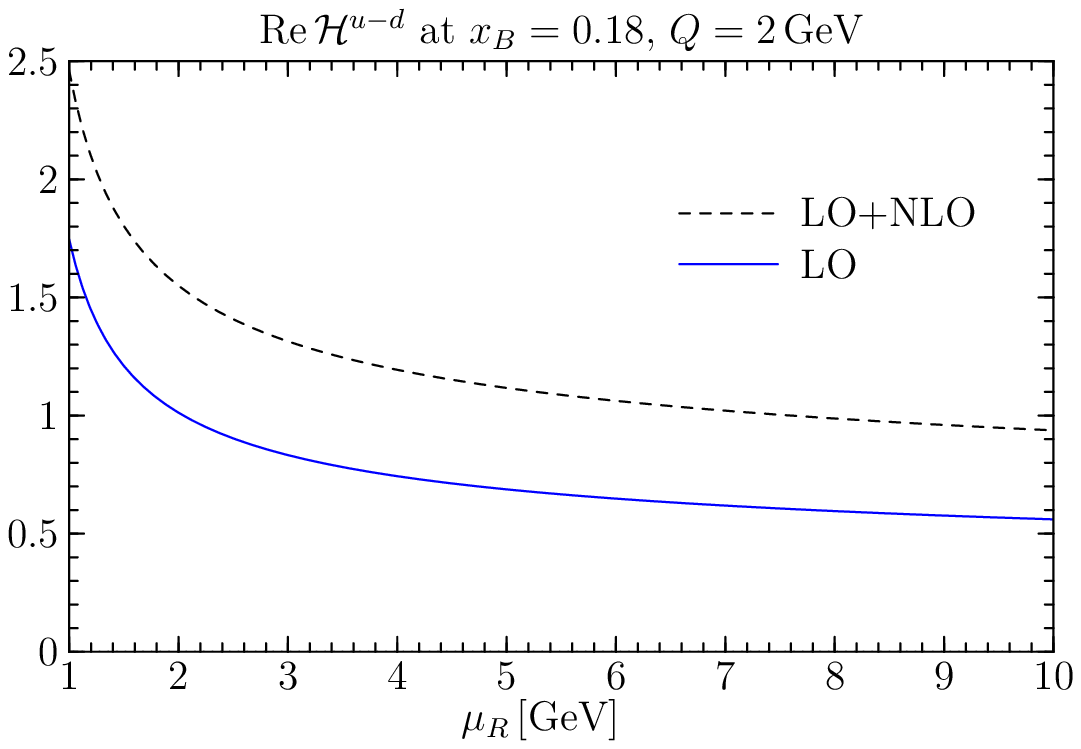}\hspace{1ex}
\includegraphics[width=\plotwidth]
{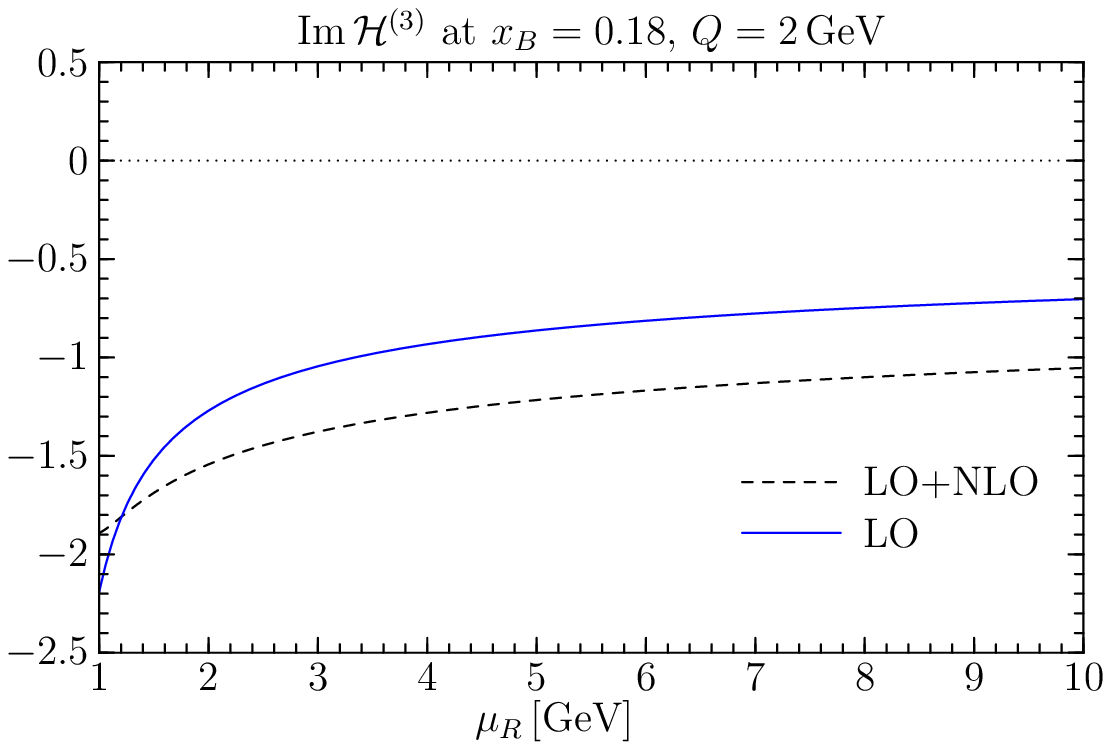}\\[1.5ex]
\includegraphics[width=\plotwidth]
{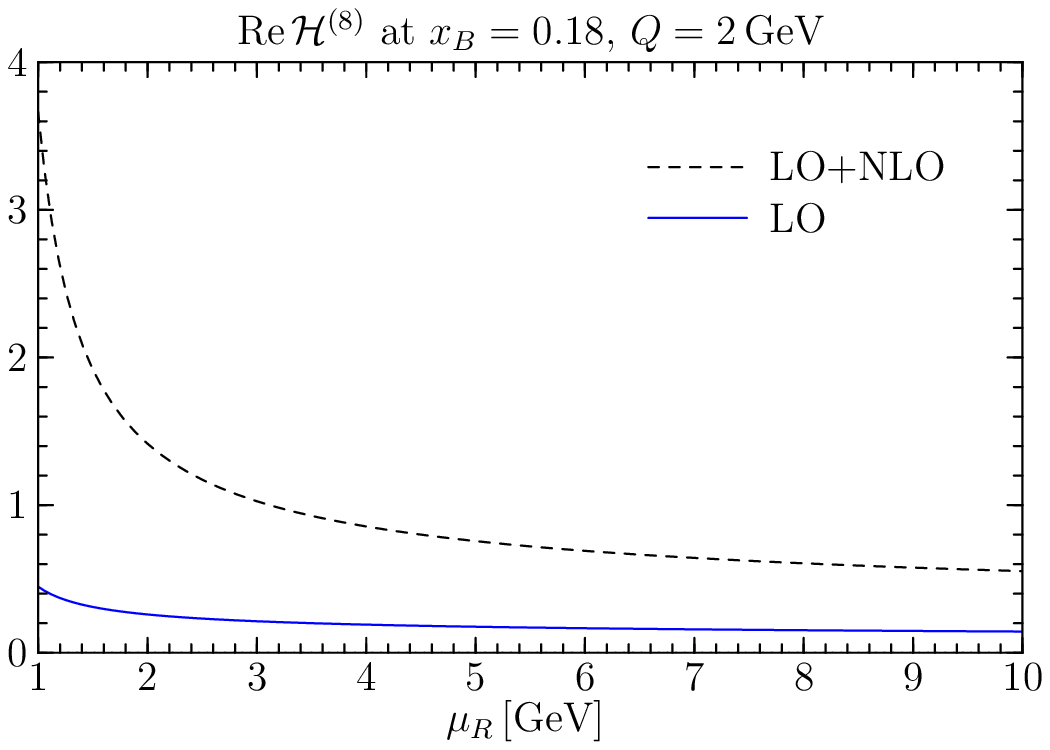}\hspace{1ex}
\includegraphics[width=\plotwidth]
{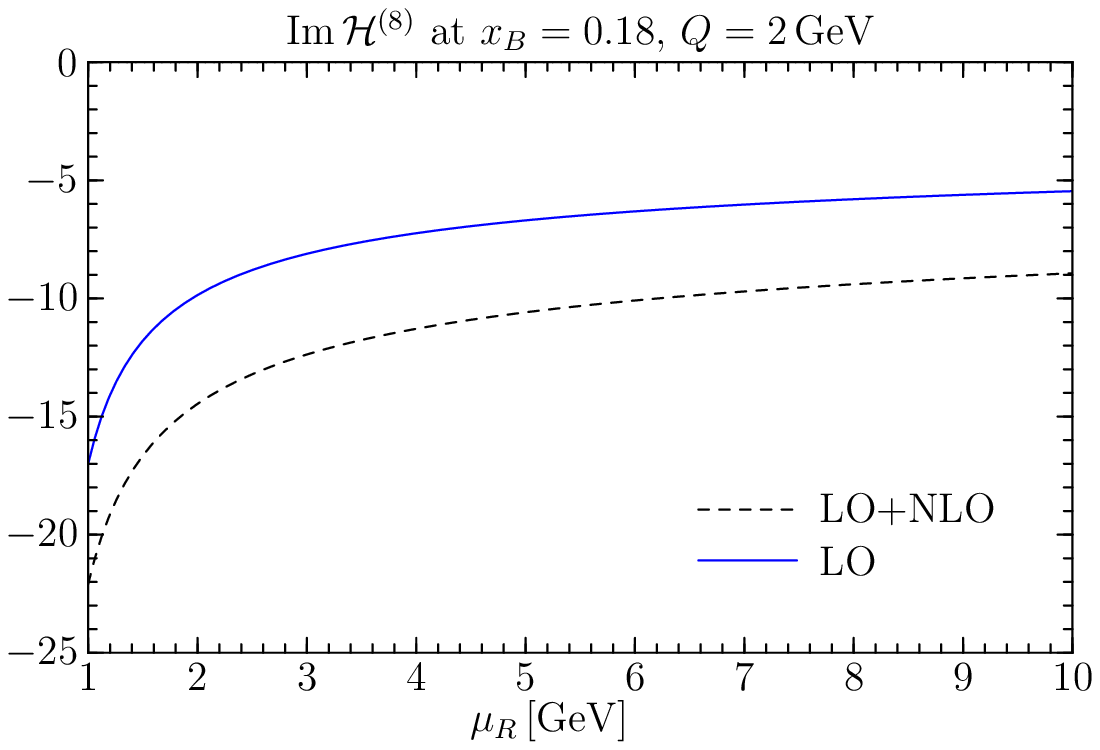}
\end{center}
\caption{\label{diff_mur_results} Dependence of the convolutions on
  $\mu_R$.  The factorization scale is held fixed at $\mu_{GPD} = Q$.}
\end{figure}

\begin{figure}
\begin{center}
\includegraphics[width=\plotwidth]
{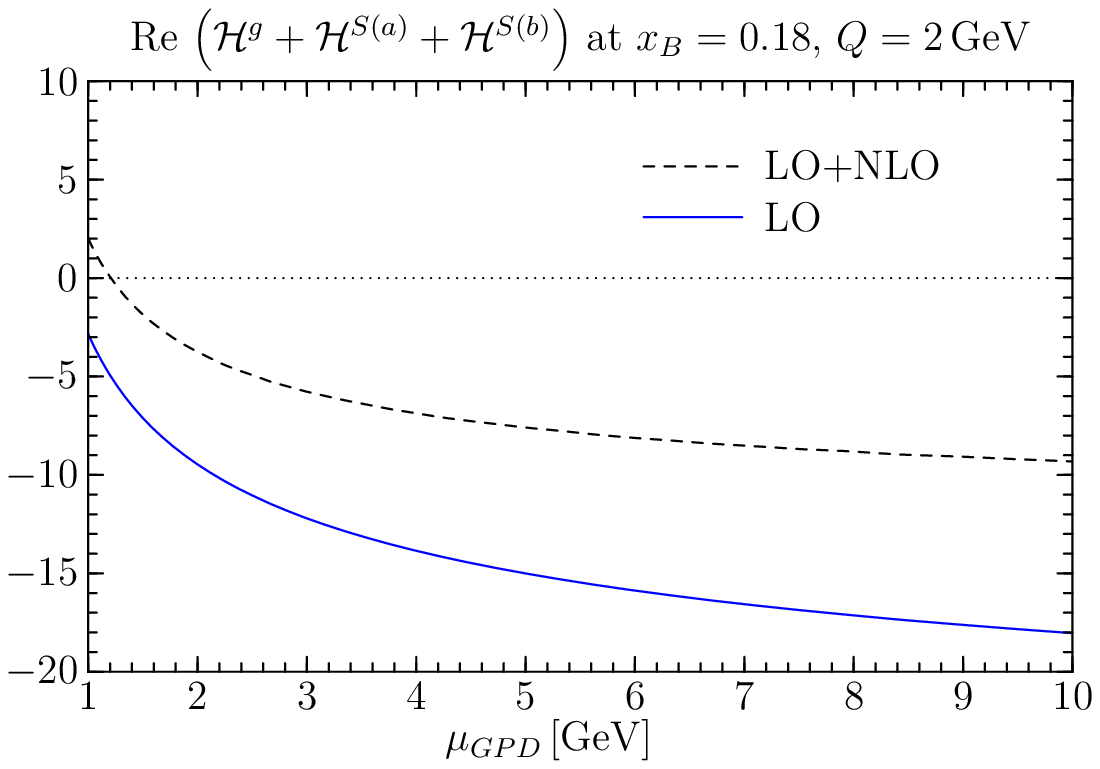}\hspace{1ex}
\includegraphics[width=\plotwidth]
{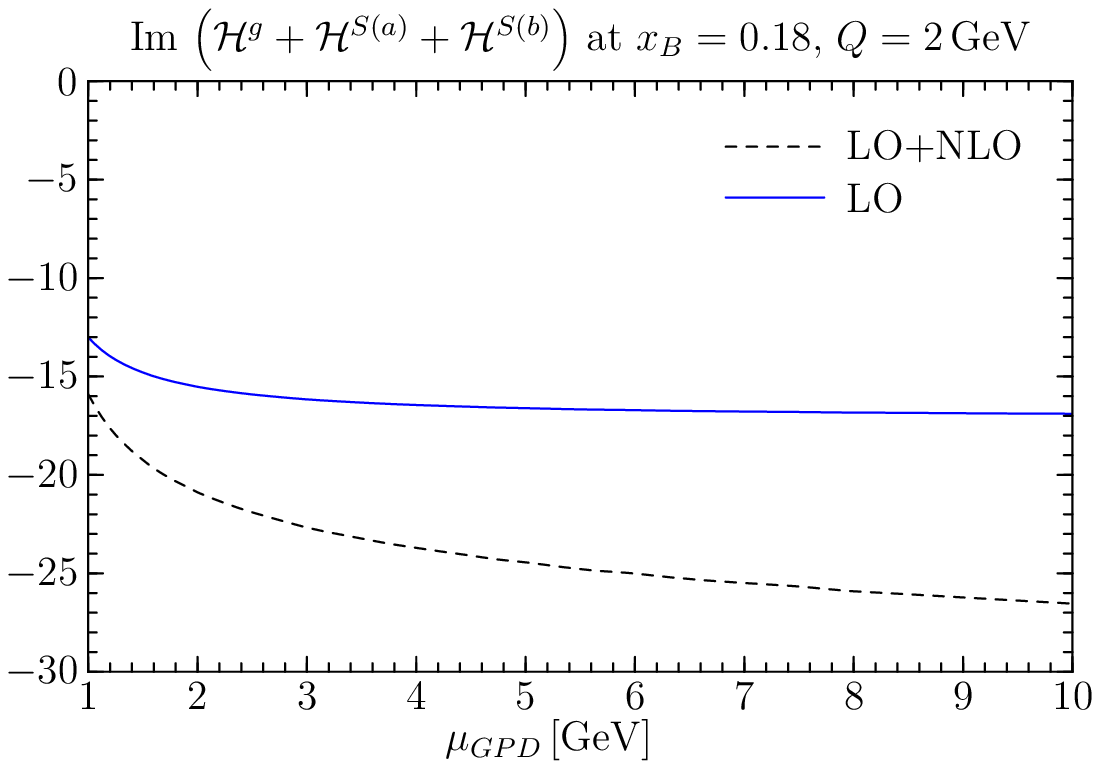}\\[1.5ex]
\includegraphics[width=\plotwidth]
{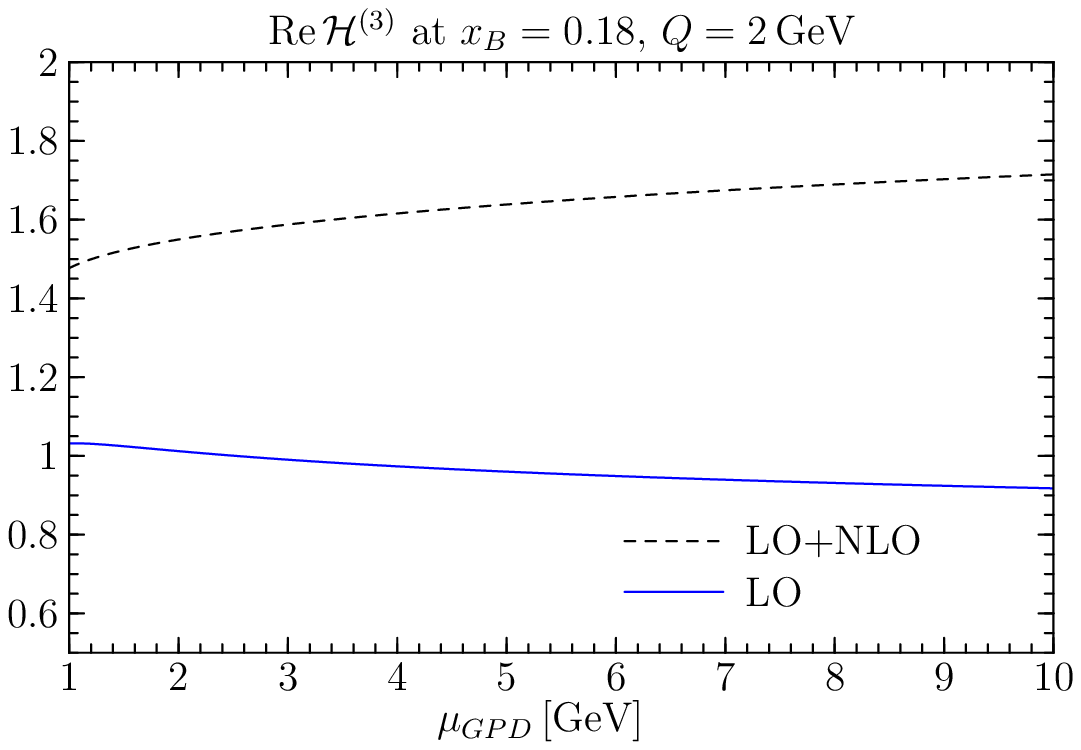}\hspace{1ex}
\includegraphics[width=\plotwidth]
{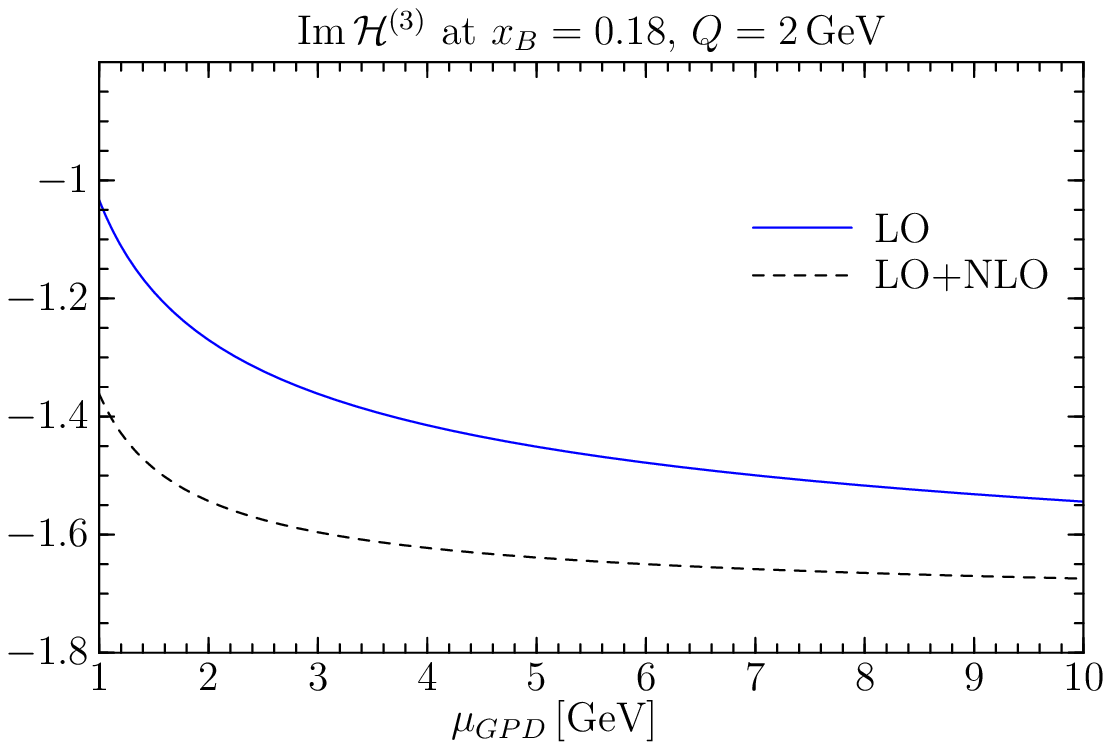}\\[1.5ex]
\includegraphics[width=\plotwidth]
{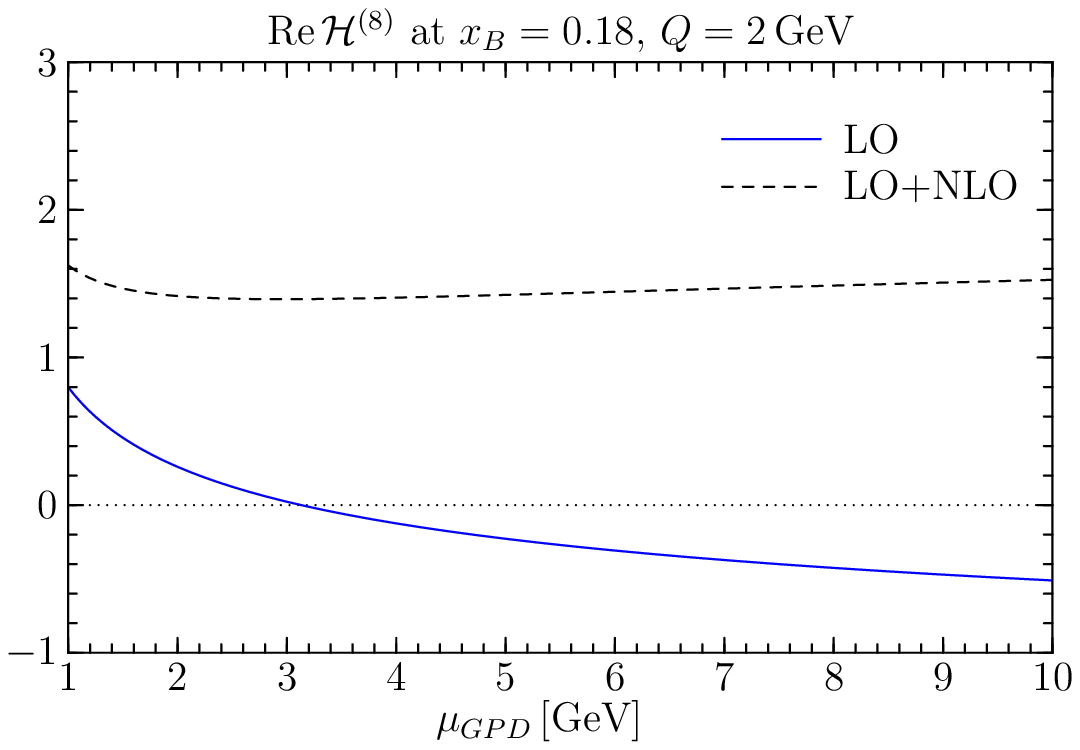}\hspace{1ex}
\includegraphics[width=\plotwidth]
{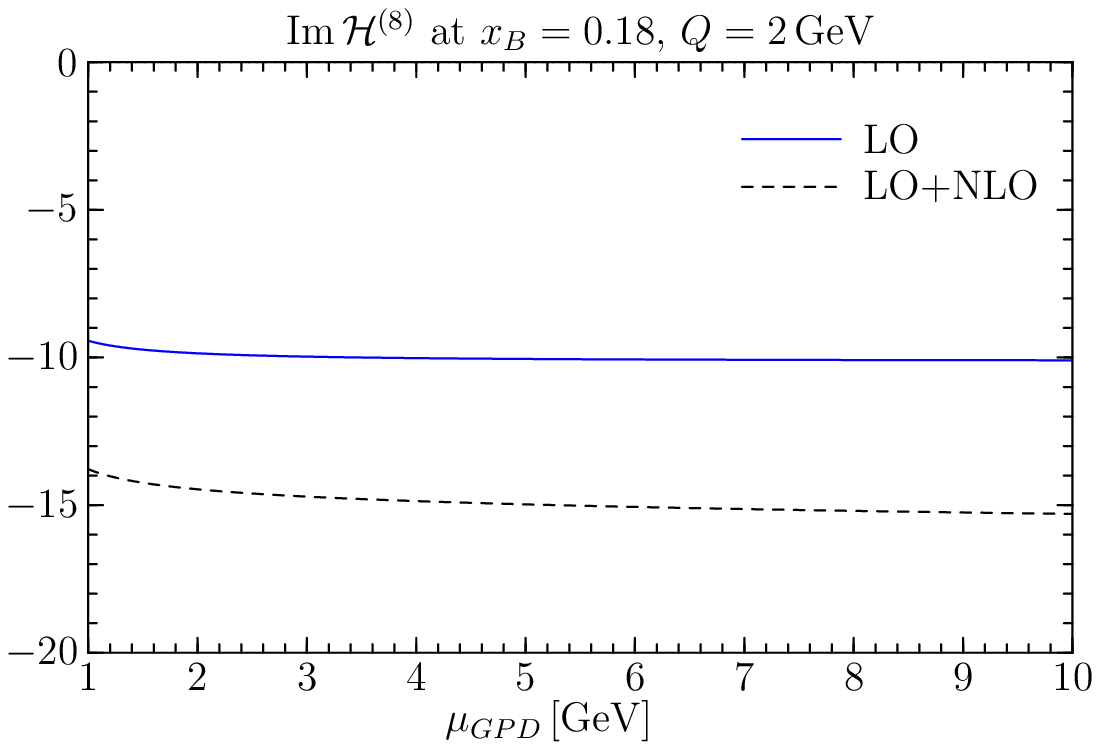}
\end{center}
\caption{\label{diff_muf_results} Dependence of the convolutions on
  $\mu_{GPD}$.  The renormalization scale is held fixed at $\mu_R =
  Q$.}
\end{figure}

\begin{figure}
\begin{center}
\includegraphics[width=\plotwidth]
{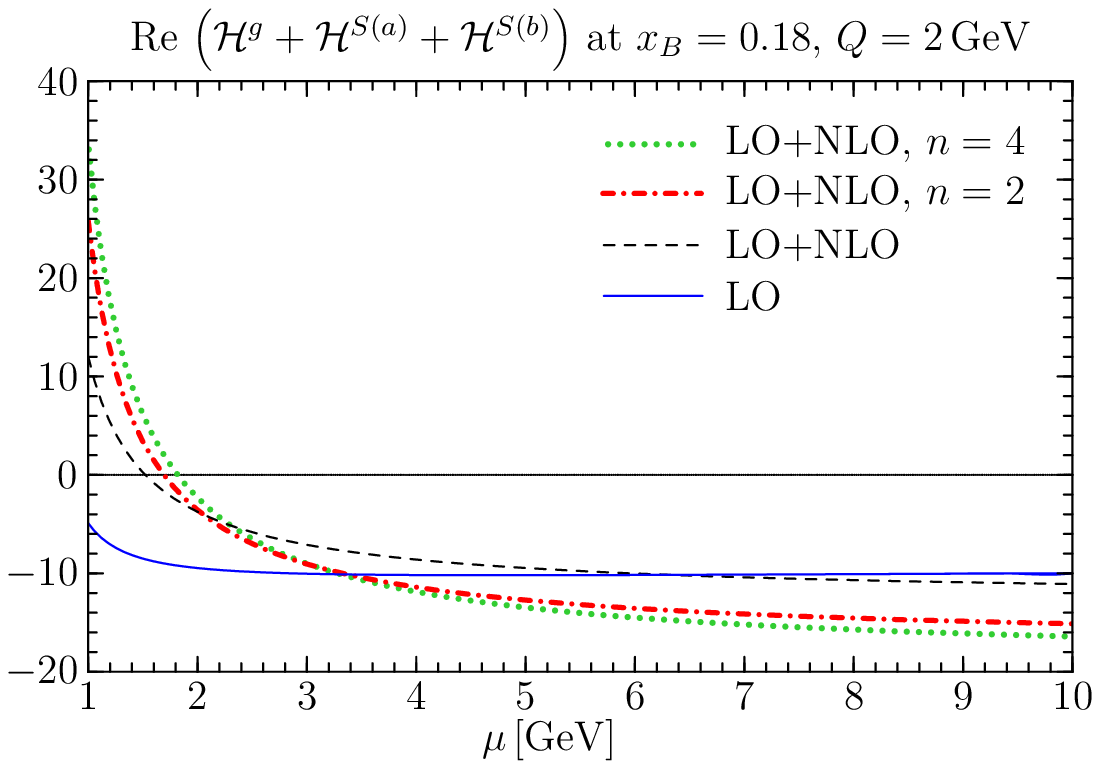}\hspace{1ex}
\includegraphics[width=\plotwidth]
{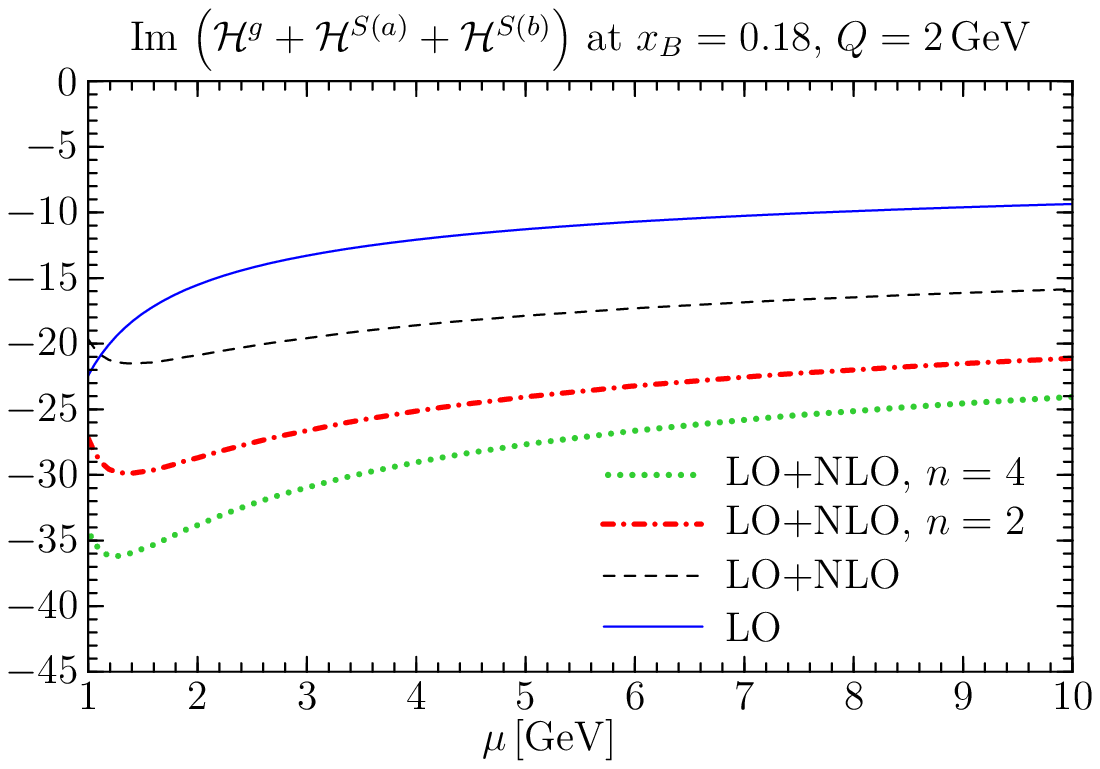}\\[1.5ex]
\includegraphics[width=\plotwidth]
{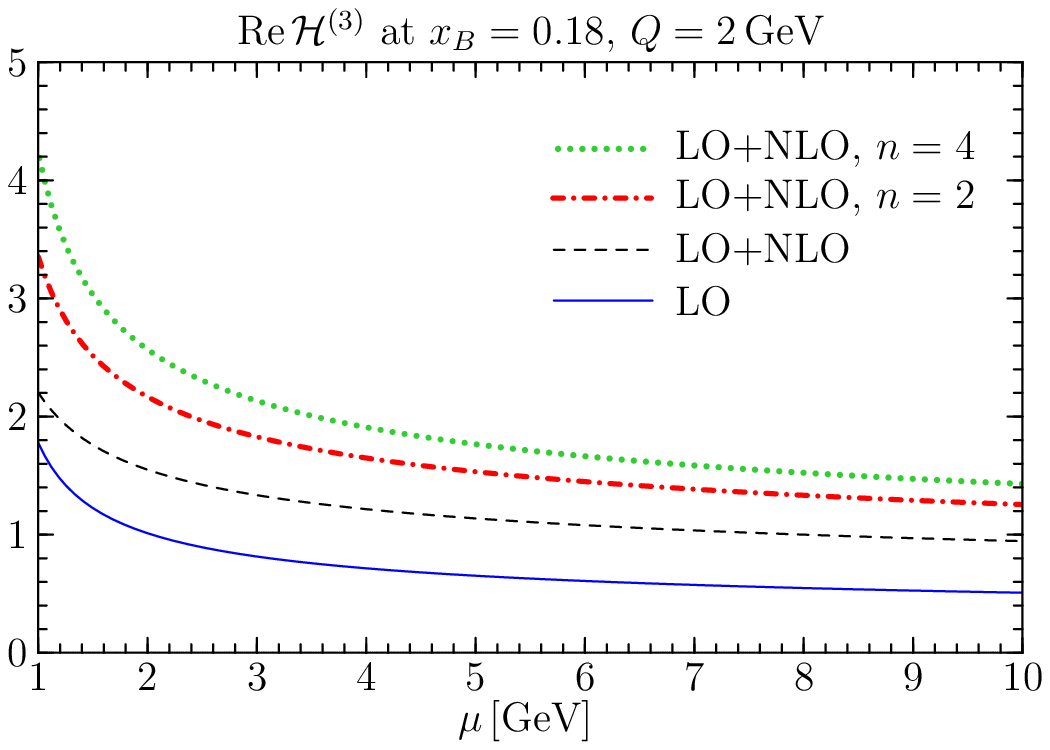}\hspace{1ex}
\includegraphics[width=\plotwidth]
{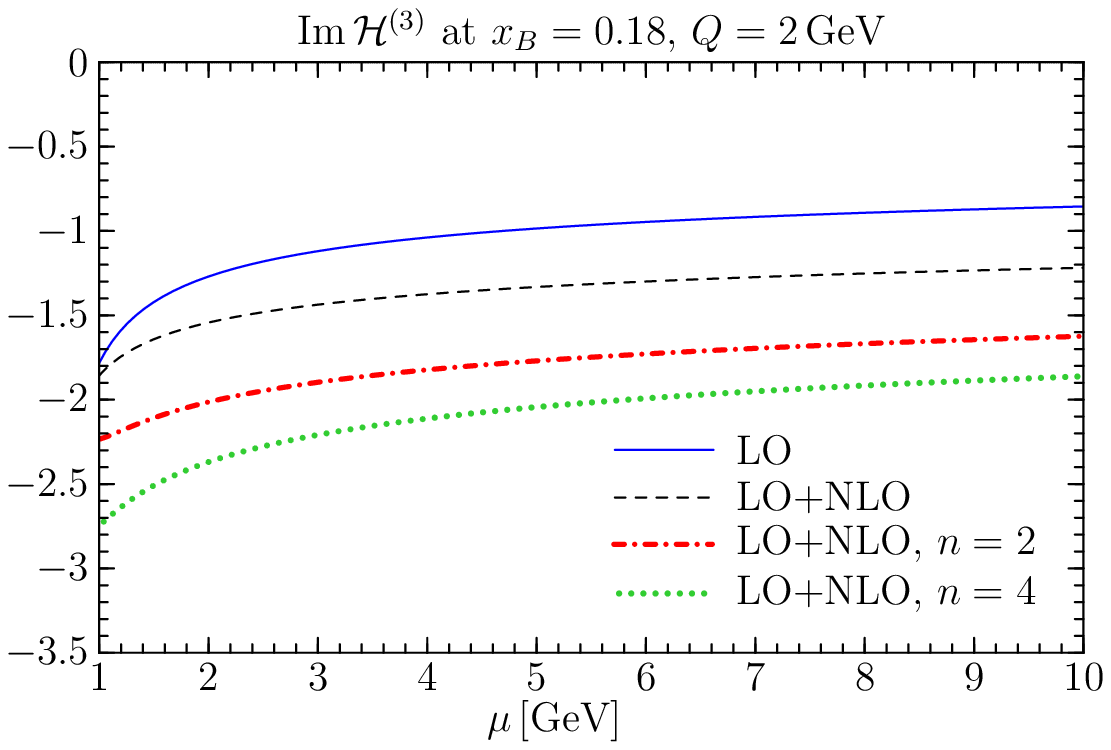}\\[1.5ex]
\includegraphics[width=\plotwidth]
{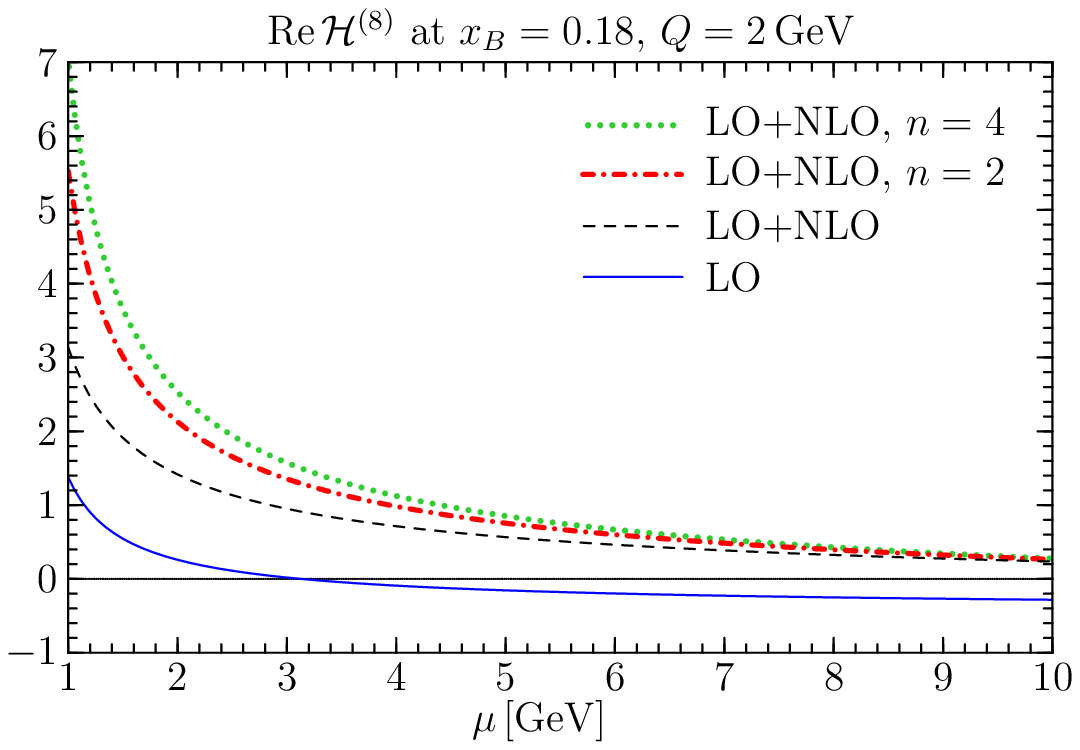}\hspace{1ex}
\includegraphics[width=\plotwidth]
{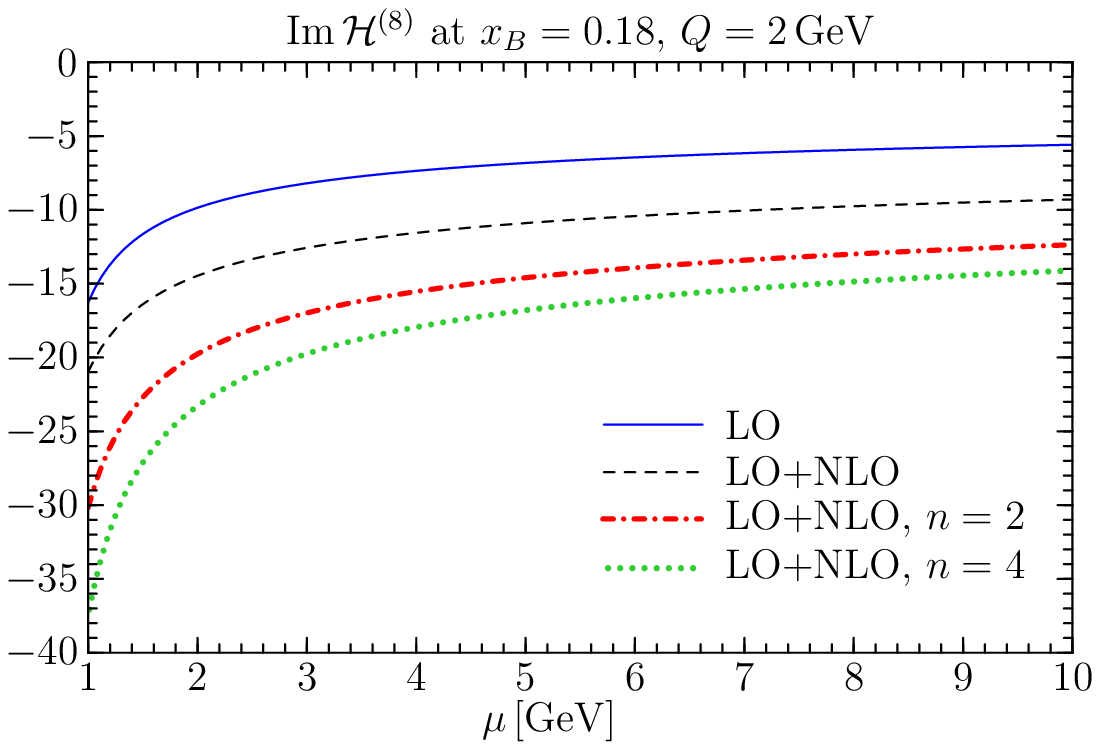}
\end{center}
\caption{\label{diff_mu_results} Dependence of the convolutions on the
  common scale $\mu = \mu_R = \mu_{GPD} = \mu_{DA}$.}
\end{figure}

Figure~\ref{diff_mur_results} shows the dependence of the convolutions
on $\mu_R$ at fixed $\mu_{GPD} = Q$.  Within the $\mu_R$ range shown
we generally find a moderate scale dependence, both at LO and at NLO.
An exception is the region $\mu_R \lsim 2\gev$, where the growth of
the LO results simply reflects the growth of $\alpha_s(\mu_R)$.  Note
that with the parameters specified at beginning of
Sect.~\ref{sec:small-x} we have $\alpha_{\smash{s}}^{(3)}(2 \gev) =
0.30$ and $\alpha_{\smash{s}}^{(3)}(1 \gev) = 0.51$.  The NLO results
further contain explicit logarithms $\ln(Q^2 /\mu_R^2)$, which in some
cases can cause corrections to grow out of control, especially for the
real parts of convolutions.  We note that for $\re\mathcal{H}^{(8)}$
the NLO correction is is unusually large compared with the LO term.
This is because of a nearby zero in $x_B$, as is seen in
Fig.~\protect\ref{ns_LO_NLO_comparison_large_x}, and should not be a
reason of particular concern.

The variation of the convolutions with $\mu_{GPD}$ at fixed $\mu_R =
Q$ is shown in Fig.~\ref{diff_muf_results}.  We again find a rather
moderate scale dependence, except when $\mu_{GPD}$ becomes small.  The
dependence on a single scale $\mu = \mu_R = \mu_{GPD}$ is shown in
Fig.~\ref{diff_mu_results}.  Note that in many cases the individual
variation of $\mu_R$ decreases the amplitude in absolute size whereas
the variation of $\mu_{GPD}$ increases it, with both tendencies
partially canceling when the scales are set equal.  Again we find that
the scale dependence becomes quite drastic below $2\gev$.

\begin{figure}
\begin{center}
\includegraphics[width=\plotwidth]
{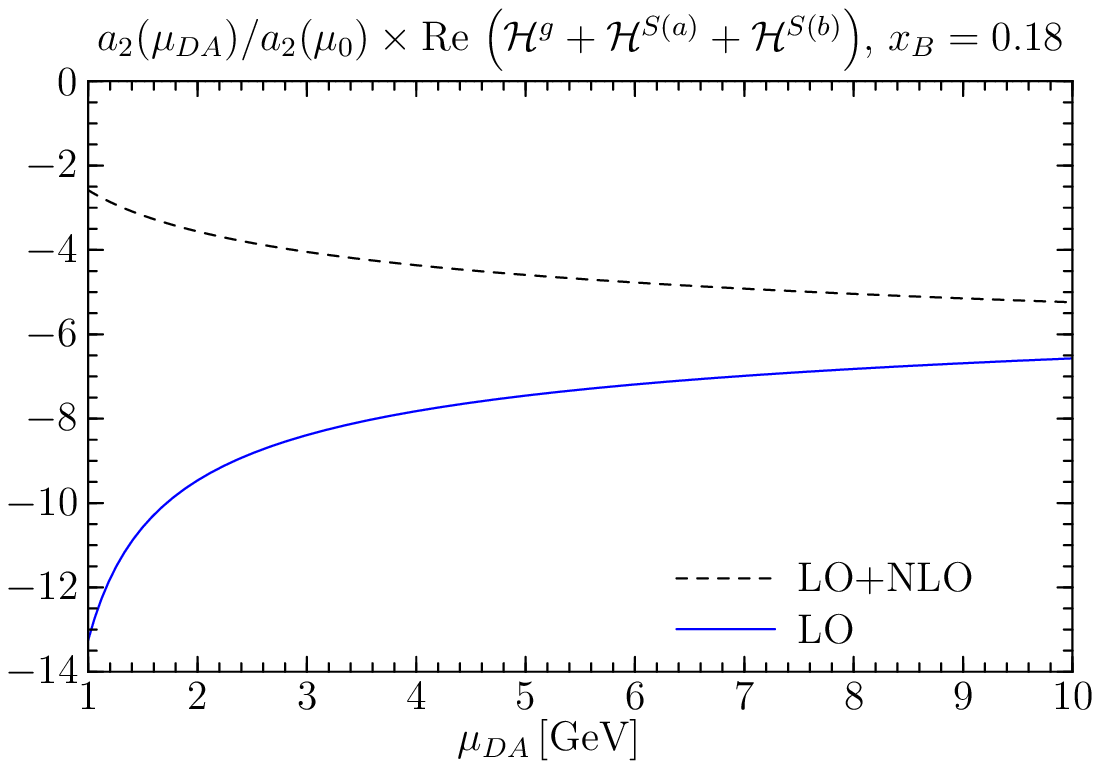}\hspace{1ex}
\includegraphics[width=\plotwidth]
{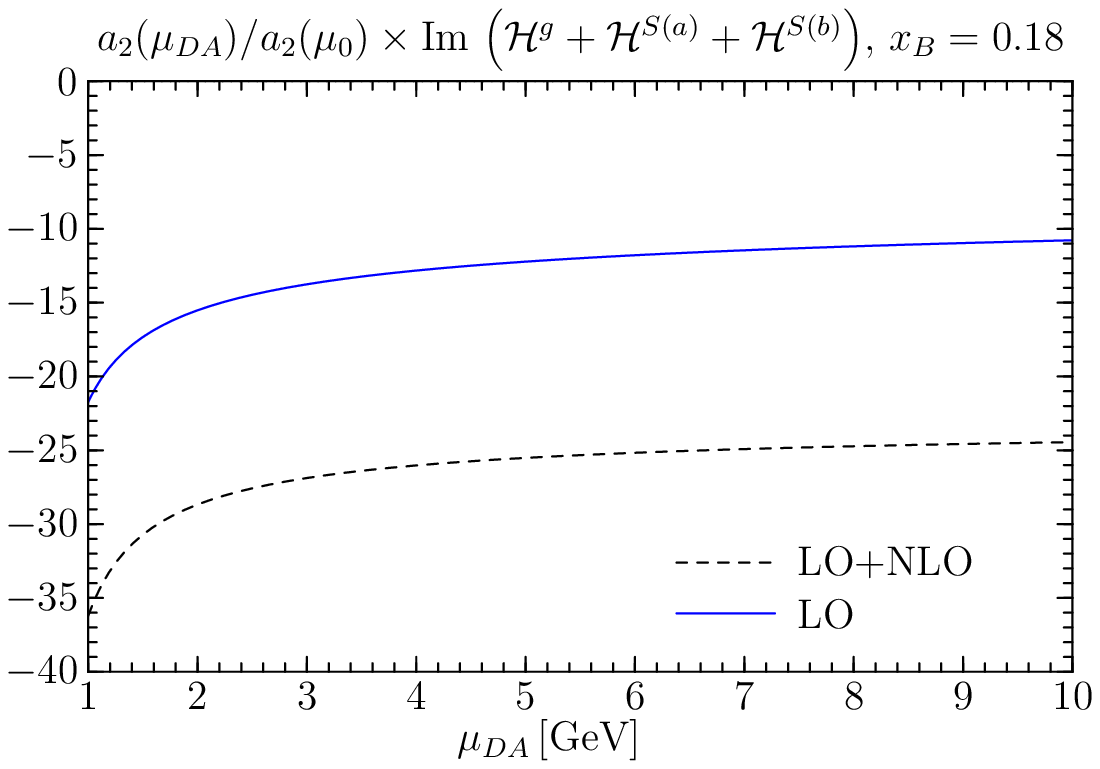}
\end{center}
\caption{\label{a2_diff_muda_results} Dependence on the factorization
  scale $\mu_{DA}$ of the convolutions in the gluon and quark singlet
  sector multiplied by the scale dependence $a_{2}(\mu_{DA})
  /a_{2}(\mu_0)$ of the corresponding Gegenbauer coefficient.  The
  reference scale for $a_{2}$ is taken as $\mu_0 = 2\gev$, and the
  other scales are set to $\mu_{GPD} = \mu_{R} = Q = 2\gev$.}
\end{figure}

We finally discuss the dependence on $\mu_{DA}$ for Gegenbauer indices
$n>0$.  According to \eqref{gegen-amp} the convolutions
$\mathcal{H}_n(\mu_{DA})$ appear multiplied by $a_n(\mu_{DA})$ in the
process amplitude, where the scale dependence of both factors
partially cancels.  In Fig.~\ref{a2_diff_muda_results} we therefore
plot convolutions multiplied with $a_2(\mu_{DA}) /a_2(\mu_0) = \bigl[
\alpha_s(\mu_{DA}) / \alpha_s(\mu_0) \bigr]{}^{\gamma_2/\beta_0}$
following the relation \eqref{DA-evol}.  The corresponding plots for
$n=4$ and for convolutions in the quark non-singlet sector look very
similar.  We find that the dependence on $\mu_{DA}$ is slightly
decreased when going to NLO.

\section{Proton helicity flip amplitudes}
\label{sec:E-convolutions}

We now turn to the convolutions of the hard-scattering kernels with
the GPDs describing proton helicity flip.  In this section we take $t=
-0.4\gev^2$, which is the value for which will present estimates for
observables in the next section.

\begin{figure}
\begin{center}
\includegraphics[width=\plotwidth]
{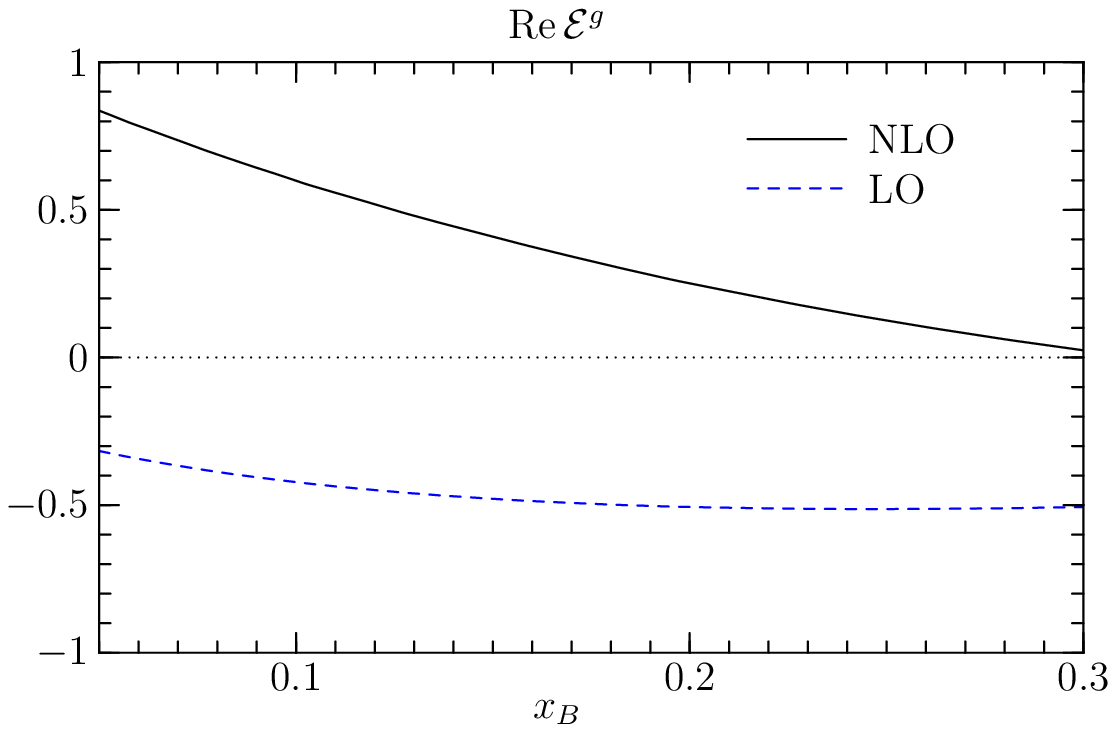}\hspace{1ex}
\includegraphics[width=\plotwidth]
{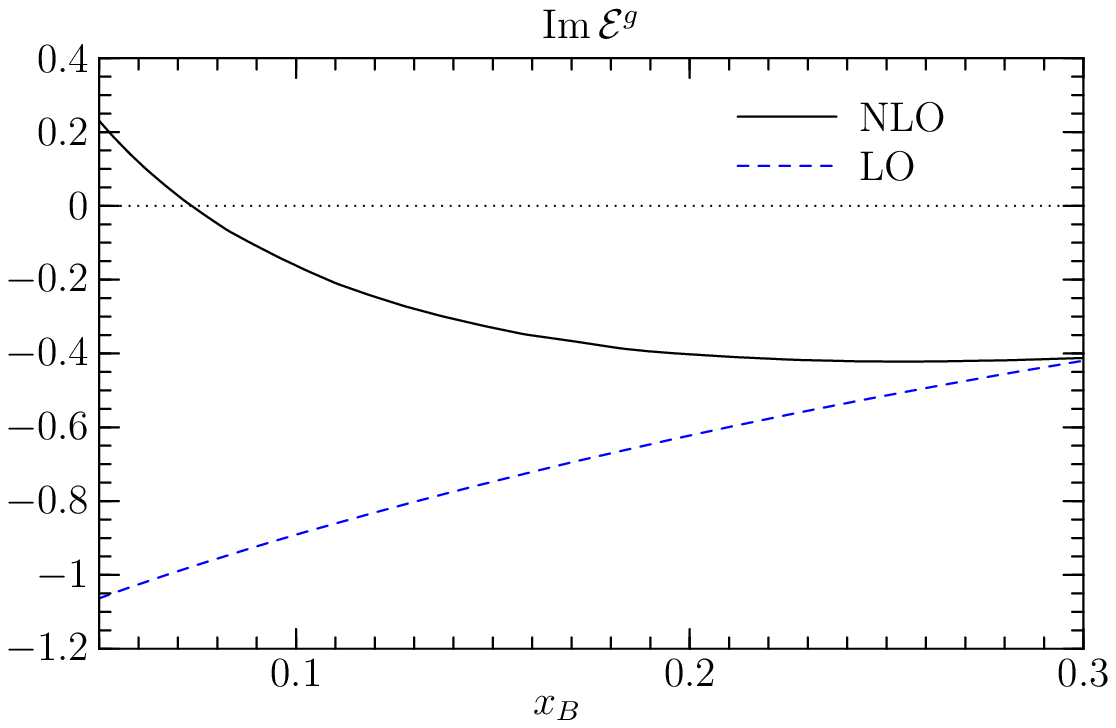}\\[0.8ex]
\includegraphics[width=\plotwidth]
{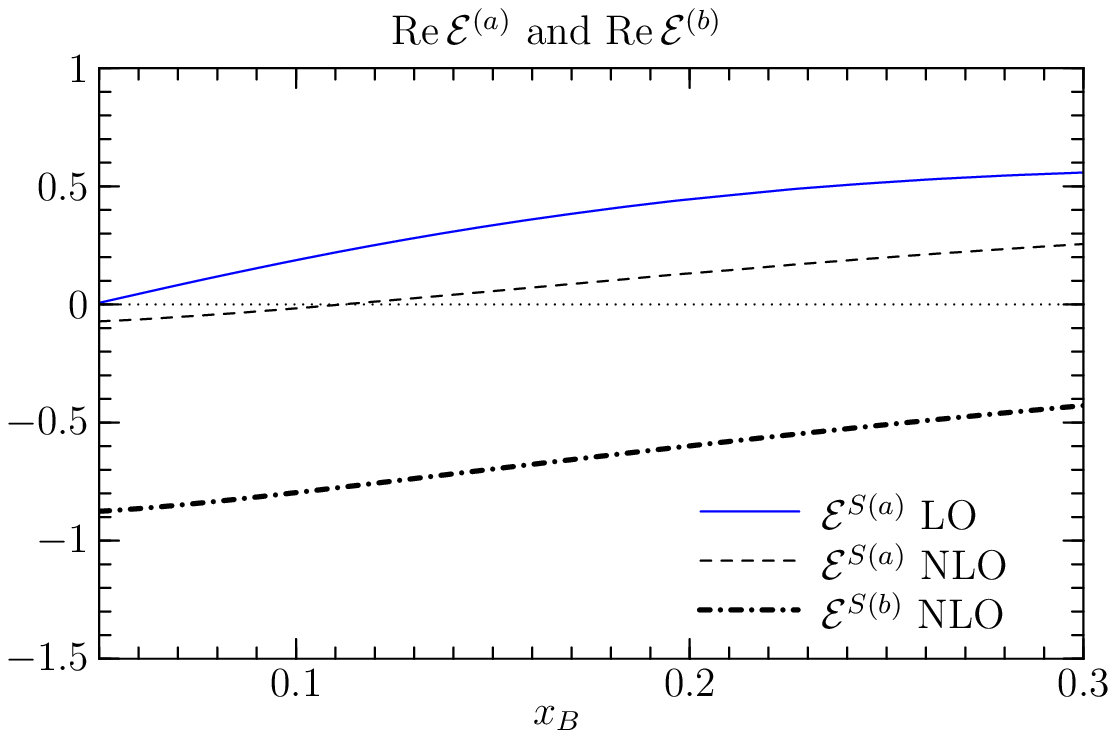}\hspace{1ex}
\includegraphics[width=\plotwidth]
{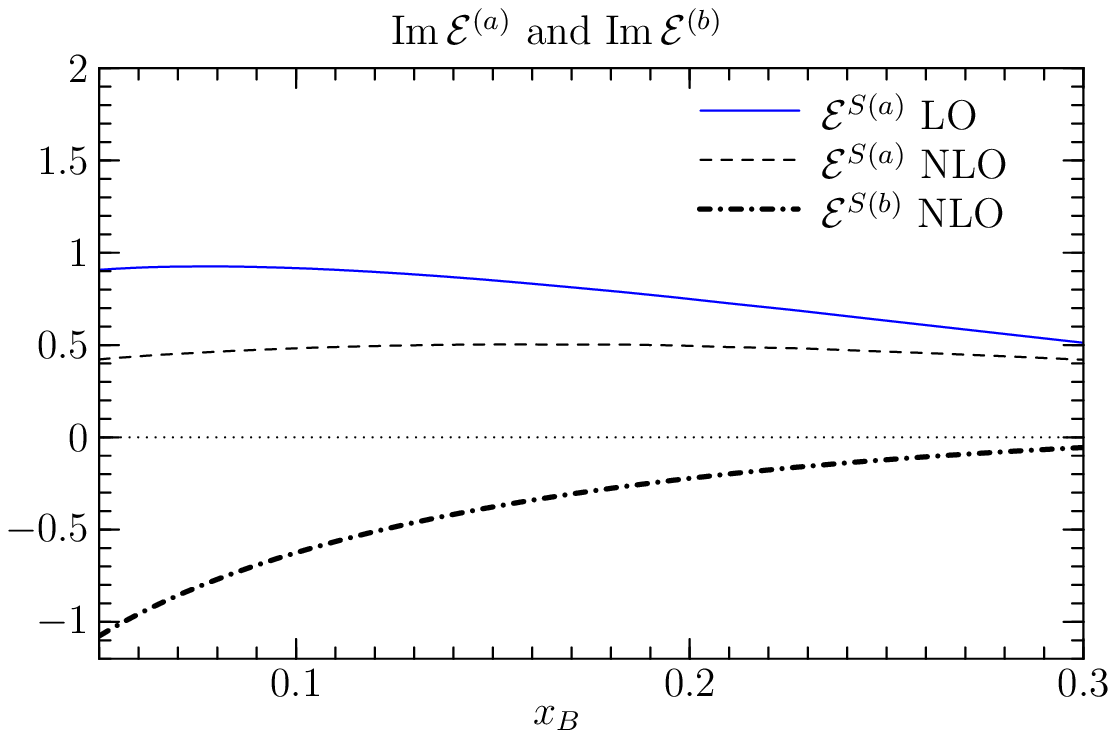}\\[0.8ex]
\includegraphics[width=\plotwidth]
{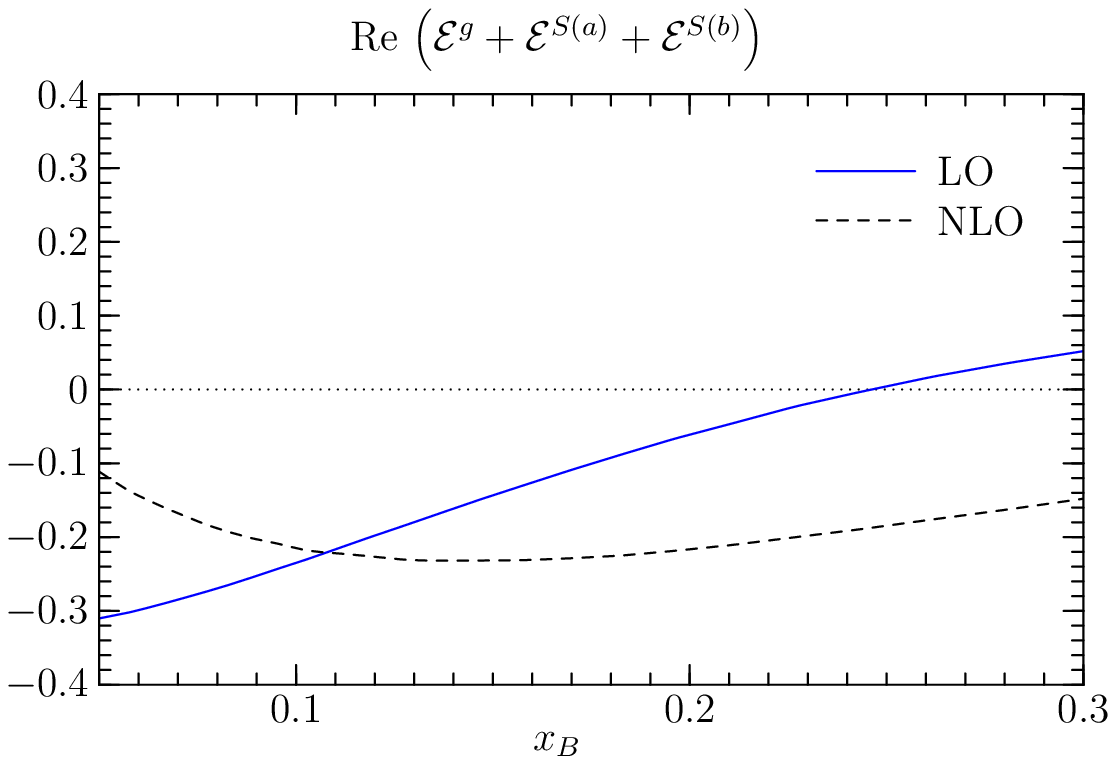}\hspace{1ex}
\includegraphics[width=\plotwidth]
{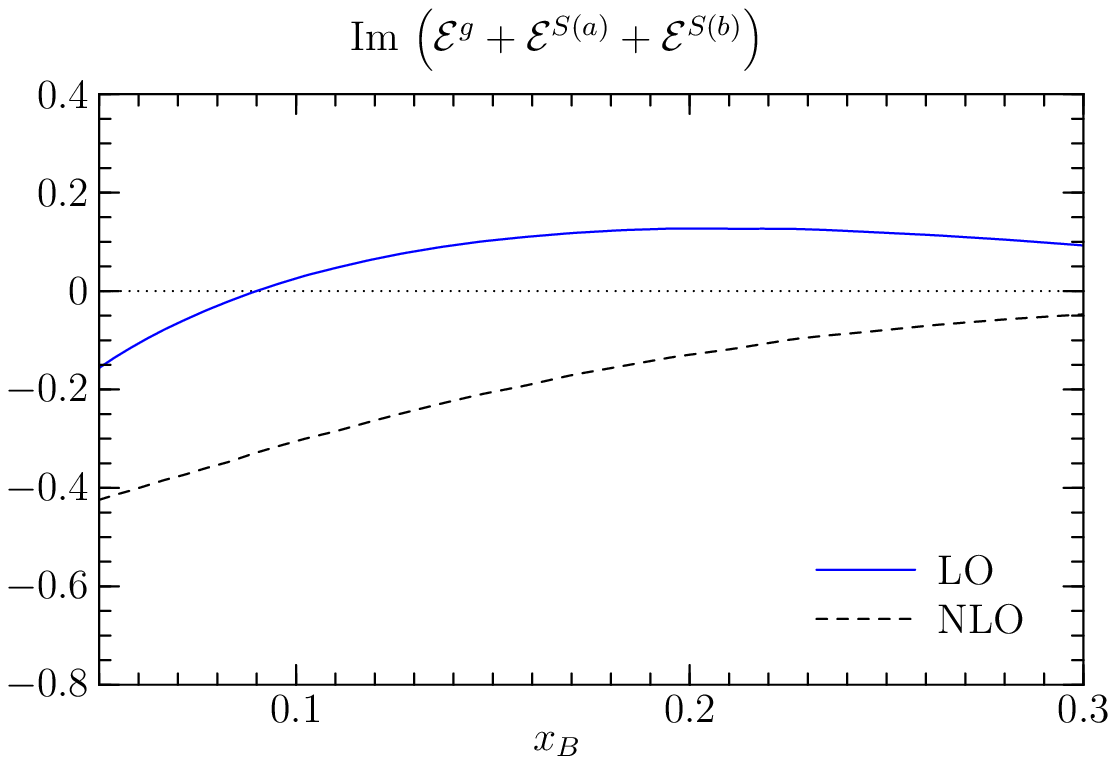}
\end{center}
\caption{\label{E-gs-mod-1} LO and NLO terms of the convolutions in
  the gluon and quark singlet sector for model 1 at $Q = 2 \gev$ and
  $t= -0.4 \gev^2$.  The scales are set to $\mu_R = \mu_{GPD} = Q$.}
\end{figure}

\begin{figure}
\begin{center}
\includegraphics[width=\plotwidth]
{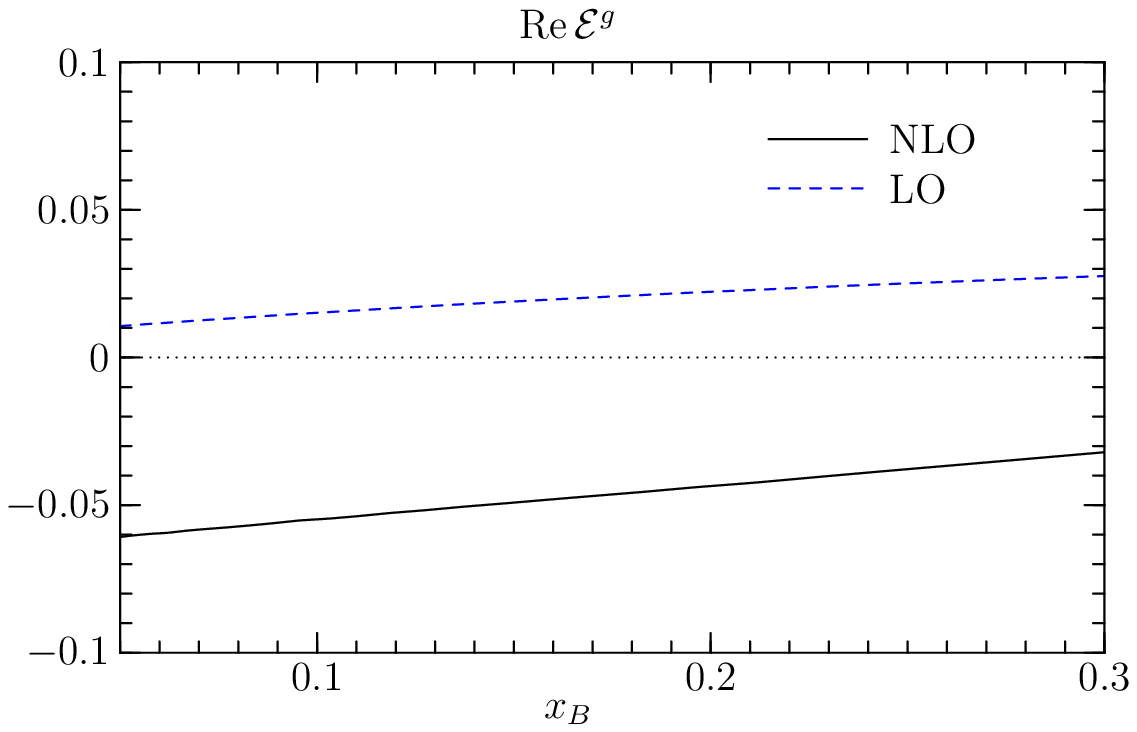}\hspace{1ex}
\includegraphics[width=\plotwidth]
{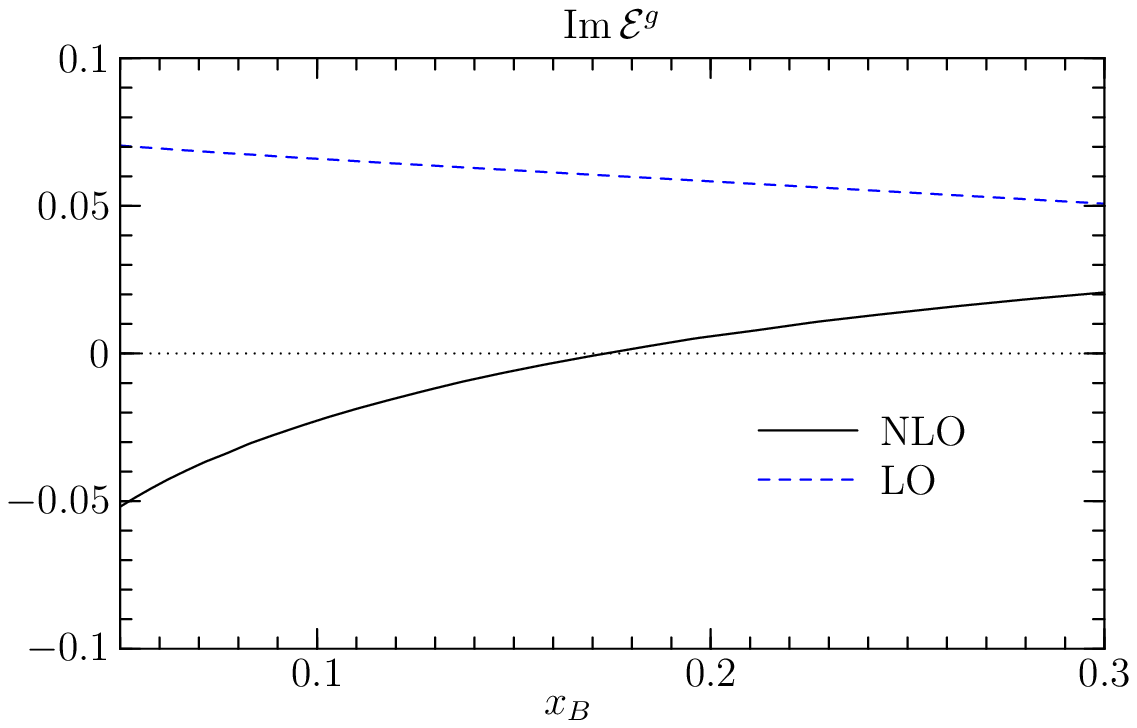}\\[0.8ex]
\includegraphics[width=\plotwidth]
{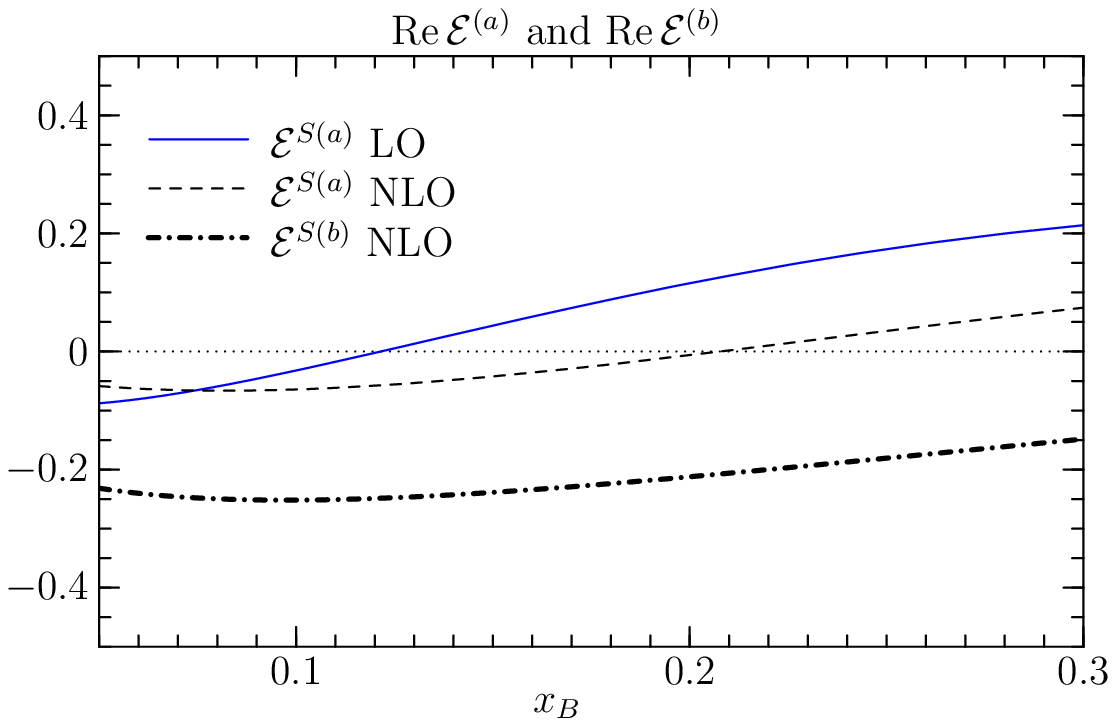}\hspace{1ex}
\includegraphics[width=\plotwidth]
{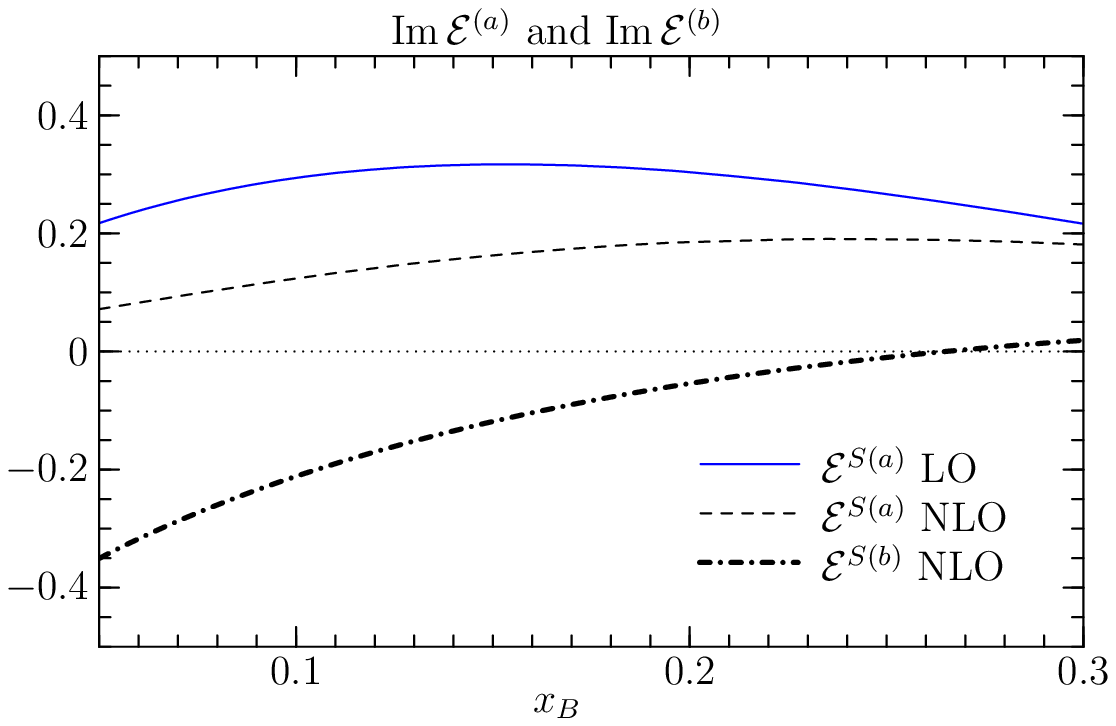}\\[0.8ex]
\includegraphics[width=\plotwidth]
{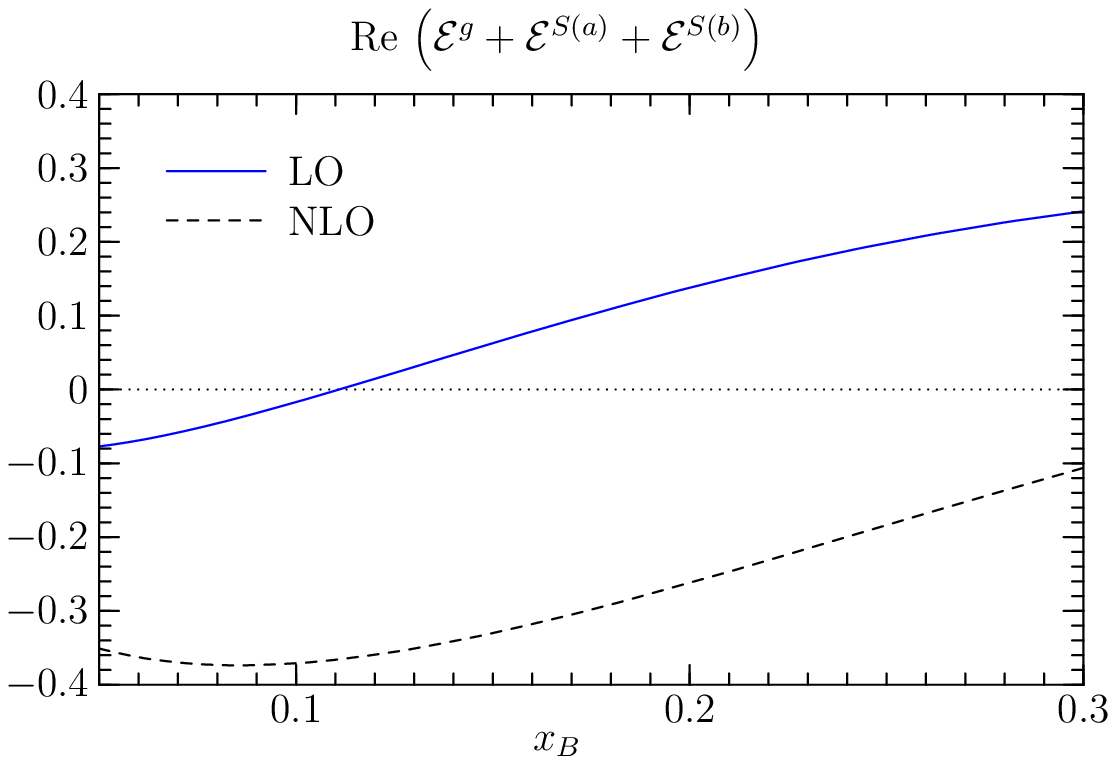}\hspace{1ex}
\includegraphics[width=\plotwidth]
{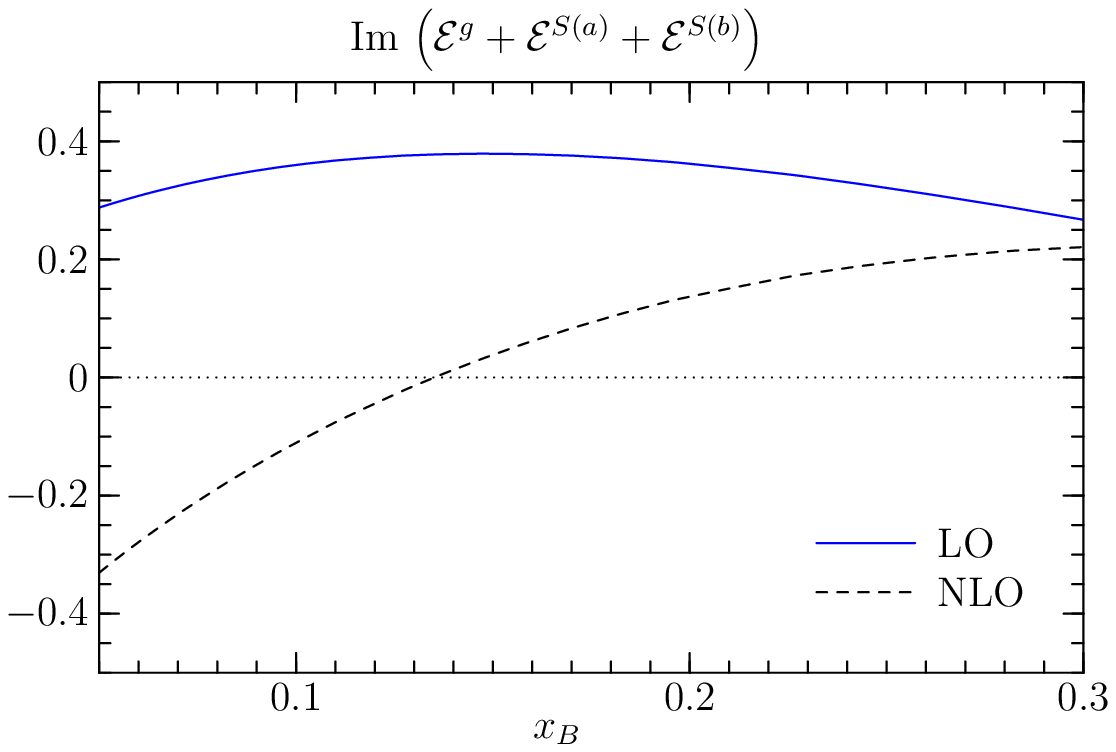}
\end{center}
\caption{\label{E-gs-mod-2} As Fig.~\protect\ref{E-gs-mod-1} but for
  model 2.}
\end{figure}

\begin{figure}
\begin{center}
\includegraphics[width=\plotwidth]
{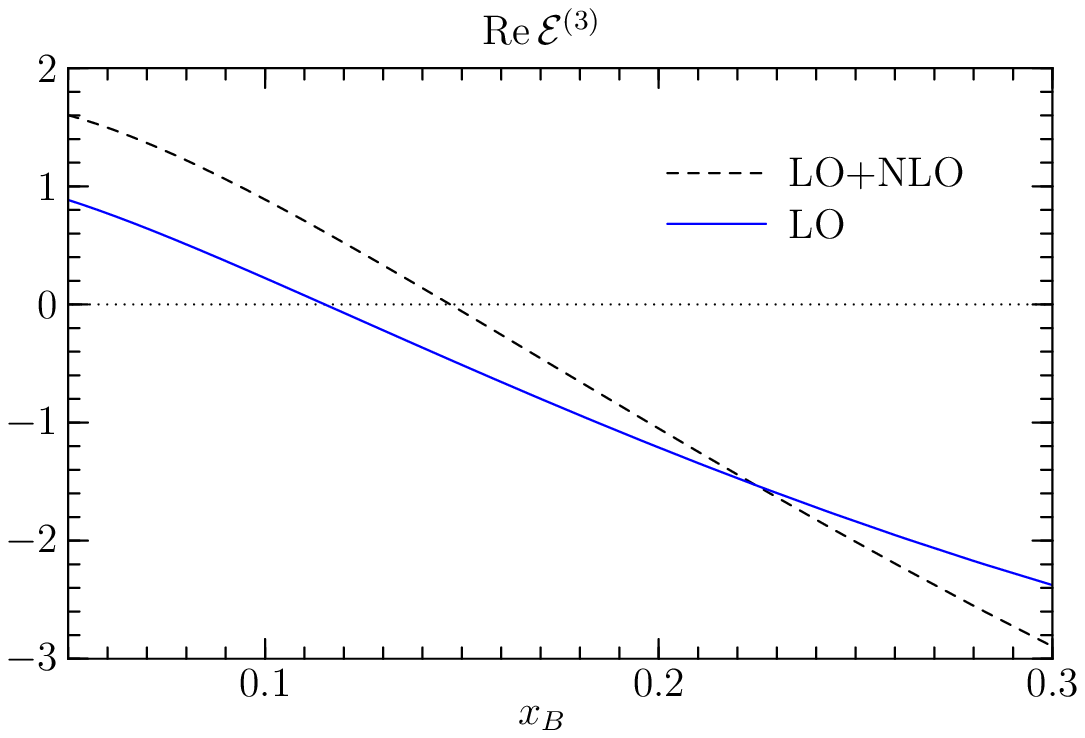}\hspace{1ex}
\includegraphics[width=\plotwidth]
{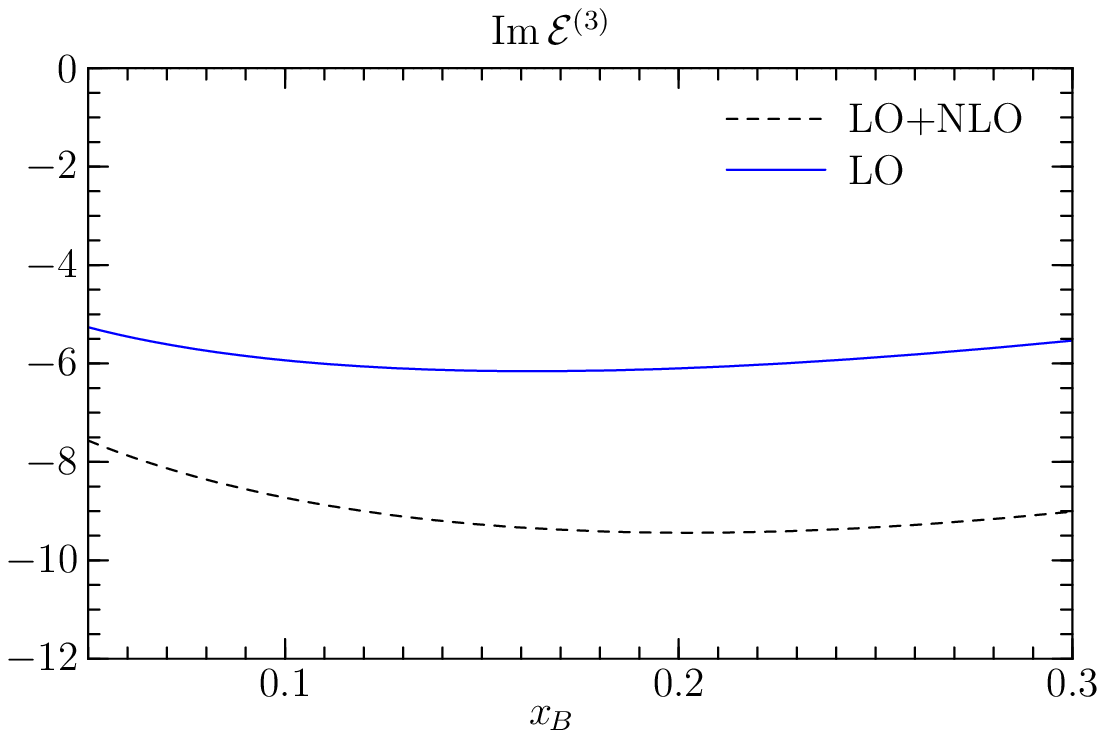}\\[0.8ex]
\includegraphics[width=\plotwidth]
{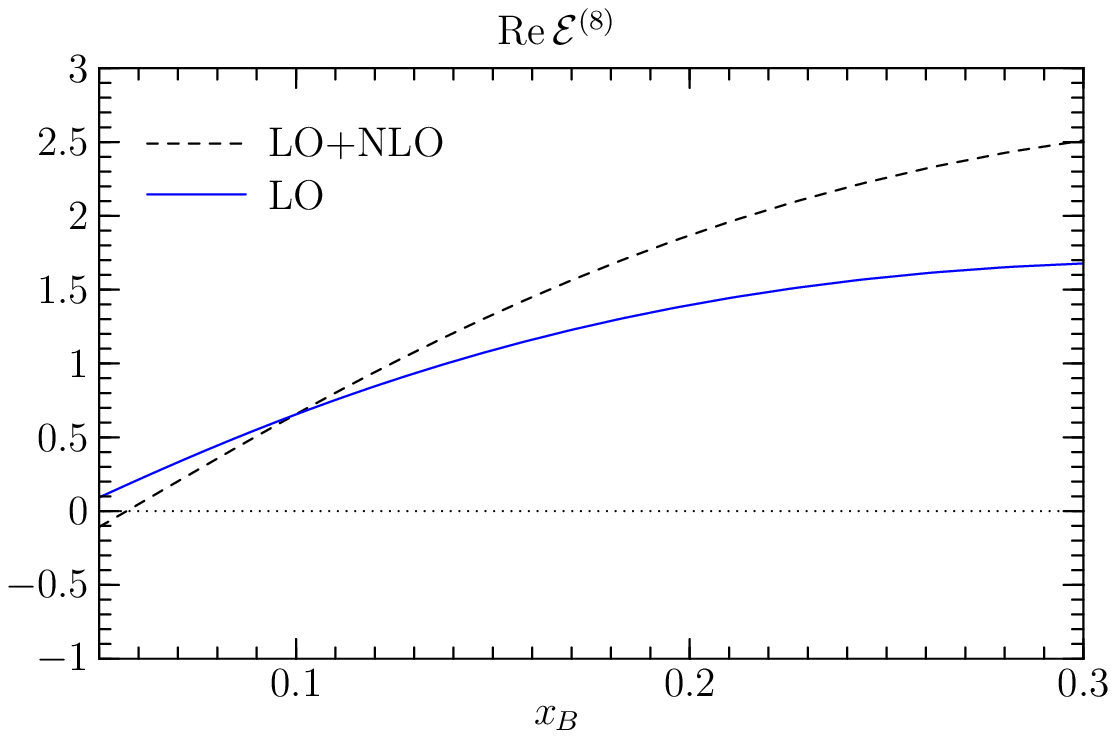}\hspace{1ex}
\includegraphics[width=\plotwidth]
{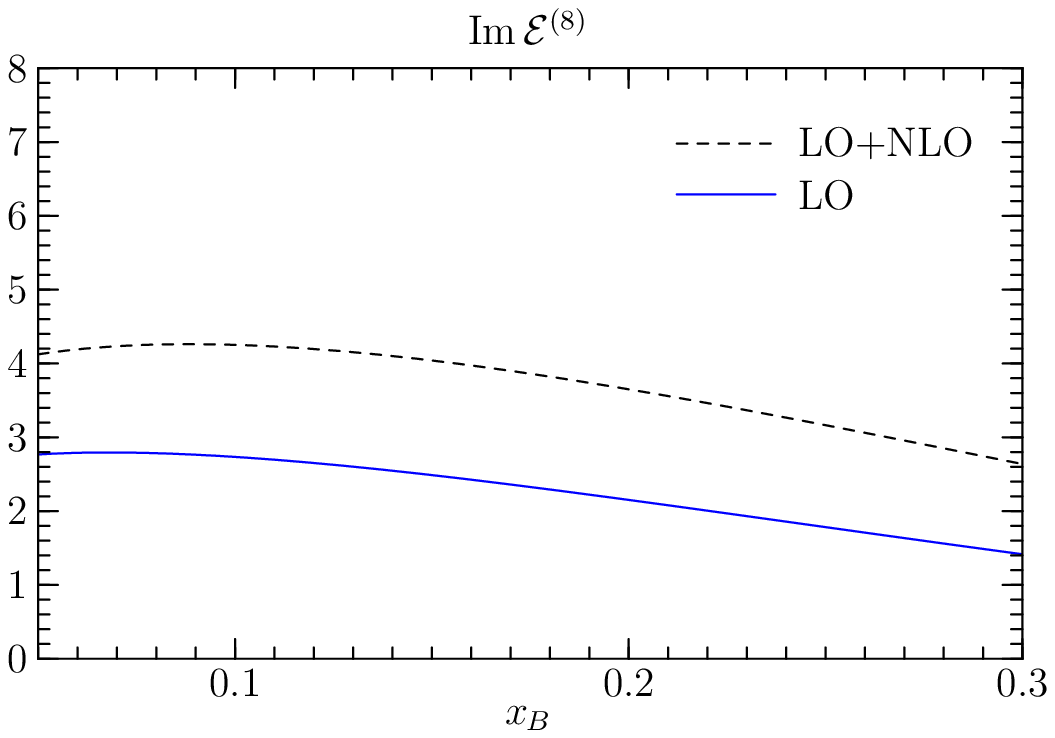}
\end{center}
\caption{\label{E-ns-mod-1} LO and NLO terms of the convolutions in
  the quark non-singlet sector for model 1 at $Q = 2 \gev$ and $t= -0.4
  \gev^2$, with $\mu_R = \mu_{GPD} = Q$.}
\end{figure}

\begin{figure}
\begin{center}
\includegraphics[width=\plotwidth]
{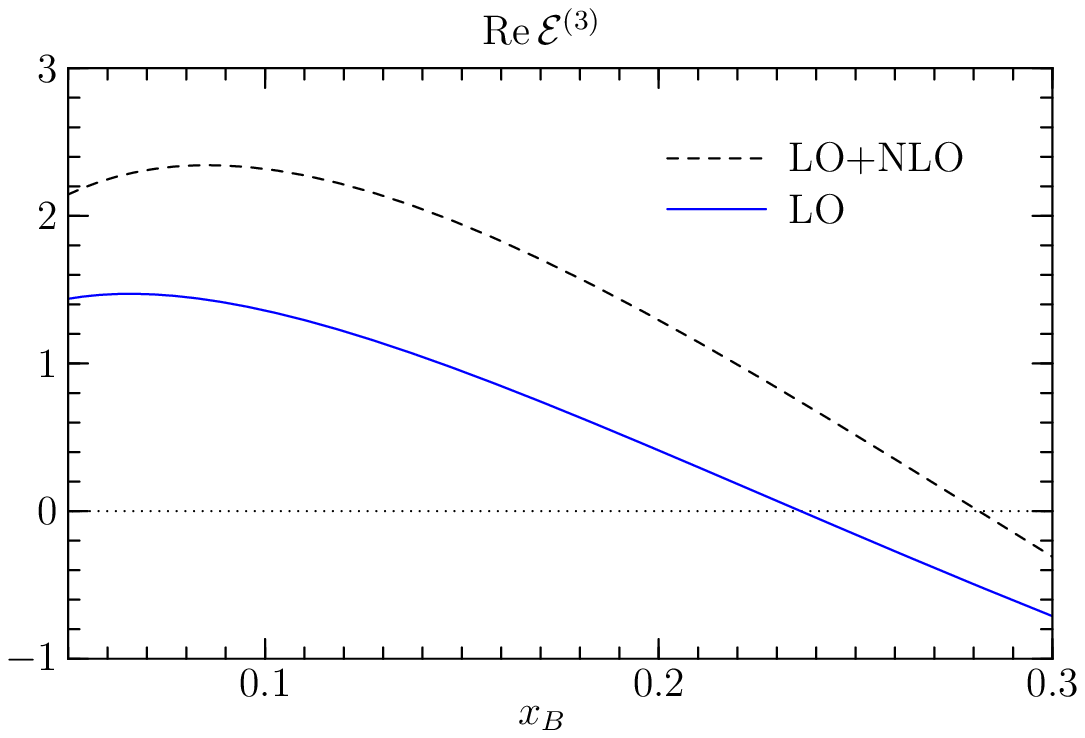}\hspace{1ex}
\includegraphics[width=\plotwidth]
{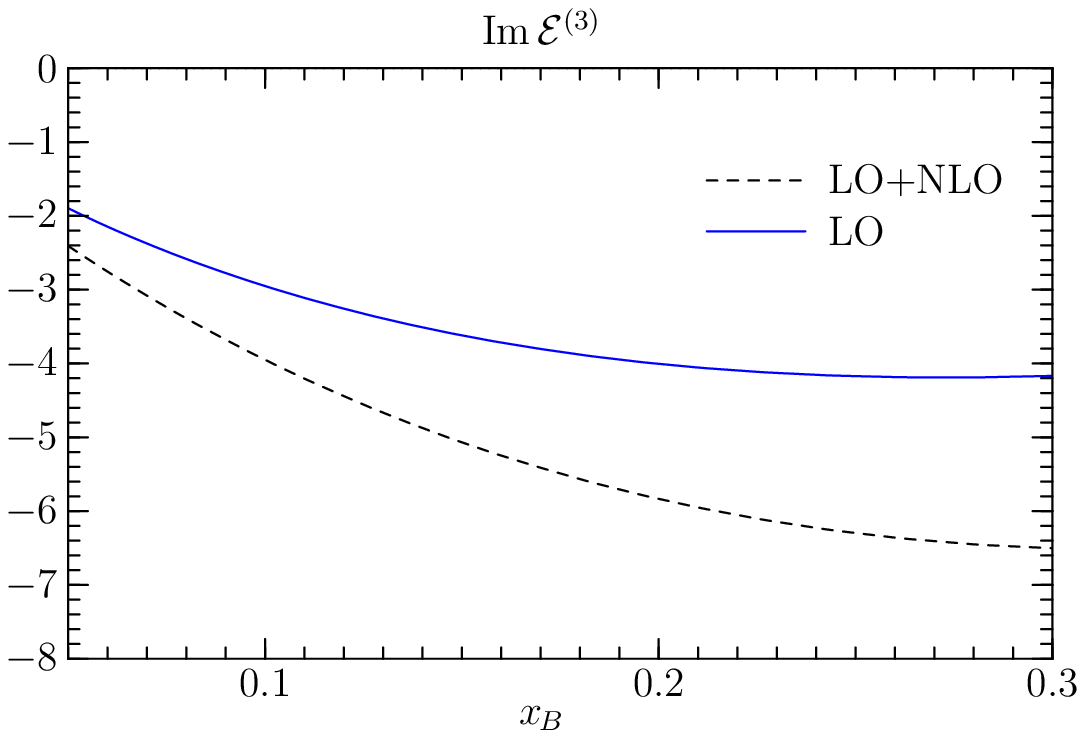}\\[0.8ex]
\includegraphics[width=\plotwidth]
{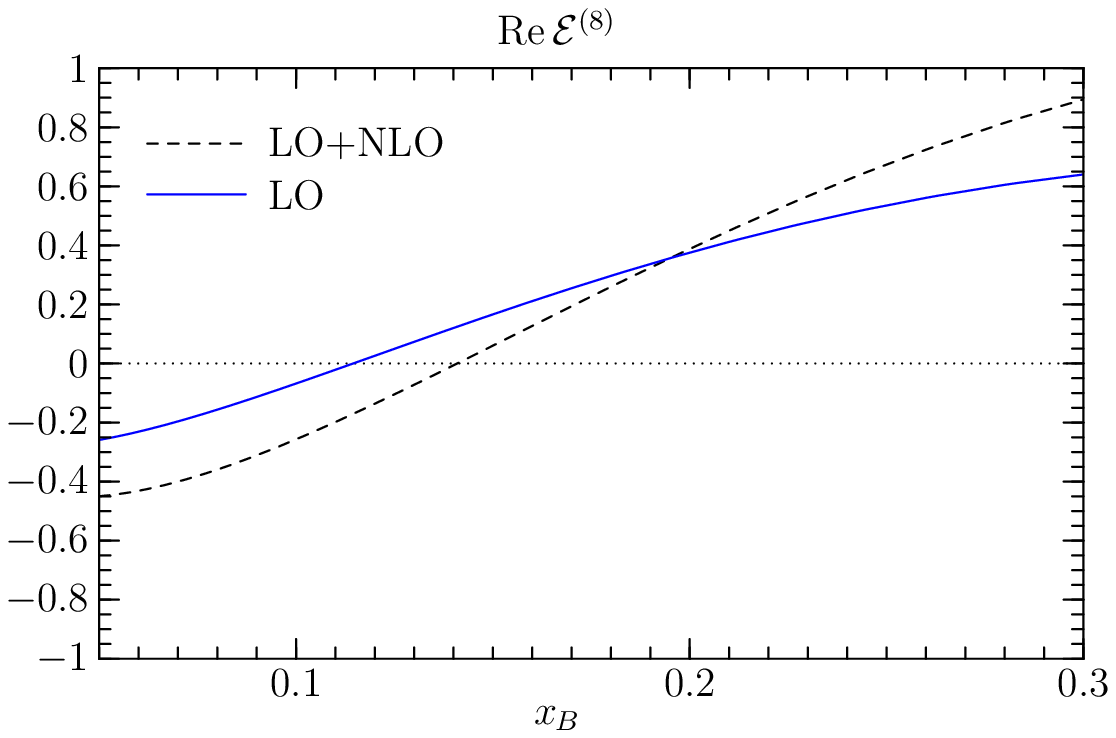}\hspace{1ex}
\includegraphics[width=\plotwidth]
{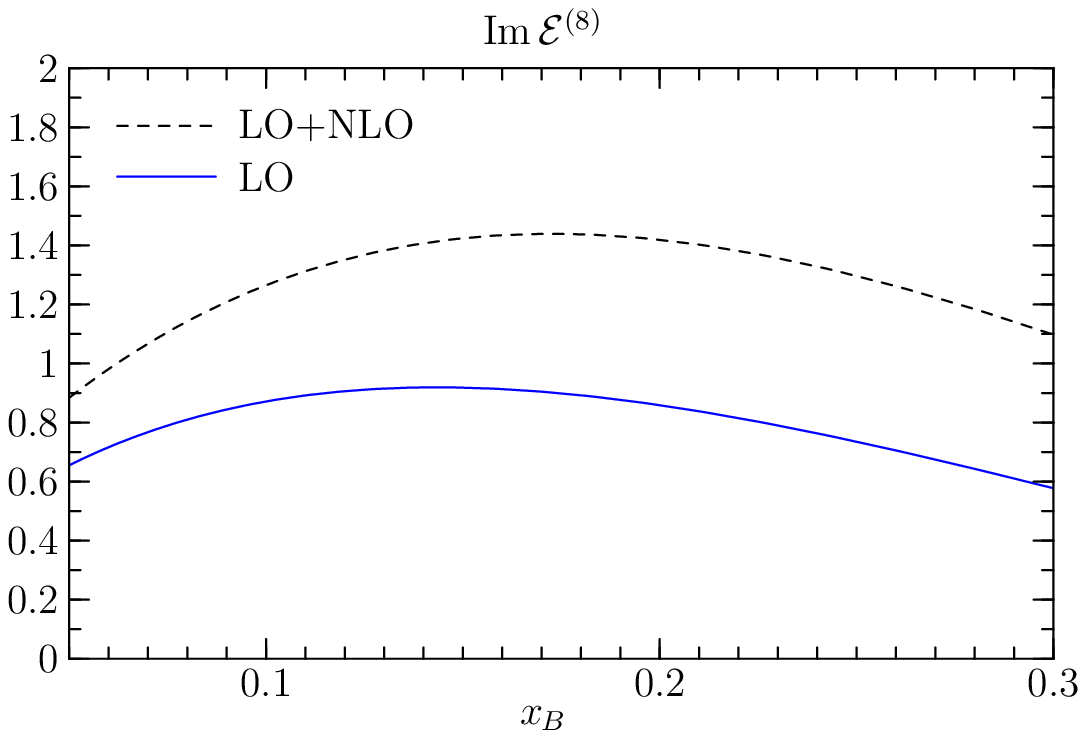}
\end{center}
\caption{\label{E-ns-mod-2} As Fig.~\protect\ref{E-ns-mod-1} but for
  model 2.}
\end{figure}

In Figs.~\ref{E-gs-mod-1} and \ref{E-gs-mod-2} we compare the LO and
NLO terms of the convolutions in the gluon and quark singlet sector
for the two models described in Sect.~\ref{sec:e-model}.  For model 1
the individual corrections for gluon and quark convolutions look quite
similar to those we saw for $\mathcal{H}$ in the previous section.
The sum of gluon and quark singlet contributions at LO is however very
small in this model because of cancellations, so that the NLO term
dominates in a wide kinematical region.  In model 2 the gluon
contributions are nearly absent, so that the quark singlet
contribution dominates in this sector.  We note that, contrary to the
individual terms, the sum of gluon and quark singlet contributions
comes out to be rather similar in the two models and is small compared
with the flavor non-singlet contributions shown in
Figs.~\ref{E-ns-mod-1} and \ref{E-ns-mod-2}.  According to our
discussion at the end of Sect.~\ref{sec:e-model} this has its origin
in the sum rule \eqref{zero-sum-rule} for the second moment of $E$ at
$t=0$, so that we expect a small net contribution from gluons and the
quark singlet in large class of models for $E$.

As shown in Figs.~\ref{E-ns-mod-1} and \ref{E-ns-mod-2}, the NLO
corrections to the quark non-singlet convolutions are relatively
moderate but not small, similarly to the case of $\mathcal{H}$.  The
size of the convolutions is quite different in the two models,
indicating the important role played at intermediate $x_B$ by sea
quarks in model 1.  Let us recall that with the double distribution
ansatz \eqref{dd-models-e} the GPDs at $x\sim \xi$ are sensitive to
forward parton distributions with momentum fractions well below $\xi$,
as discussed in Sect.~4.3.3 of~\cite{Diehl:2003ny}.


\section{Cross sections and asymmetries}
\label{sec:cross}

Having studied in detail the building blocks of the scattering
amplitude for vector meson production, we now combine them to
observables.  We recall that to leading order in $1/Q$ there are just
two of these: the unpolarized $\gamma^* p$ cross section and the
asymmetry for a transversely polarized target, both referring to
longitudinal polarization of virtual photon and produced meson.  The
$ep$ cross section in the leading $1/Q$ approximation can be written
as
\begin{equation}
  \label{AUT-def}
\frac{d\sigma(ep\to epV)}{dt\, dQ^2\, dy\, d\phi\, d\phi_S}
 = \frac{\alpha}{4\pi^3}\, \frac{1-x_B}{Q^2}\,
   \frac{1-y}{y}\, \frac{d\sigma_L}{dt}
   \Bigl[ 1 + S_T \sin(\phi-\phi_S)\ms A_{UT} \Bigr]
\end{equation}
where $y$ is the usual inelasticity variable for deep inelastic
scattering and $S_T$ denotes the transverse component of the target
polarization.  $\phi$ is the azimuthal angle between lepton plane and
hadron plane, and $\phi_S$ is the azimuthal angle between lepton plane
and target spin vector, both defined according to the Trento
convention \cite{Bacchetta:2004jz}.  The $\gamma^* p$ cross section
$d\sigma_L/dt$ and the polarization asymmetry $A_{UT}$ depend on
$x_B$, $Q^2$ and $t$.  To leading order in $1/Q$ they are given by
\begin{align}
  \label{sigmaL}
\frac{d\sigma_L}{dt} &=
\frac{\pi^2}{9}\, \frac{\alpha}{Q^6}\,
\frac{(2-x_B)^2}{1-x_B}\, f_V^2\,
\Bigl[\ms (1-\xi^2)\, | \mathcal{H}_V^{} |^2
   - \bigl(\ms t/(4m_p^2) + \xi^2 \ms\bigr)\, | \mathcal{E}_V^{} |^2
   - 2\xi^2 \re \bigl( \mathcal{E}_V^*\ms \mathcal{H}_V^{} \bigr)
\ms \Bigr]
\intertext{and}
  \label{AUT}
A_{UT} &= \frac{\sqrt{t_0-t\rule{0pt}{1.8ex}}}{m_p}\,
   \frac{\sqrt{1-\xi^2}\,
     \im\bigl( \mathcal{E}_V^*\ms
               \mathcal{H}_V^{} \bigr)}{\rule{0pt}{1em}
   (1-\xi^2)\, | \mathcal{H}_V^{} |^2
   - \bigl(\ms t/(4m_p^2) + \xi^2 \ms\bigr)\, | \mathcal{E}_V^{} |^2
   - 2\xi^2 \re \bigl( \mathcal{E}_V^*\ms \mathcal{H}_V^{} \bigr)} \,,
\end{align}
where $t_{0} = -4 m_p^2\ms \xi^2 /(1-\xi^2)$.  Here we have combined
the convolutions \eqref{F-def} into
\begin{equation}
  \label{F_V-def}
\mathcal{F}_V = Q_V \sum_{n=0}^\infty a_n^{}\ms
\biggl[ \mathcal{F}^g_n + \mathcal{H}^{S(a)}_n + \mathcal{F}^{S(b)}_n
      + e_{V}^{(3)}\, \mathcal{F}^{(3)}_{n\phantom{V}}
      + e_{V}^{(8)}\, \mathcal{F}^{(8)}_{n\phantom{V}} \biggr] \,,
\end{equation}
with analogous combinations for $\mathcal{H}_V$ and $\mathcal{E}_V$.
In the remainder of this section we take the asymptotic form of the
meson distribution amplitude, i.e.\ we set $a_n=0$ for $n\ge 2$.  As
long as $\mathcal{E}_V$ is not much larger than $\mathcal{H}_V$, the
cross section \eqref{sigmaL} is dominated by the term with
$|\mathcal{H}_V|^2$ in a wide range of kinematics, where the
prefactors $\xi^2$ and $t/(4m_p^2)$ of the other terms are small.  The
asymmetry \eqref{AUT} is then approximately given by
\begin{equation}
  \label{asy-approx}
A_{UT} \approx \frac{\sqrt{t_0-t\rule{0pt}{1.8ex}}}{m_p} \;
  \frac{\im\bigl( \mathcal{E}_V^{*}\ms
                  \mathcal{H}_V^{} \bigr)}{| \mathcal{H}_V^{} |^2}
= \frac{\sqrt{t_0-t\rule{0pt}{1.8ex}}}{m_p} \;
  \biggl| \frac{\mathcal{E}_V}{\mathcal{H}_V}\, \biggr|
  \sin\delta_V \,,
\end{equation}
where $\delta_V = \arg (\mathcal{H}_V /\mathcal{E}_V)$ is the
relative phase between $\mathcal{H}_V$ and $\mathcal{E}_V$.

\begin{figure}
\begin{center}
\includegraphics[width=\plotwidth]
{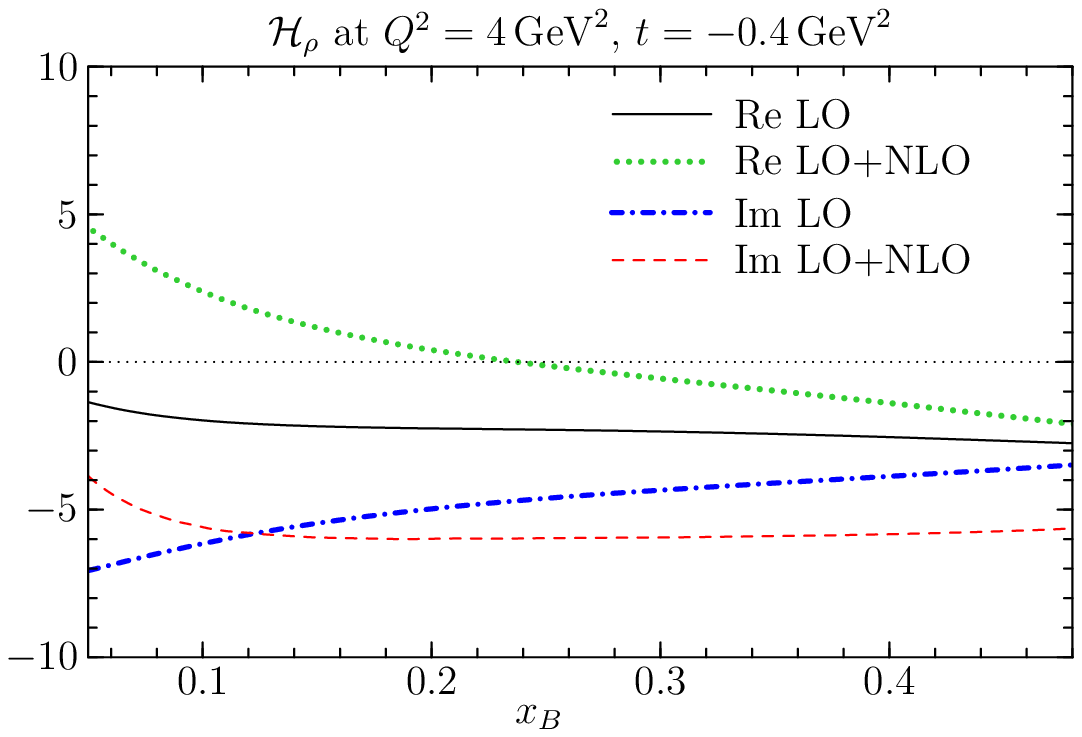}\hspace{1ex}
\includegraphics[width=\plotwidth]
{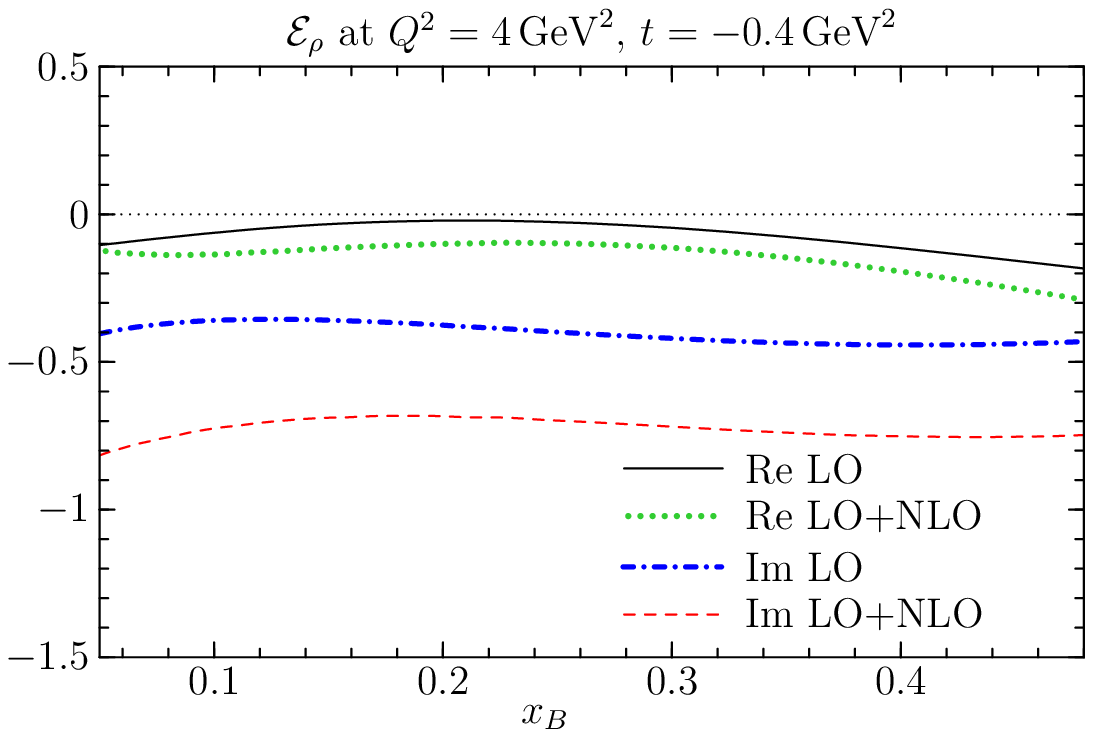}
\end{center} 
\caption{\label{rho_re_im} Real and imaginary parts of the
  convolutions \protect\eqref{F_V-def} for $\rho$ production at $Q=2
  \gev$ and $t= -0.4\gev^2$, with model 1 taken for the proton
  helicity-flip distributions.  The scales are set to $\mu_R=
  \mu_{GPD} =Q$.}
%
\begin{center}
\includegraphics[width=\plotwidth]
{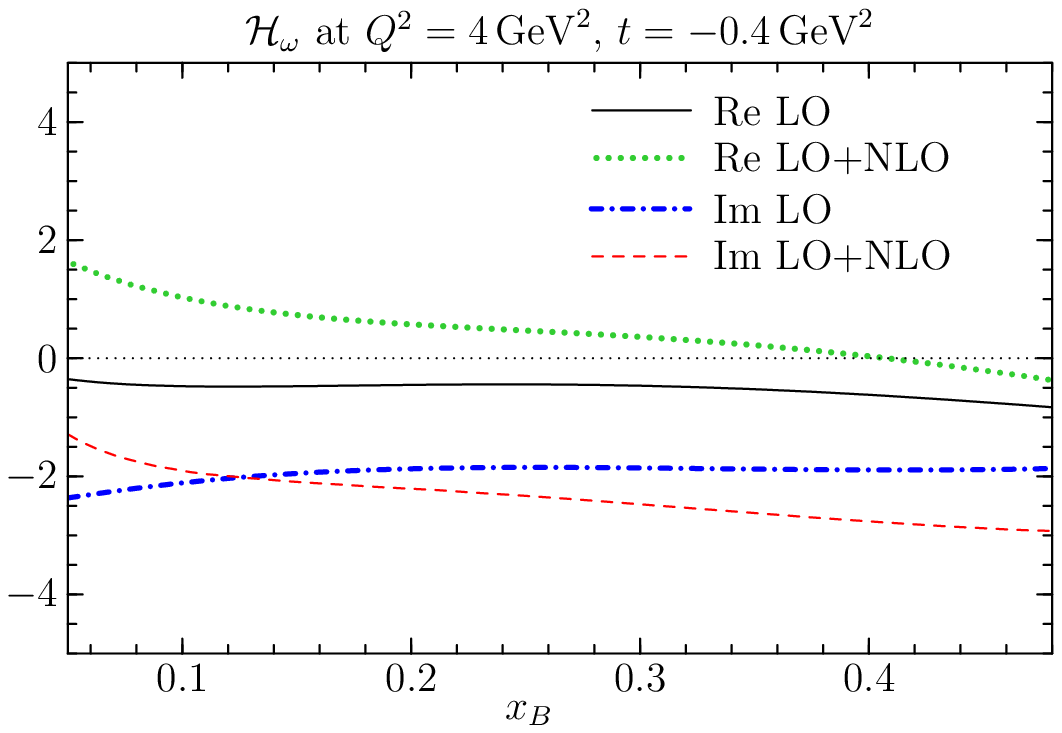}\hspace{1ex}
\includegraphics[width=\plotwidth]
{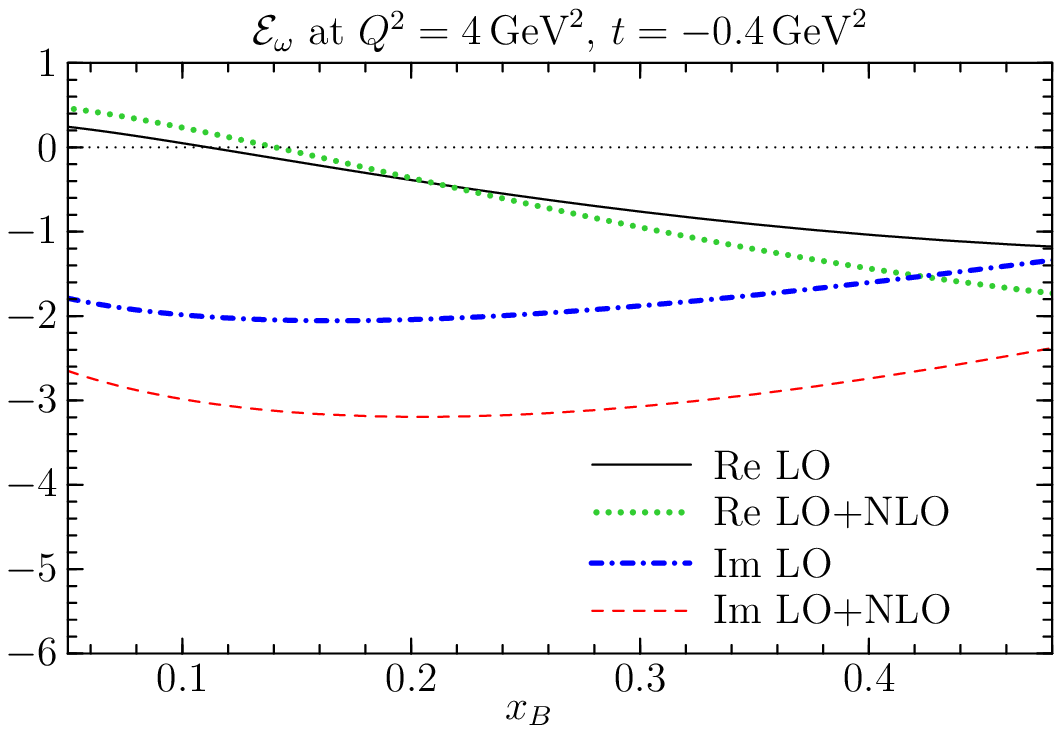}
\end{center} 
\caption{\label{omega_re_im} As Fig.~\protect\ref{rho_re_im} but for
  $\omega$ production.}
\end{figure}

Figure~\ref{rho_re_im} shows the real and imaginary parts of the
convolutions appearing in \eqref{sigmaL} and \eqref{AUT}.
$\mathcal{H}_\rho$ is dominated by the gluon and quark singlet part,
and in line with our discussion in Sect.~\ref{sec:large-x} we find
rather moderate corrections for the imaginary part but a very unstable
real part in a wide range of $x_B$.
As for $\mathcal{E}_\rho$, its real part is very small and subject to
large relative corrections, whereas its imaginary part is much larger
and receives corrections of order $100\%$ .  As we see in
Fig.~\ref{E-ns-mod-1}, the individual flavor non-singlet combinations
$\mathcal{E}^{(3)}$ and $\mathcal{E}^{(8)}$ are less affected by
corrections, but they have opposite sign and partially cancel in the
sum relevant for $\rho$ production. The rather small but unstable
contribution from the gluon and quark singlet terms is hence important
in this channel and largely responsible for the NLO corrections seen
in Fig.~\ref{rho_re_im}.
As a further consequence of the cancellations just mentioned, the size
of $\mathcal{E}_\rho$ is tiny compared with $\mathcal{H}_\rho$.  The
quark flavor combination relevant for $\rho$ production is $2u + d$,
where in our model the flavor combinations add for $H$ but largely
cancel for $E$.  

For $\omega$ production we see in Fig.~\ref{omega_re_im} that at
smaller $x_B$ the convolution $\mathcal{H}_\omega$ is about $1/3$ of
$\mathcal{H}_\rho$, which follows from the dominance of the gluon
contribution as seen in \eqref{charge_factors} and
\eqref{quark_charge_factors}, whereas at larger $x_B$ differences
between the two channels appear.  $\mathcal{E}_\omega$ is much bigger
than $\mathcal{E}_\rho$ since in the combination $2u - d$ the
contributions of $u$ and $d$ quarks add for $E$, and the size of its
radiative corrections reflects the one of $\mathcal{E}^{(3)}$ in
Fig.~\ref{E-ns-mod-1}.  We note that the dominance of the imaginary
over the real part in $\mathcal{E}_\omega$ and $\mathcal{E}_\rho$ is
less pronounced in model 2, as can be anticipated by comparing
Figs.~\ref{E-ns-mod-1} and \ref{E-ns-mod-2}.

\begin{figure}
\begin{center}
\includegraphics[width=\plotwidth]
{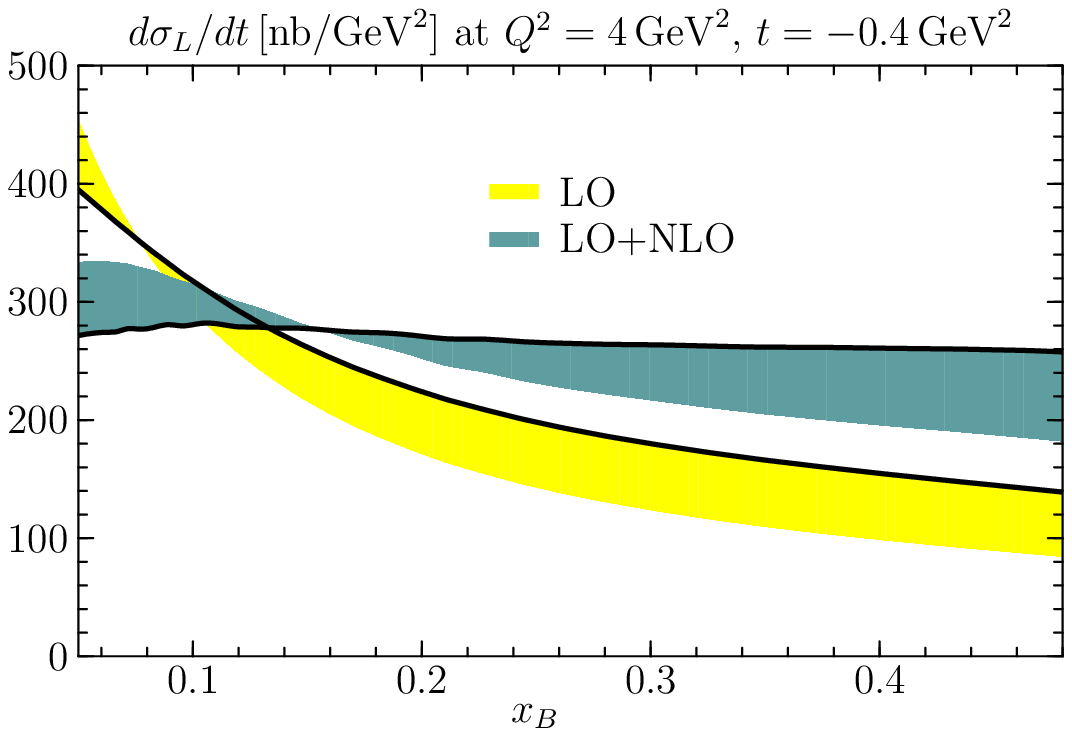}\hspace{1ex}
\includegraphics[width=\plotwidth]
{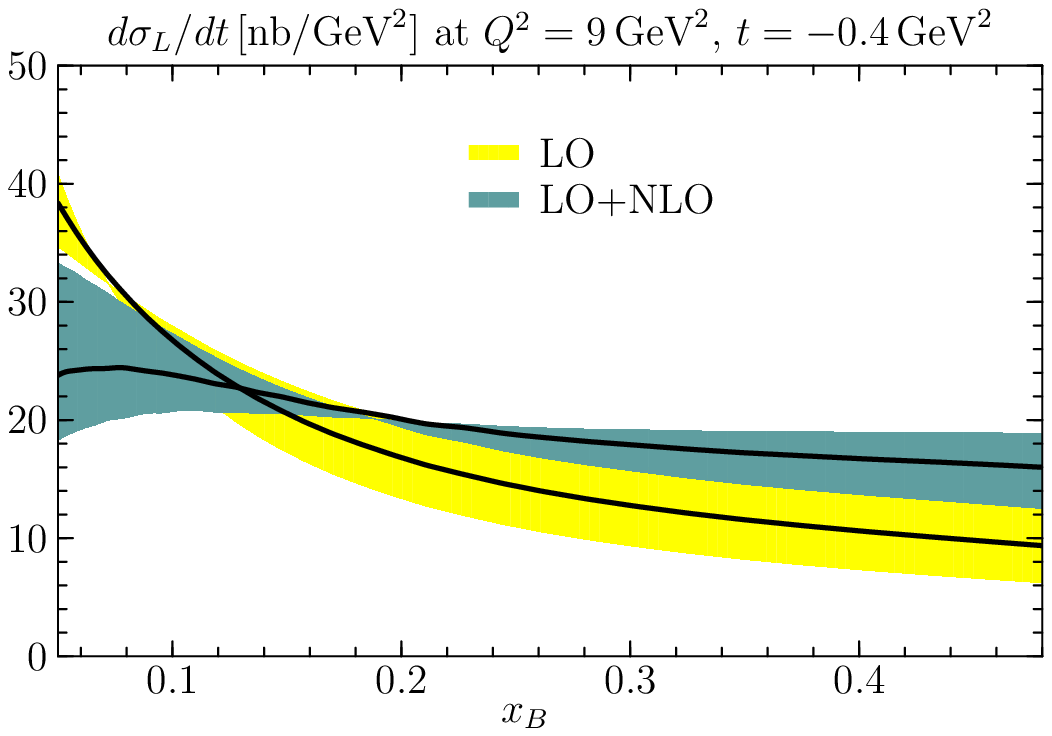}
\end{center} 
\vspace{-1em}
\caption{\label{cs_rho} Longitudinal cross section for $\gamma^* p\to
  \rho\ms p$ in model 1.  Bands correspond to the range $2\gev < \mu <
  4\gev$ in the left and to $2\gev < \mu < 6\gev$ in the right plot,
  and solid lines to $\mu=Q$ in both cases.}
%
\vspace{1em}
%
\begin{center}
\includegraphics[width=\plotwidth]
{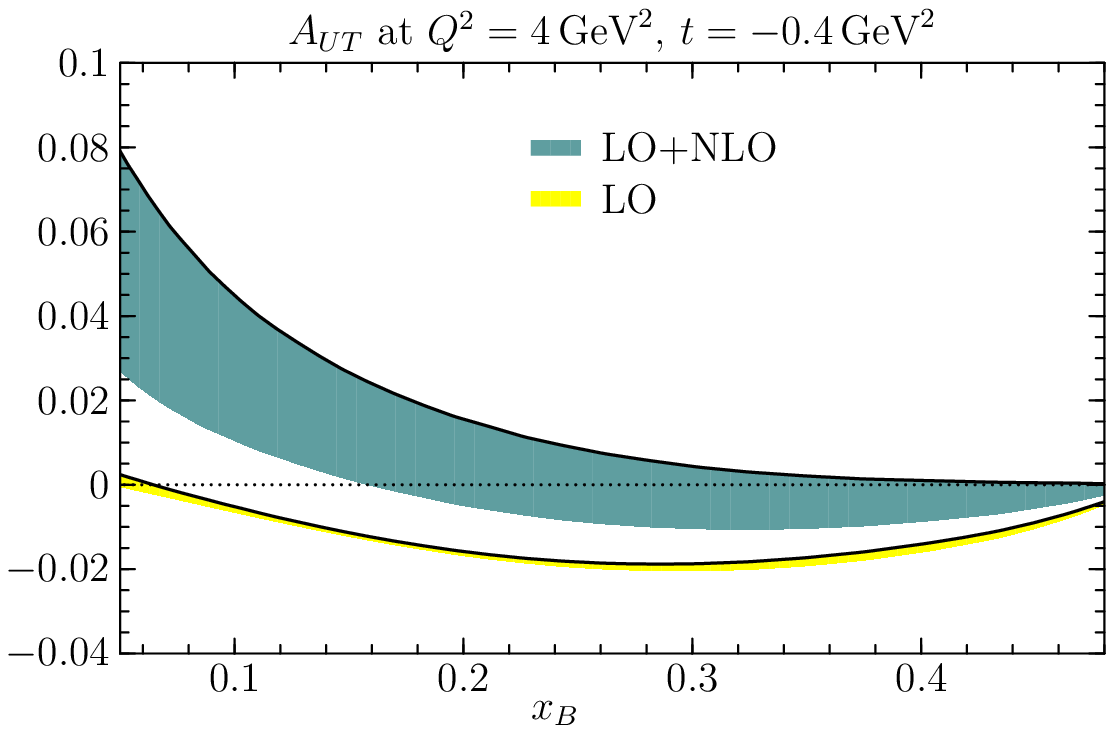}\hspace{1ex}
\includegraphics[width=\plotwidth]
{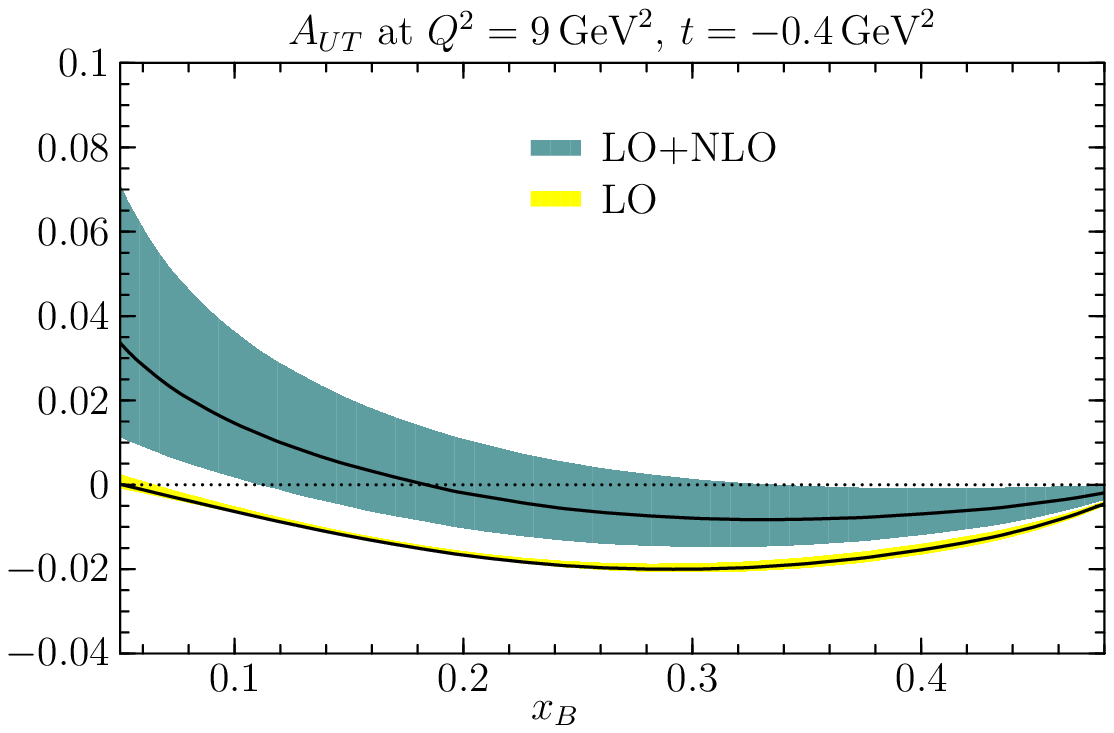}
\end{center}
\vspace{-1em}
\caption{\label{rho_asym_1} The transverse target spin asymmetry
  defined in \protect\eqref{AUT-def}, calculated for model 1.  The
  meaning of the bands and solid lines is as in
  Fig.~\protect\ref{cs_rho}.}
%
\vspace{1em}
%
\begin{center}
\includegraphics[width=\plotwidth]
{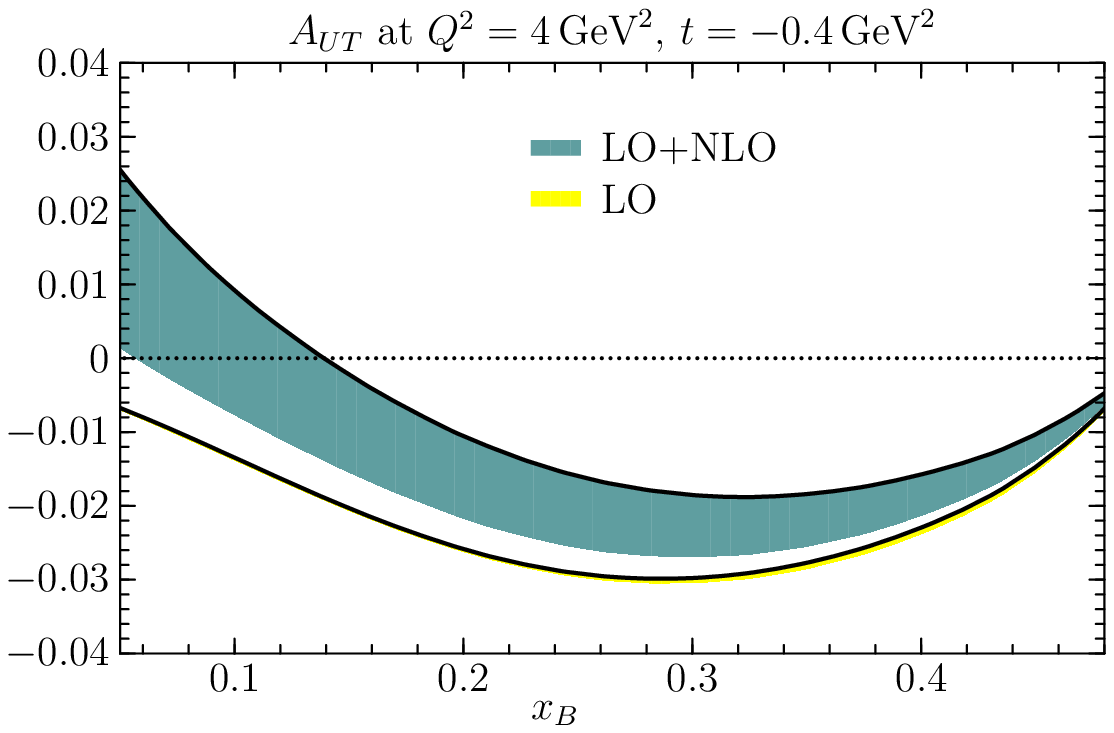}\hspace{1ex}
\includegraphics[width=\plotwidth]
{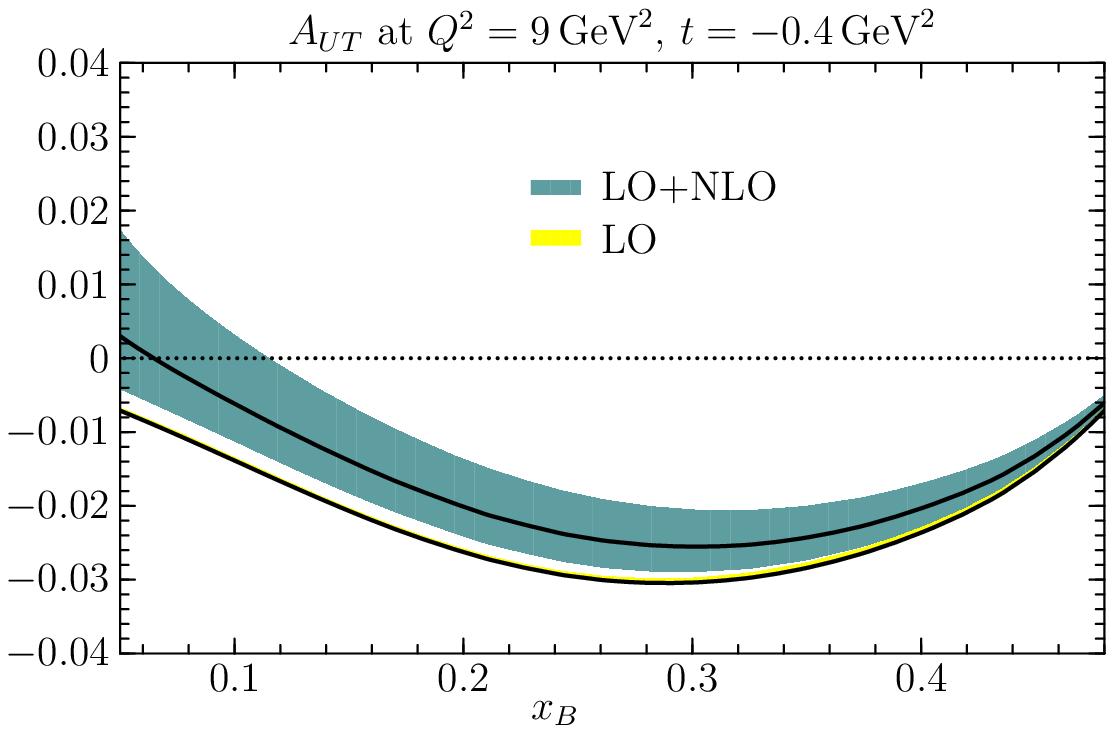}
\end{center}
\vspace{-1em}
\caption{\label{rho_asym_2} As Fig.~\protect\ref{rho_asym_1} but for
  model 2.}
\end{figure}

The cross section $d\sigma_L/dt$ for $\rho$ production is dominated by
$(\im\mathcal{H}_\rho)^2$, except for contributions from
$(\re\mathcal{H}_\rho)^2$ at small $x_B$ for NLO and at large $x_B$
for LO.  Given the size of corrections to $\im\mathcal{H}_\rho$ in
Fig.~\ref{rho_re_im} we thus have quite substantial NLO effects in the
cross section at $Q^2= 4\gev^2$, as shown in Fig.~\ref{cs_rho}.  For
$Q^2= 9\gev^2$ and $x_B >0.1$ the relative corrections decrease.
The plot has been calculated with model 1 for $E$, but since its
contribution to $d\sigma_L/dt$ is negligible the corresponding curves
for model 2 look very similar.  To obtain an estimate of scale
uncertainties, we show bands corresponding to $\mu= \mu_R =\mu_{GPD}$
between $2\gev$ and $2Q$.  Given our discussion in the previous
section, we do not consider it meaningful to go to scales below
$2\gev$, so that the bands in the figure are strongly asymmetric.  For
$Q^2=4 \gev^2$ they go only in one direction, and the band of the LO
result does not provide an estimate for the size of the NLO
corrections, which turn out to go in the other direction.

We have a very peculiar situation for the polarization asymmetry
$A_{UT}$ in $\rho$ production, which as shown in
Figs.~\ref{rho_asym_1} and \ref{rho_asym_2} is very small in both
models 1 and 2 due to the cancellations in $\mathcal{E}_\rho$
discussed above.  $A_{UT}$ changes quite dramatically from LO to NLO
in a wide range of kinematics, clearly because of the NLO corrections
in the numerator.  A closer look at Fig.~\ref{rho_re_im} reveals that
the large perturbative corrections in $\im\bigl( \mathcal{E}_\rho^*\ms
\mathcal{H}_\rho^{} \bigr)$ are mainly due to the large corrections to
both $\re\mathcal{H}_\rho$ and $\re\mathcal{E}_\rho$.  These hardly
affect the unpolarized cross section, which is strongly dominated by
$\im\mathcal{H}_\rho$.  At higher $Q^2$ the instability of $A_{UT}$ is
less pronounced, and in model 2 we even have quite small corrections.
We note that the bands from the scale variation at LO order are
extremely narrow in Figs.~\ref{rho_asym_1} and \ref{rho_asym_2}.  This
is because the scale variation of $\alpha_s(\mu_R)$ cancels in the
ratio $A_{UT}$ at LO and because in the kinematics we are looking at,
the $\mu_{GPD}$ dependence of $\mathcal{H}_\rho$ and
$\mathcal{E}_\rho$ is rather weak.  In this situation, the scale
uncertainty of the LO result does obviously not provide a good
estimate for the size of higher-order corrections.  Let us finally
remark that at $t= -0.4\gev^2$ the asymmetry $A_{UT}$ must go to zero
as $x_B$ tends to $0.484$ because of the prefactor
$\sqrt{t_0-t\rule{0pt}{1.8ex}}$ in \eqref{AUT}.

\begin{figure}
\begin{center}
\includegraphics[width=\plotwidth]
{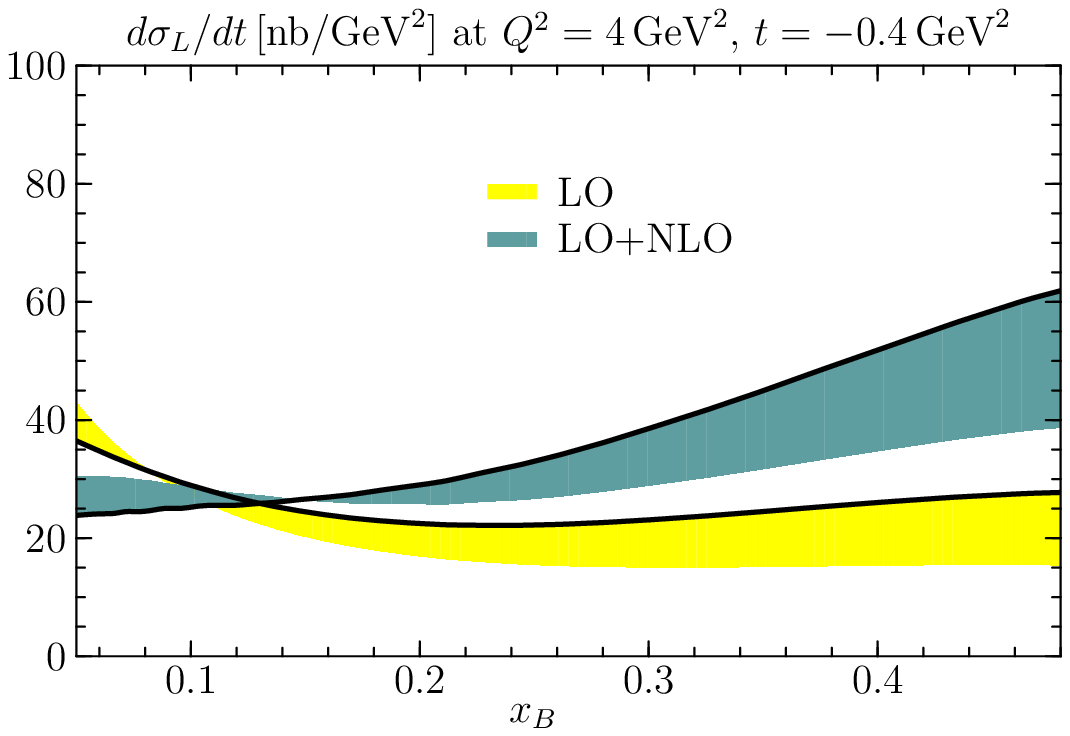}\hspace{1ex}
\includegraphics[width=\plotwidth]
{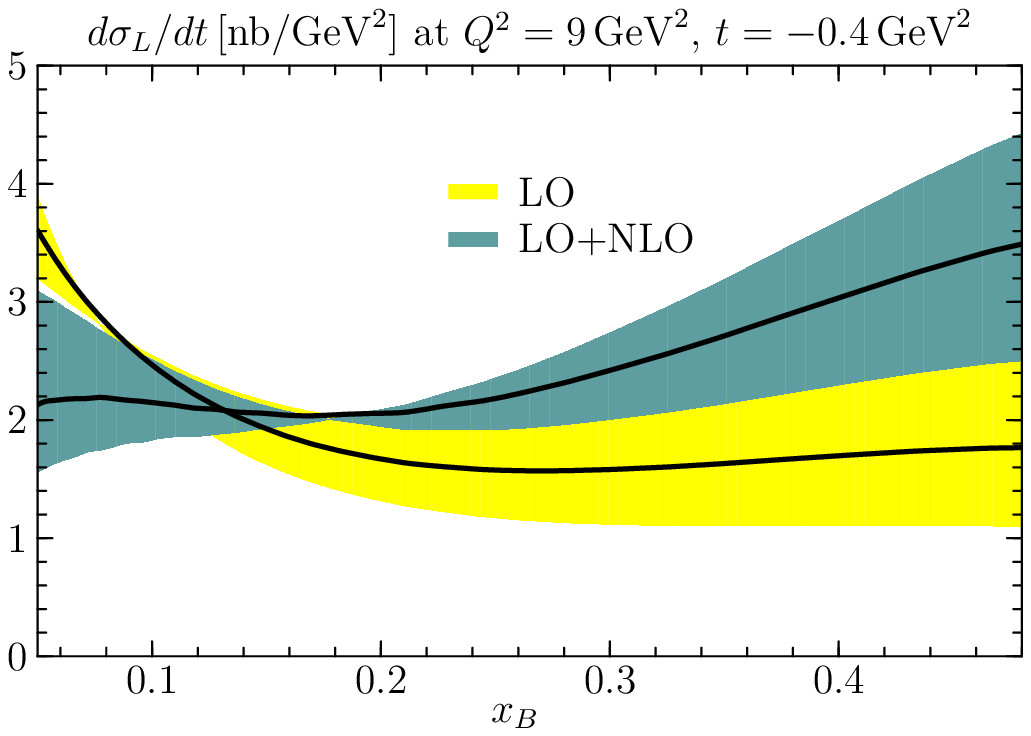}
\end{center} 
\vspace{-1em}
\caption{\label{cs_omega} As Fig.~\protect\ref{cs_rho} but for
  $\gamma^* p\to \omega\ms p$.}
%
\vspace{1em}
%
\begin{center}
\includegraphics[width=\plotwidth]
{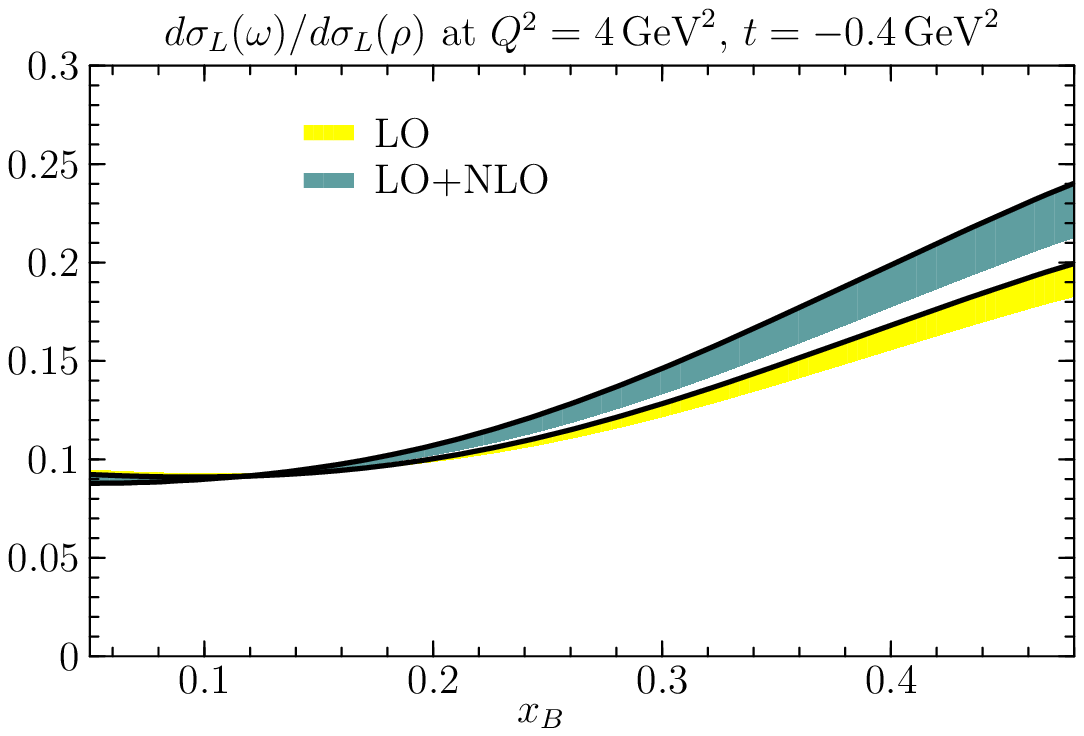}\hspace{1ex}
\includegraphics[width=\plotwidth]
{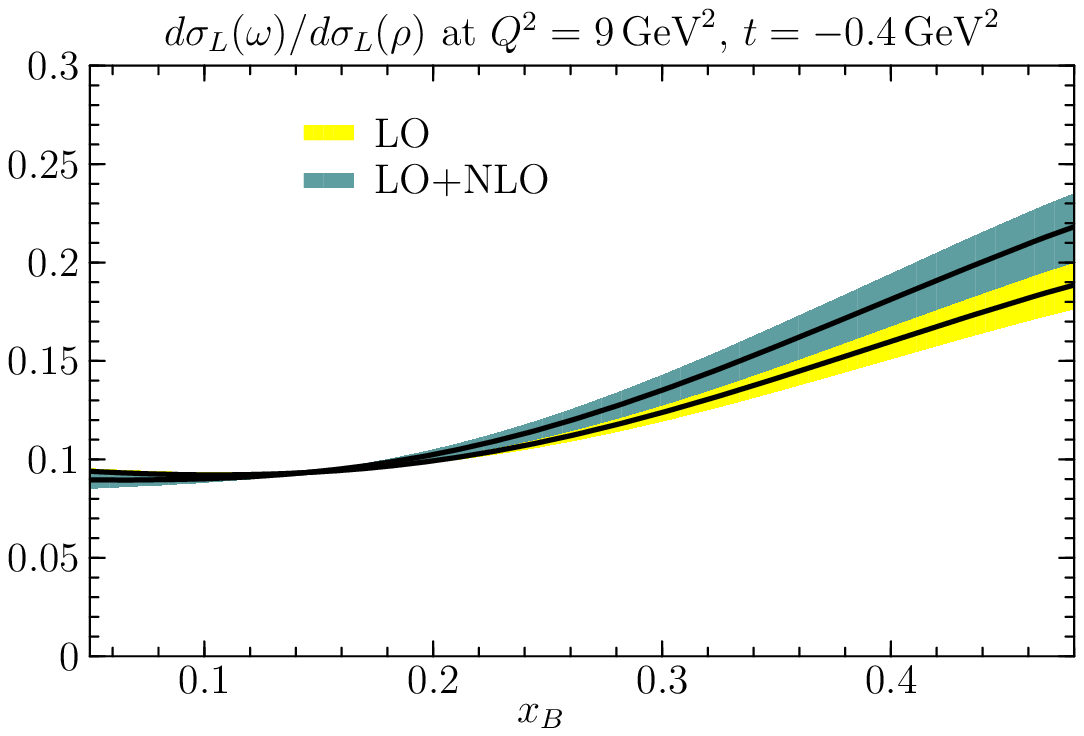}
\end{center}
\vspace{-1em}
\caption{\label{omega_rho_ratio} Ratio of cross sections
  $d\sigma_L/dt$ for $\omega$ and $\rho$ production in model 1.  The
  meaning of the bands and solid lines is as in
  Fig.~\protect\ref{cs_rho}.}
%
\vspace{1em}
%
\begin{center}
\includegraphics[width=\plotwidth]
{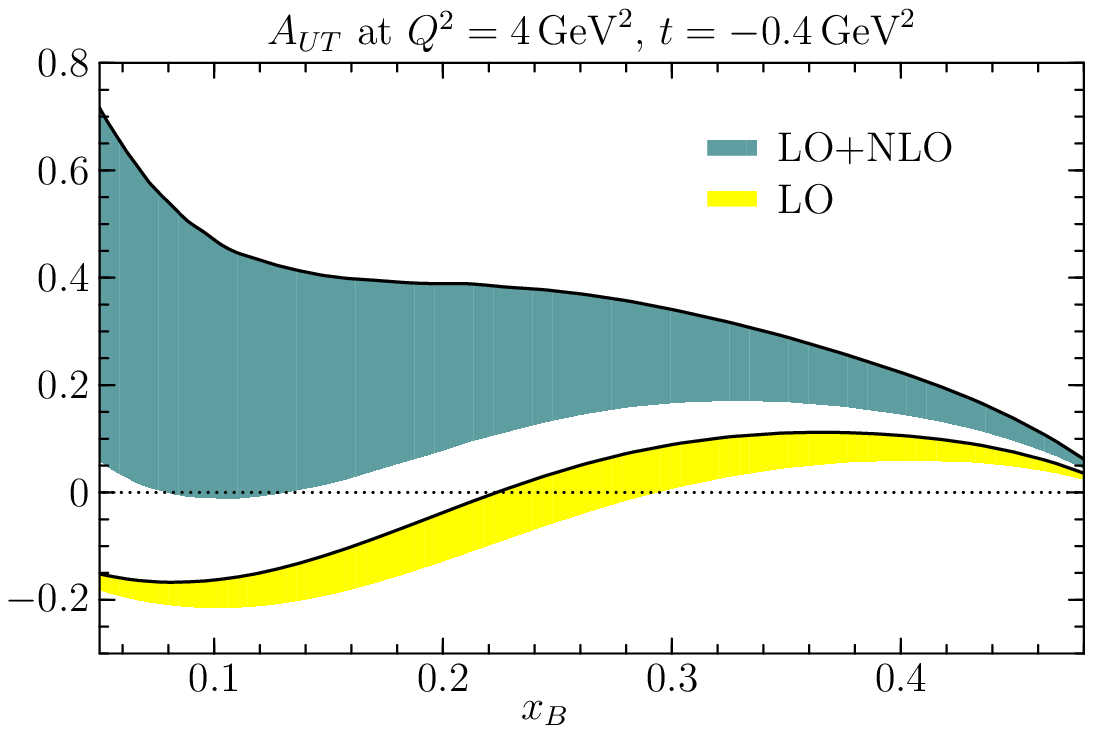}\hspace{1ex}
\includegraphics[width=\plotwidth]
{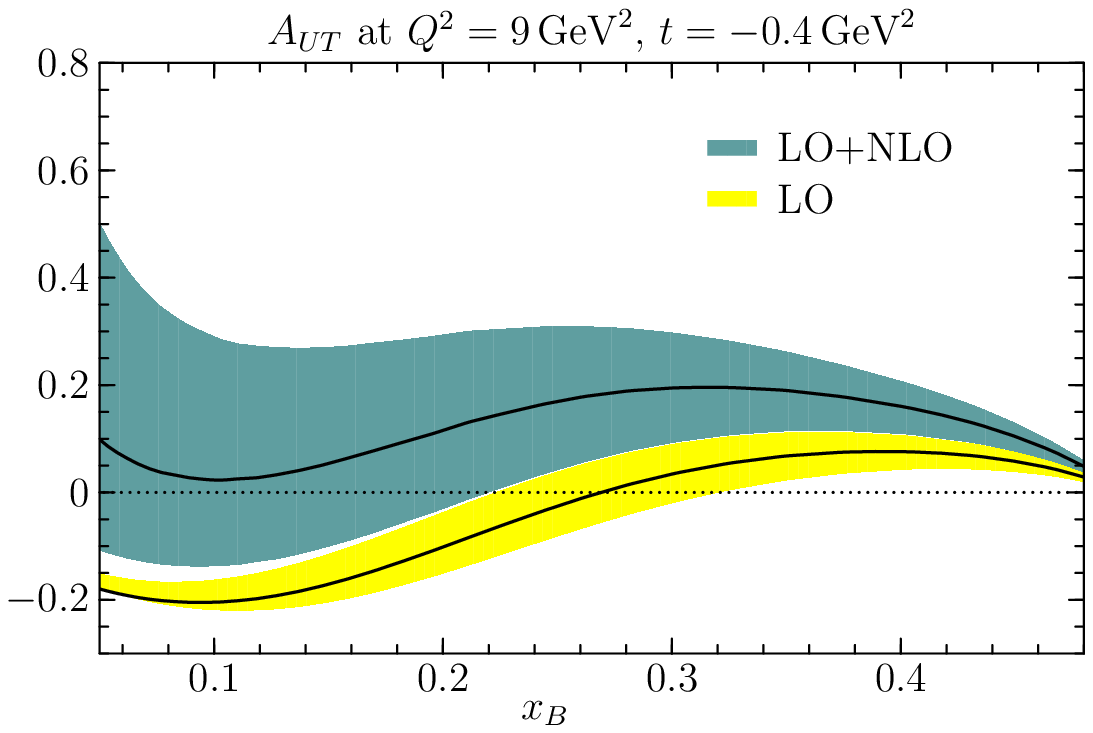}
\end{center}
\vspace{-1em}
\caption{\label{omega_asym} As Fig.~\protect\ref{rho_asym_1} but for
  $\gamma^* p\to \omega\ms p$.}
\end{figure}

The cross section for $\omega$ production is shown in
Fig.~\ref{cs_omega} and shows a similar pattern of NLO corrections to
the one in $\rho$ production, reflecting the similar pattern of
corrections we have seen for $\im\mathcal{H}_\rho$ and
$\im\mathcal{H}_\omega$.  As a result the ratio of cross sections
$d\sigma_L/dt$ in the two channels is quite stable under radiative
corrections, as seen in Fig.~\ref{omega_rho_ratio}.  The target
polarization asymmetry, shown in Fig.~\ref{omega_asym} for model 1,
changes however drastically between LO and NLO at small to
intermediate $x_B$.  This is because $\im\mathcal{E}_\omega$ then
dominates over $\re\mathcal{E}_\omega$, so that its product with the
unstable convolution $\re\mathcal{H}_\omega$ controls the numerator of
the asymmetry.  The absolute size of $A_{UT}$ can be large in this
channel since $|\mathcal{E}_\omega| \sim |\mathcal{H}_\omega|$ in our
model.  According to Fig.~\ref{omega_re_im}, the relative phase
$\delta_\omega$ is close to zero at LO for $x_B \lsim 0.3$, so that
the factor $\sin\delta_\omega$ in \eqref{asy-approx} makes $A_{UT}$
small and prone to large radiative corrections.

\begin{figure}
\begin{center}
\includegraphics[width=\plotwidth]
{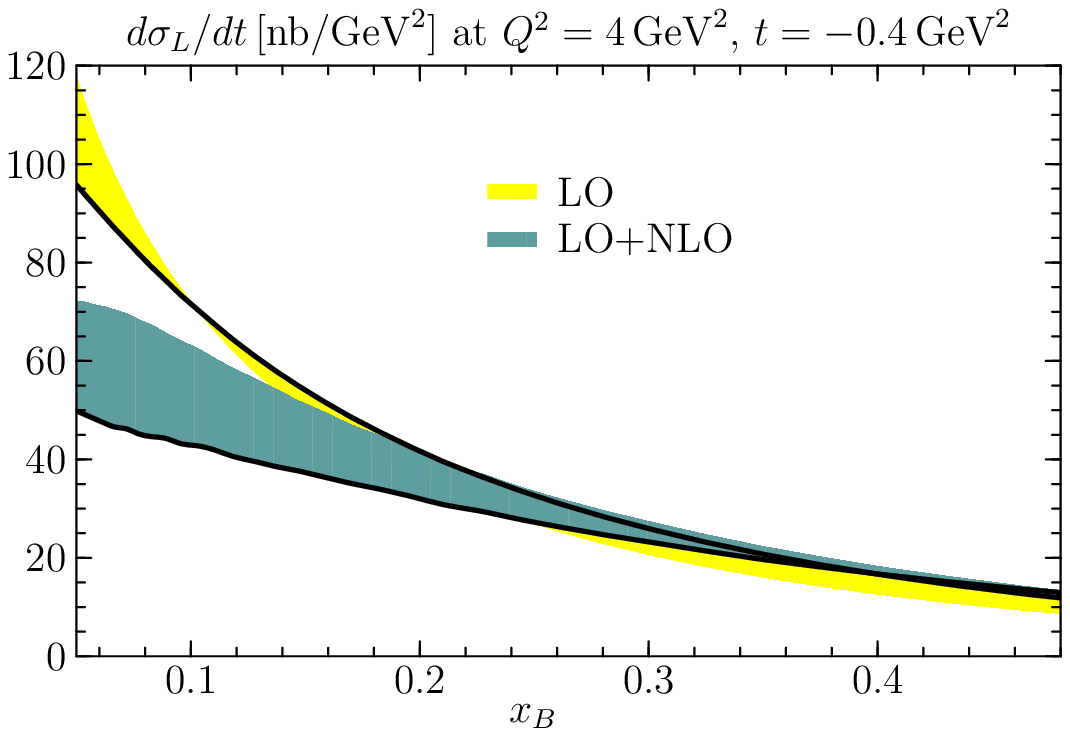}\hspace{1ex}
\includegraphics[width=\plotwidth]
{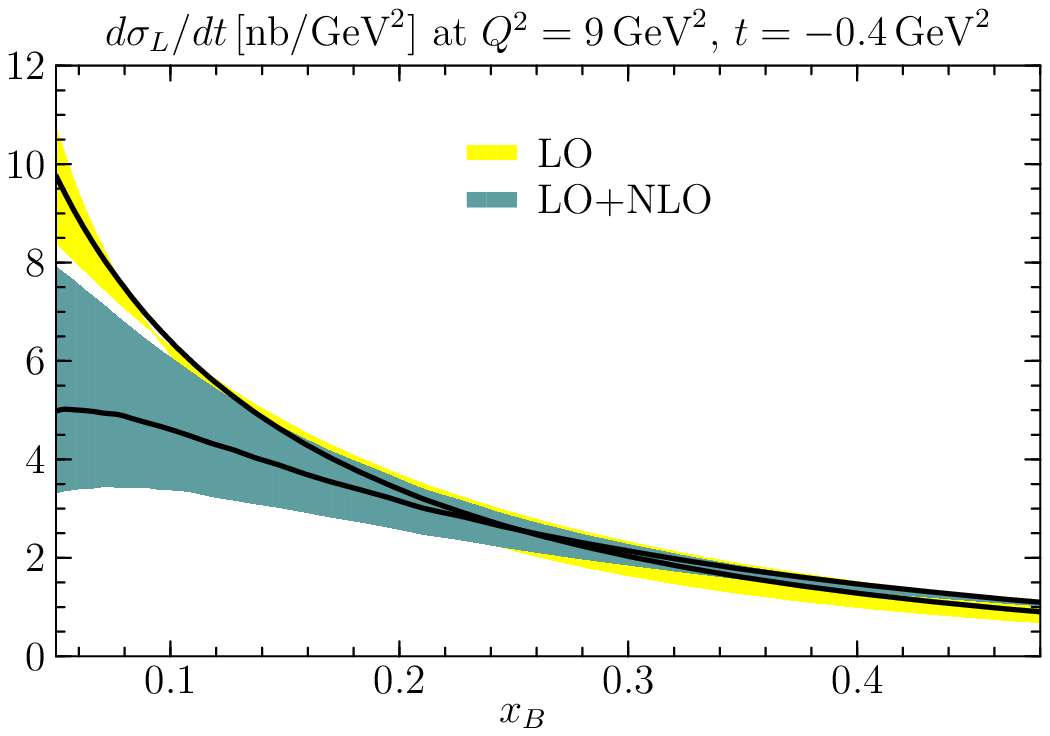}
\end{center} 
\vspace{-1em}
\caption{\label{cs_phi} As Fig.~\protect\ref{cs_rho} but for
  $\gamma^* p\to \phi\ms p$.}
%
\vspace{1em}
%
\begin{center}
\includegraphics[width=\plotwidth]
{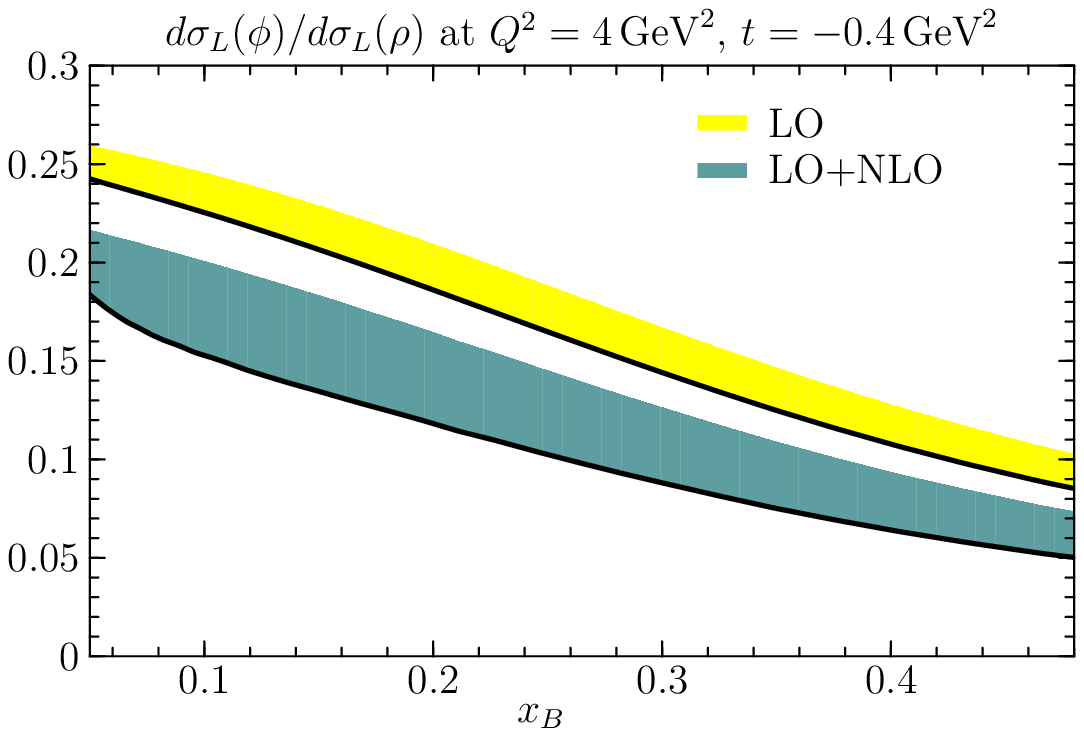}\hspace{1ex}
\includegraphics[width=\plotwidth]
{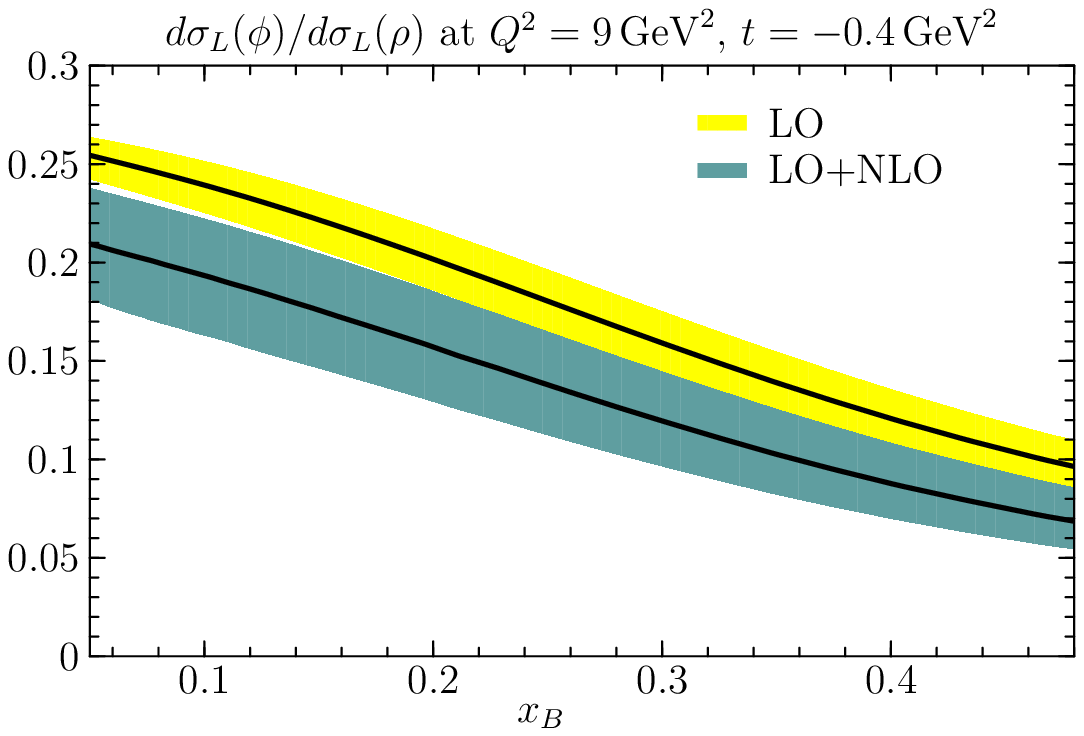}
\end{center}
\vspace{-1em}
\caption{\label{phi_rho_ratio} Ratio of cross sections $d\sigma_L/dt$
  for $\phi$ and $\rho$ production in model 1.  The meaning of the
  bands and solid lines is as in Fig.~\protect\ref{cs_rho}.}
%
\vspace{1em}
%
\begin{center}
\includegraphics[width=\plotwidth]
{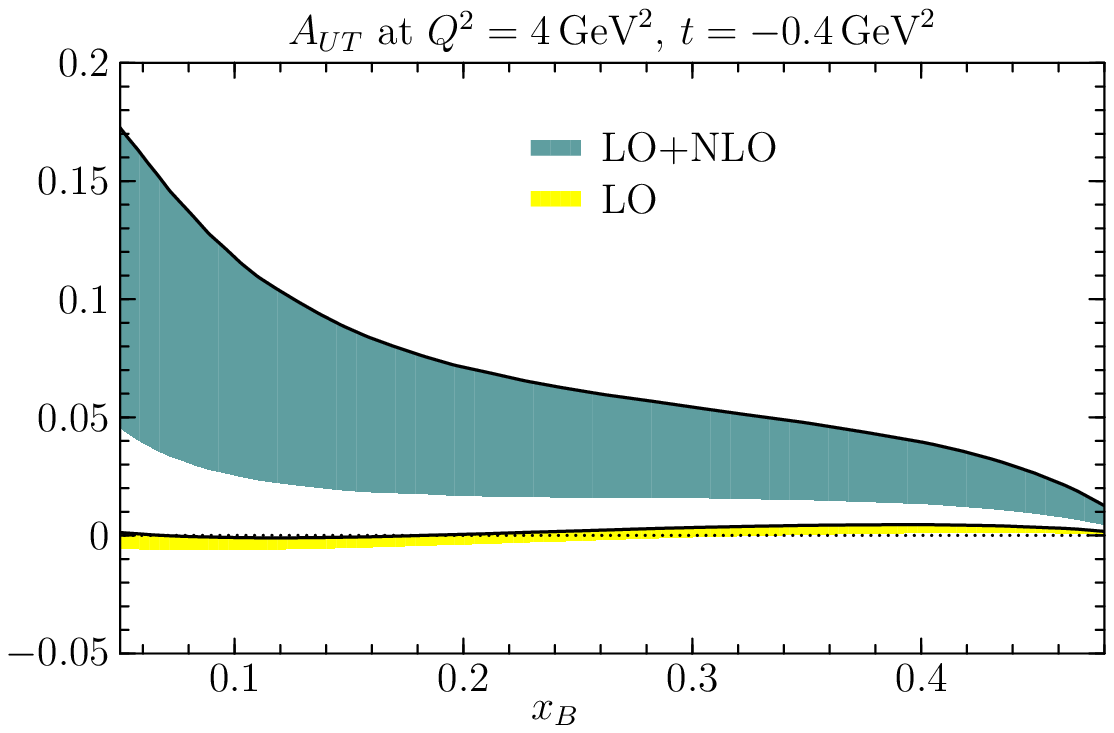}\hspace{1ex}
\includegraphics[width=\plotwidth]
{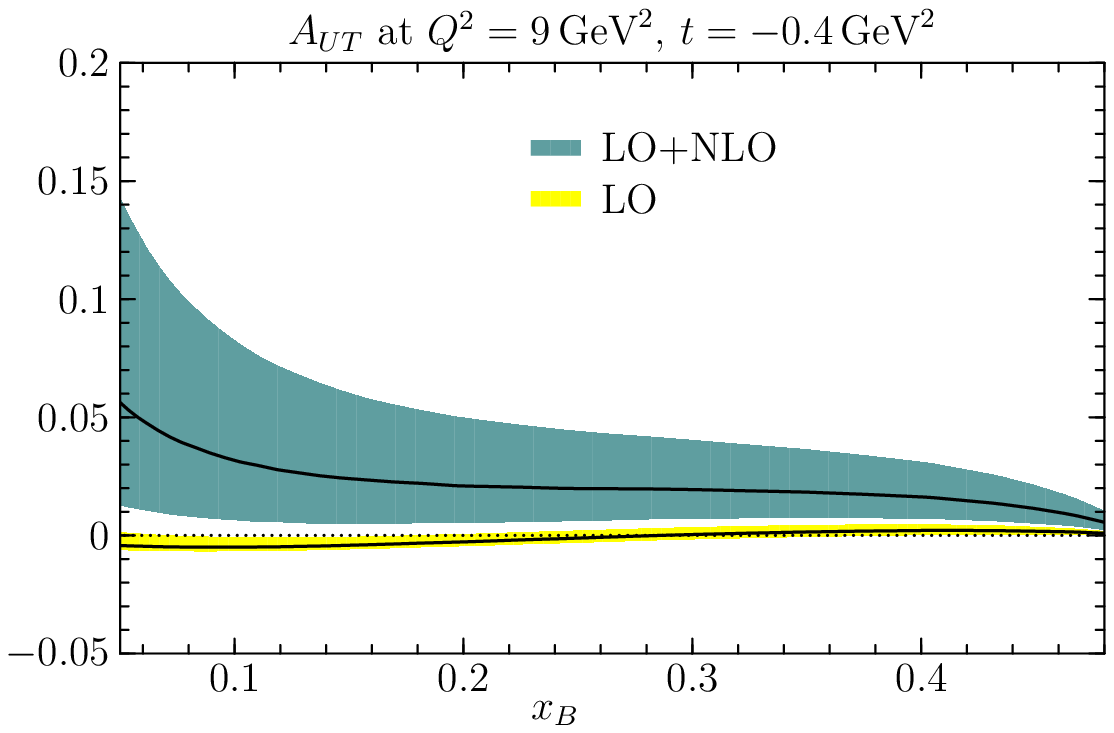}
\end{center}
\vspace{-1em}
\caption{\label{phi_asym} As Fig.~\protect\ref{rho_asym_1} but for
  $\gamma^* p\to \phi\ms p$.}
\end{figure}

Let us finally take a look at $\phi$ production.  At LO this channel
is strongly dominated by gluon exchange, since in our models strange
quark distributions are small for $H$ and even more so for $E$.  At
NLO we have further contributions from the pure singlet terms
$\mathcal{H}^{S(b)}$ and $\mathcal{E}^{S(b)}$, which are not
negligible.  We see in Fig.~\ref{cs_phi} that the NLO corrections to
the cross section are large at small $x_B$ and slowly decrease with
$x_B$.  Except for the region of small $x_B$, this pattern is quite
different from the one in $\rho$ production, so that the cross section
ratio for the two channels receives important corrections at larger
$x_B$ as we see in Fig.~\ref{phi_rho_ratio}.  The asymmetry $A_{UT}$
is essentially zero at LO, because in our model the relative phase
$\delta_\phi$ between $\mathcal{H}_\phi$ and $\mathcal{E}_\phi$ is
very close to zero.  This changes at NLO, where in model~1 we obtain a
small to moderate $A_{UT}$, as shown in Fig.~\ref{phi_asym}.


\clearpage


\section{Pseudoscalar meson production}
\label{sec:pseudo}

Having studied in detail the production of vector mesons, let us
finally take a look at pseudoscalar production.  We will only consider
$\gamma^* p\to \pi^+ n$, which was already studied at NLO in
\cite{Belitsky:2001nq}.  Gluon distributions do not contribute in this
channel.

In the collinear approximation the amplitude for this process can be
written as
\begin{align} 
  \label{pion_amp_NLO}
\mathcal{M} &=
\frac{4\pi\sqrt{4\pi\alpha}}{\xi\ms Q N_c}\, f_\pi
\int_0^1 dz\, \phi_\pi(z)
\int_{-1}^1 dx\,
\Bigl[ e_u\ms T_a(\zb,x,\xi) - e_d\ms T_a(z,-x,\xi) \Bigr] \,
\Bigl[ \widetilde{F}^{u}(x,\xi,t) - \widetilde{F}^{d}(x,\xi,t) \Bigr]
\nonumber \\
&= \frac{4\pi\sqrt{4\pi\alpha}}{\xi\ms Q N_c}\, f_\pi
\sum_{n=0}^\infty a_n^{}\ms \widetilde{\mathcal{F}}^{\ms\pi}_n
\end{align}
with $e_u = 2/3$, $e_d = -1/3$ and $f_\pi = 131 \mev$.  $\phi_\pi(z)$
is the twist-two distribution amplitude of the pion and has a
Gegenbauer decomposition as in \eqref{gegen-phi}.  The convolutions
$\widetilde{\mathcal{F}}^{\ms\pi}_n$ are defined as
\begin{equation}
  \label{Ftilde-def}
\widetilde{\mathcal{F}}^{\ms\pi}_n =
\int_{-1}^1 dx\, 
\Bigl[ e_u\ms T_{a,n}(x,\xi) - e_d\ms T_a(-x,\xi) \Bigr] \,
\Bigl[ \widetilde{F}^{u}(x,\xi,t) - \widetilde{F}^{d}(x,\xi,t) \Bigr] \,,
\end{equation}
and the kernels $T_a(\bar{z},x,\xi)$ and $T_{a,n}(x,\xi)$ are the same
as in Sect.~\ref{sec:kernels}.
The matrix elements $\widetilde{F}^q$ are the counterparts of $F^q$
for polarized quarks and given by
\begin{equation}
\widetilde{F}^q(x,\xi,t) =
    \frac{1}{(p+p')\cdot n} \left[
    \widetilde{H}^q(x,\xi,t)\, \bar u(p') \ms\slashed{n}
       \gamma_5\ms u(p)
  + \widetilde{E}^q(x,\xi,t)\, \bar u(p')\,
      \frac{(p'-p)\cdot n}{2 m_p}\, \gamma_5\ms u(p) \right]
\end{equation}
in terms of the generalized parton distributions $\widetilde{H}$ and
$\widetilde{E}$, where as in the unpolarized case we use the
conventions of \cite{Diehl:2003ny}.
Since the hard-scattering kernel in \eqref{Ftilde-def} is neither even
nor odd in $x$, the convolution involves both the charge-conjugation
even and odd combinations
\begin{align}
  \label{def-Ftilde-pm}
\widetilde{F}^{q(+)}(x,\xi,t) &=
  \widetilde{F}^q(x,\xi,t) + \widetilde{F}^q(-x,\xi,t) \,,
&
\widetilde{F}^{q(-)}(x,\xi,t) &=
  \widetilde{F}^q(x,\xi,t) - \widetilde{F}^q(-x,\xi,t) \,.
\end{align}

We model the distributions $\widetilde{H}$ in close analogy to the
unpolarized case and set
\begin{align}
  \label{dd-models-pol}
\widetilde{H}^{q(+)}(x,\xi,t) &= 
  \int_{-1}^1 d\beta \int_{-1+|\beta|}^{1-|\beta|} d\alpha\;
  \delta(x-\beta-\xi\alpha)\, h^{(2)}(\beta,\alpha)\,
  \widetilde{H}^{q(+)}(\beta,0,t) \,,
\nonumber \\
\widetilde{H}^{q(-)}(x,\xi,t) &= 
  \int_{-1}^1 d\beta \int_{-1+|\beta|}^{1-|\beta|} d\alpha\;
  \delta(x-\beta-\xi\alpha)\, h^{(2)}(\beta,\alpha)\,
  \widetilde{H}^{q(-)}(\beta,0,t) \,,
\end{align}
with $h^{(2)}(\beta,\alpha)$ as in \eqref{profile} and
\begin{align}
\widetilde{H}^{q(+)}(x,0,t)
&= \Delta q_v(x) \exp\bigl[ t f_{q_v}(x) \bigr]
 + 2 \Delta \bar{q}(x) \exp\bigl[ t f_{\bar{q}}(x) \bigr] \,,
\nonumber \\
\widetilde{H}^{q(-)}(x,0,t)
&= \Delta q_v(x) \exp\bigl[ t f_{q_v}(x) \bigr]
\end{align}
for $x>0$.  The values for $x<0$ are determined by the symmetry
properties following from \eqref{def-Ftilde-pm}.  For the polarized
valence and antiquark densities $\Delta q_v$ and $\Delta \bar{q}$ we
use the NLO parameterization from \cite{Bluemlein:2002be} at $\mu=
2\gev$, and for the $t$ dependence we take the same functions
$f_{q_v}(x)$ as in \eqref{DFJK4-f}, \eqref{DFJK-params} and
furthermore set $f_{\bar{q}}(x) = f_{q_v}(x)$.  As was shown in
\cite{Diehl:2004cx}, this gives a good description of the isovector
axial form factor via the sum rule
\begin{equation}
F_A(t) = \int_0^1 dx\, \bigl[ \widetilde{H}^{u(+)}(x,0,t)
                            - \widetilde{H}^{d(+)}(x,0,t) \bigl] \,.
\end{equation}
For the nucleon helicity-flip distribution $\widetilde{E}$ we take a
pion exchange ansatz
\begin{equation}
  \label{Etilde-model}
\widetilde{E}^{u}(x,\xi,t) = -\widetilde{E}^{d}(x,\xi,t) = 
\frac{\theta\bigl( \xi-|x| \bigr)}{2\xi}\,
\phi_\pi\left(\frac{x+\xi}{2\xi}\right)
\frac{2m_p^2\, g_A}{m_\pi^2 -t}\, 
\frac{\Lambda^2-m_\pi^2}{\Lambda^2-t} ,
\end{equation}
with the nucleon axial charge $g_A \approx 1.26$ and a cutoff
parameter $\Lambda = 0.8 \gev$ \cite{Koepf:1995yh} to suppress large
off-shellness of the exchanged pion in the $t$ channel.


\subsection{Results}
\label{sec:pi-results}

\begin{figure}[t]
\begin{center}
\includegraphics[width=\plotwidth]
{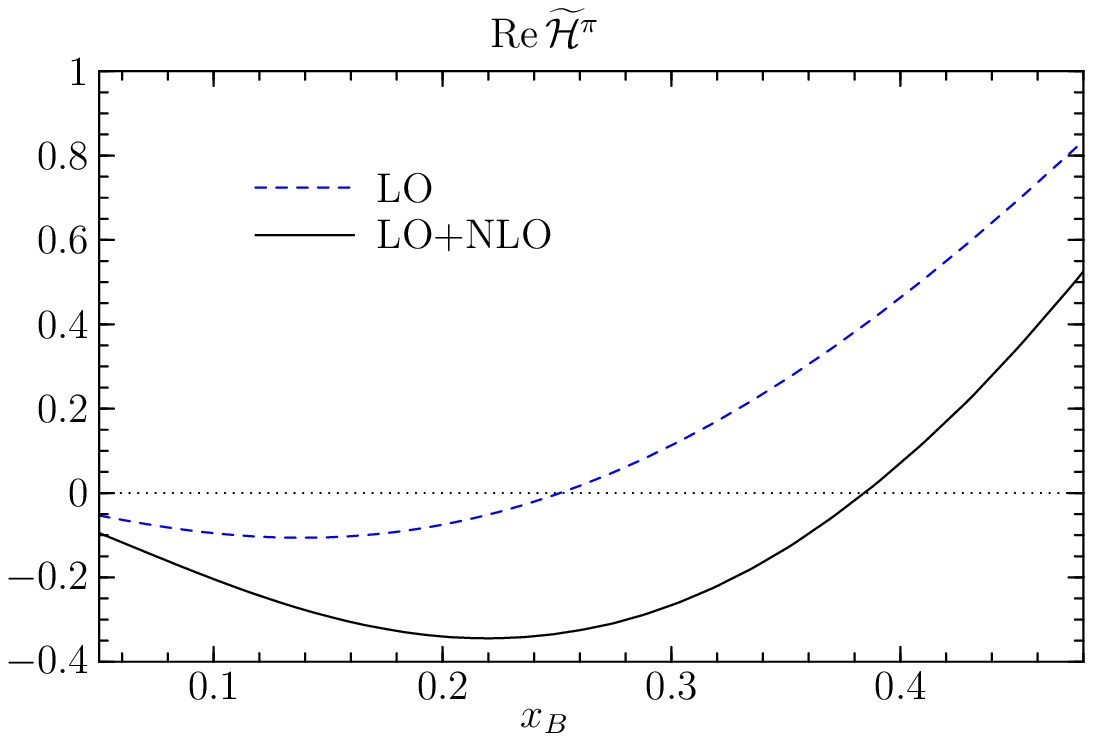}\hspace{1ex}
\includegraphics[width=\plotwidth]
{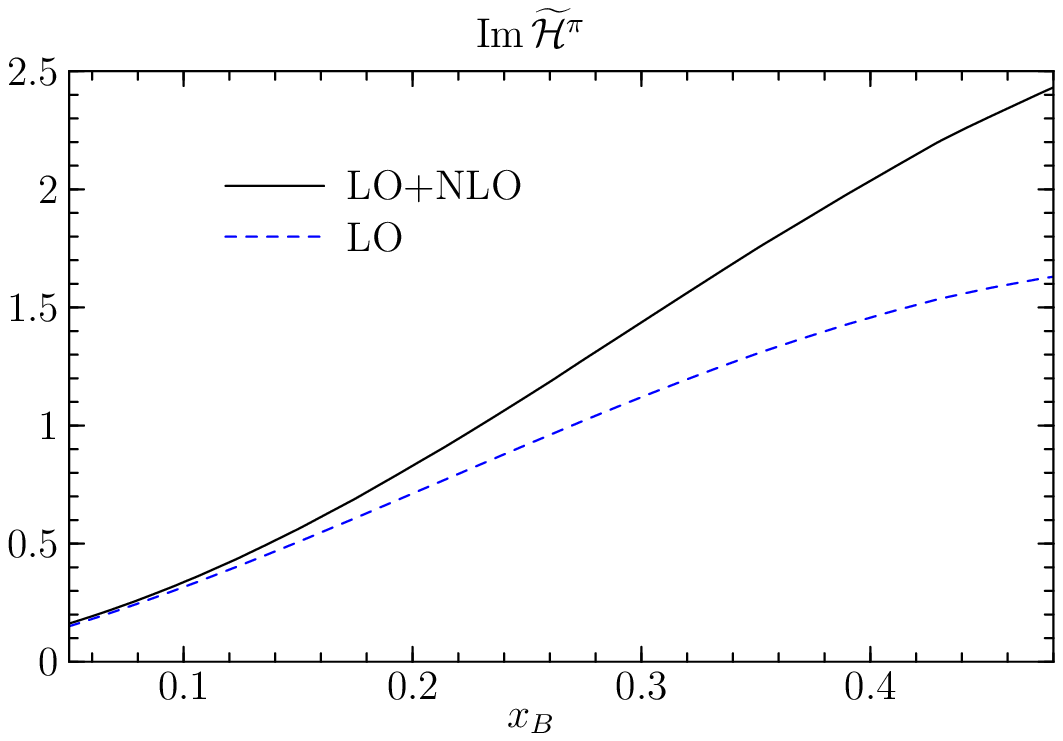}
\end{center} 
\caption{\label{Htilde} The convolution
  $\widetilde{\mathcal{H}}^{\ms\pi}_{\ms 0}$ defined as in
  \protect\eqref{Ftilde-def}, evaluated at $Q= 2\gev$ and $t=
  -0.4\gev^2$.  The scales are set to $\mu_R = \mu_{GPD} = Q$.}
\end{figure}

The convolution $\widetilde{\mathcal{H}}^{\ms\pi}_{n}$ at LO and NLO
is shown in Fig.~\ref{Htilde} for $n=0$.  We find moderate corrections
for the imaginary part and larger ones for the real part.
For $\widetilde{\mathcal{E}}^{\ms\pi}_{n}$ we can easily give the
analytic form of the NLO result.  The scale dependent terms admit a
closed expression,
\begin{align}
  \label{E-tilde-res-gen}
\sum_{n} a_n^{}\ms \widetilde{\mathcal{E}}^{\ms\pi}_n
\,\propto\,
\sum_{m,n} a_m^{} & a_n^{}\ms
\biggl\{ 1 + \frac{\alpha_s}{4\pi} 
\nonumber \\
 & \; \times \biggl[\,
  \beta_0\ms \biggl( \frac{14}{3} + \frac{\gamma_m+\gamma_n}{2C_F}
                   - \ln\frac{Q^2}{\mu_R^2} \biggr)
  - \gamma_m \ln\frac{Q^2}{\mu^2_{GPD}}
  - \gamma_n \ln\frac{Q^2}{\mu^2_{DA}} + \ldots \,\biggr]
\biggr\} \,,
\end{align}
where the $\ldots$ denote contributions which depend neither on $Q^2$
and the scales nor on $\beta_0$.  Including these terms we can write
\begin{align}
  \label{Etilde-res}
\sum_{n} & a_n^{}\ms \widetilde{\mathcal{E}}^{\ms\pi}_n
\,\propto\,
(1 + a_2 + a_4)^2 + \frac{\alpha_s(\mu_R)}{\pi}
\biggl[
  \frac{79}{12} + 25.0\ms a_2 + 32.8\ms a_4
    + 53.4\ms a_2\ms a_4 + 21.4\ms a_2^2 + 32.6\ms a_4^2
\nonumber \\
& - \frac{9}{4}\, (1+a_2+a_4)^2\ms \ln\frac{Q^2}{\mu_R^2}
 - (1+a_2+a_4)
  \biggl( \frac{25}{18}\, a_2 + \frac{91}{45}\, a_4 \biggl)
  \biggl( \ln\frac{Q^2}{\mu^2_{GPD}} + \ln\frac{Q^2}{\mu^2_{DA}}
  \biggr) \biggr]
+ \ldots \,,
\end{align}
where we have set $n_f=3$ in $\beta_0$ and where we approximated
numerically the coefficients written with a decimal point.  Here the
$\ldots$ denote terms with higher Gegenbauer coefficients.  Note that
these coefficients appear twice, once for the produced pion and once
for the pion exchange ansatz of the distribution $\widetilde{E}$.  Up
to a global factor, the expression \eqref{Etilde-res} also gives the
NLO result for the electromagnetic pion form factor $F_\pi(Q^2)$ at
large spacelike momentum transfer $Q^2$, and it agrees with the result
in the detailed study \cite{Melic:1998qr}.
Let us first discuss the case $m=n=0$ relevant for the asymptotic form
of the pion distribution amplitude, where the convolution has no
dependence on $\mu_{GPD}$ and $\mu_{DA}$.  We then have the rather
large coefficient $79/12 \approx 6.6$ in square brackets, so that with
the scale choice $\mu_R = Q$ there are quite large NLO corrections.
The corrections are zero for $\mu_R^2 = e^{-79/27}\ms Q^2 \approx
0.05\, Q^2$, which is outside the perturbative region for most cases
relevant in practice.  The BLM scale for this case is yet smaller:
with \eqref{E-tilde-res-gen} we reproduce the well-known result
$\mu_R^2 = e^{-14/3}\ms Q^2 \approx 0.01\ms Q^2$
\cite{Belitsky:2001nq}.  The coefficient of $\alpha_s/\pi$ in
\eqref{Etilde-res} is then $-47/12 \approx -3.9$ and thus again rather
large, but of course the scale $\mu_R$ is outside the perturbative
region for all experimentally relevant kinematics.  We finally see in
\eqref{Etilde-res} that for higher Gegenbauer moments the correction
terms are larger than for $m=n=0$.  The reason for this is the same
which we discussed in Sect.~\ref{sec:large-x} for the convolutions
$\mathcal{H}$.  In \eqref{E-tilde-res-gen} we also see that the BLM
scale becomes smaller for higher $m$ and $n$.

The observables for exclusive pion production at leading order in
$1/Q$ are the same as for vector meson production, and the $ep$ cross
section is given as in \eqref{AUT-def}.  The cross section for a
longitudinal photon and the transverse target asymmetry are now
respectively given by
\begin{align}
  \label{pi-cs}
\frac{d\sigma_L}{dt} &=
\frac{\pi^2}{9}\, \frac{\alpha}{Q^6}\,
\frac{(2-x_B)^2}{1-x_B}\, (2 f_\pi)^2\,
\Bigl[\ms (1-\xi^2)\, | \widetilde{\mathcal{H}}_\pi^{} |^2
   - \xi^2\ms t/(4m_p^2)\, | \widetilde{\mathcal{E}}_\pi^{} |^2
   - 2\xi^2 \re \bigl( \widetilde{\mathcal{E}}_\pi^*\ms
                       \widetilde{\mathcal{H}}_\pi^{} \bigr)
\ms \Bigr]
\intertext{and}
A_{UT} &= - \frac{\sqrt{t_0-t\rule{0pt}{1.8ex}}}{m_p}\,
   \frac{\xi\ms \sqrt{1-\xi^2}\,
   \im\bigl( \widetilde{\mathcal{E}}_\pi^*\ms
             \widetilde{\mathcal{H}}_\pi^{} \bigr)}{\rule{0pt}{1.05em}
   (1-\xi^2)\, | \widetilde{\mathcal{H}}_\pi^{} |^2
   - \xi^2\ms t/(4m_p^2)\, | \ms\widetilde{\mathcal{E}}_\pi^{} |^2
   - 2\xi^2 \re \bigl( \widetilde{\mathcal{E}}_\pi^*\ms
                       \widetilde{\mathcal{H}}_\pi^{} \bigr)} \,,
\end{align}
with
\begin{align}
\widetilde{\mathcal{H}}_\pi &= \sum_{n=0}^\infty a_n^{}\ms
  \widetilde{\mathcal{H}}^{\ms\pi}_n \,,
&
\widetilde{\mathcal{E}}_\pi &= \sum_{n=0}^\infty a_n^{}\ms
  \widetilde{\mathcal{E}}^{\ms\pi}_n \,.
\end{align}
For numerical estimates we take the asymptotic pion distribution
amplitude in the following, setting $a_n=0$ for $n\ge 2$.  We note
that the recent lattice study \cite{Braun:2006dg} obtained a rather
moderate value $a_2(\mu_0) = 0.201(114)$ at $\mu_0 = 2\gev$.

\begin{figure}
\begin{center}
\includegraphics[width=\plotwidth]
{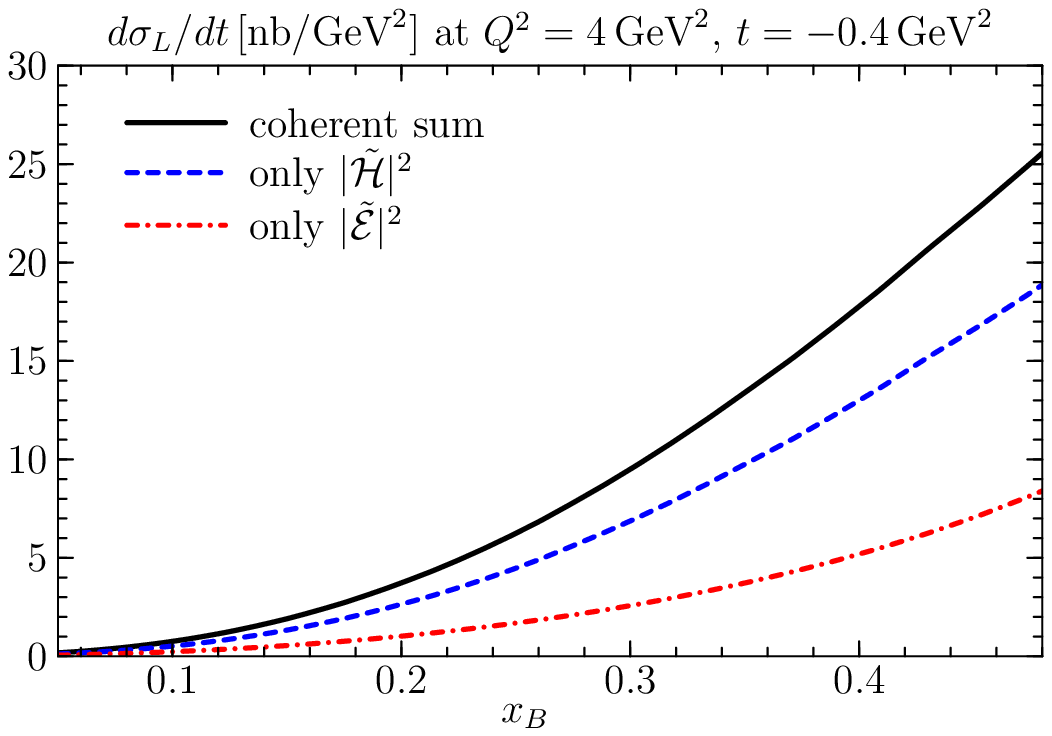}
\end{center}
\vspace{-1em}
\caption{\label{cs_pi_HE} The longitudinal cross section for $\gamma^*
  p\to \pi^+ n$, evaluated at NLO.  Shown are the separate
  contributions from the terms with $|\widetilde{\mathcal{H}}_\pi|^2$
  and with $|\widetilde{\mathcal{E}}_\pi|^2$ in \protect\eqref{pi-cs},
  as well as the complete expression.}
%
\vspace{1em}
%
\begin{center}
\includegraphics[width=\plotwidth]
{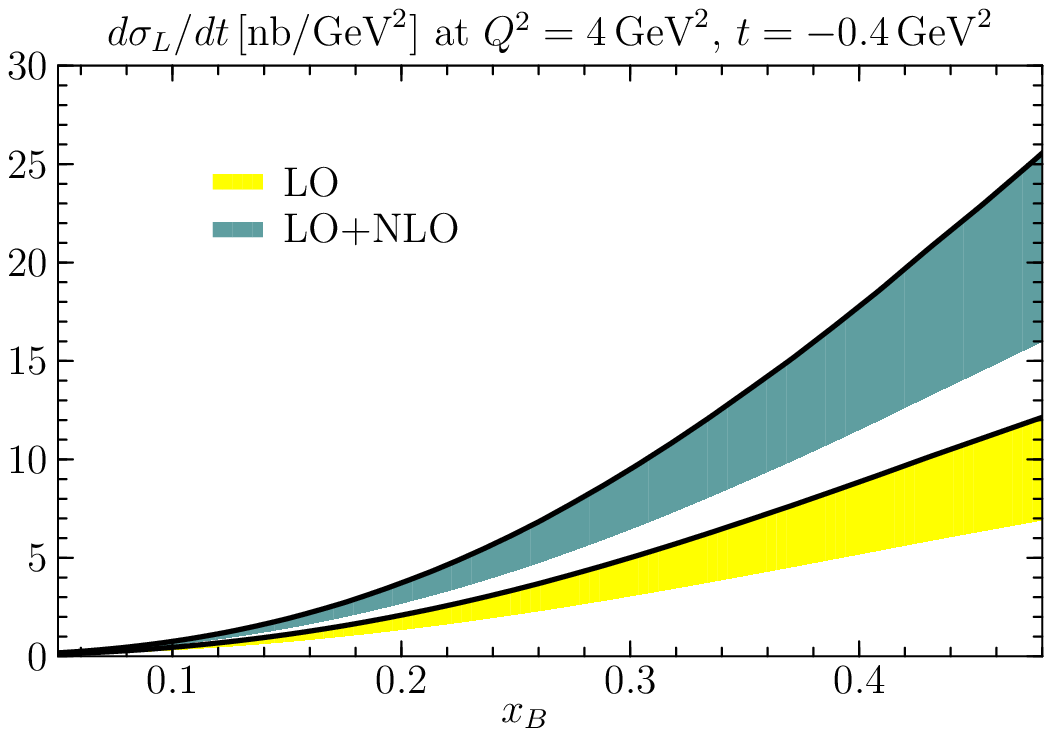}\hspace{1ex}
\includegraphics[width=\plotwidth]
{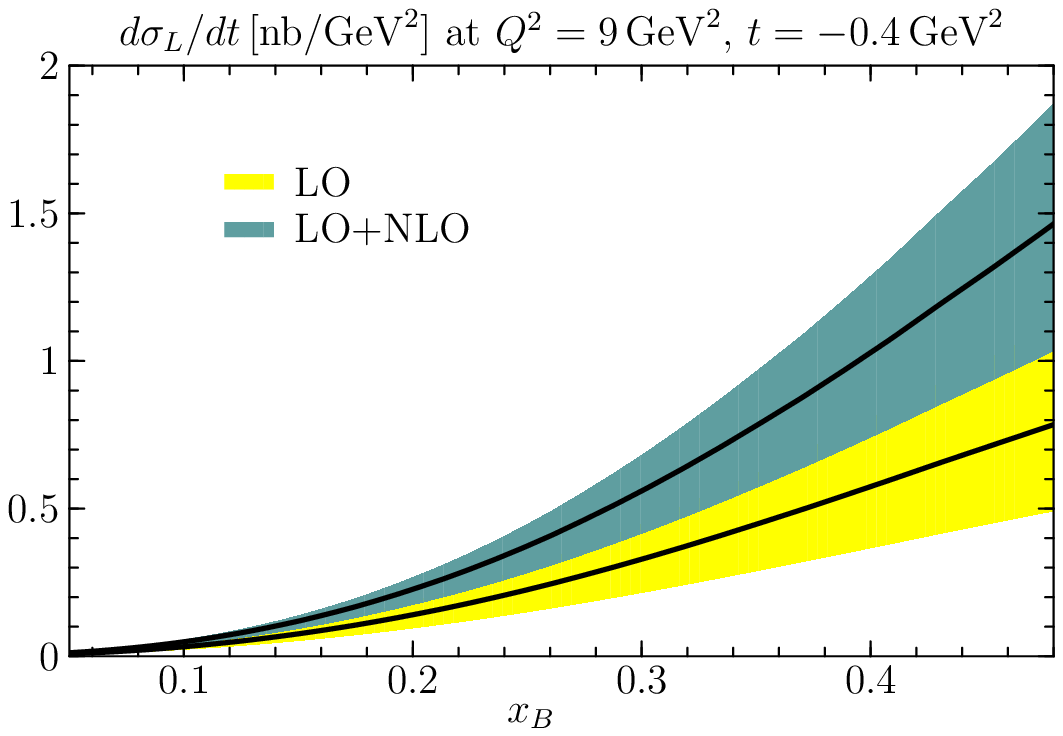}
\end{center} 
\vspace{-1em}
\caption{\label{cs_pi} Longitudinal cross section for $\gamma^* p\to
  \pi^+ n$.  Bands correspond to the range $2\gev < \mu < 4\gev$ in
  the left and to $2\gev < \mu < 6\gev$ in the right plot, and solid
  lines to $\mu=Q$ in both cases.}
%
\vspace{1em}
%
\begin{center}
\includegraphics[width=\plotwidth]
{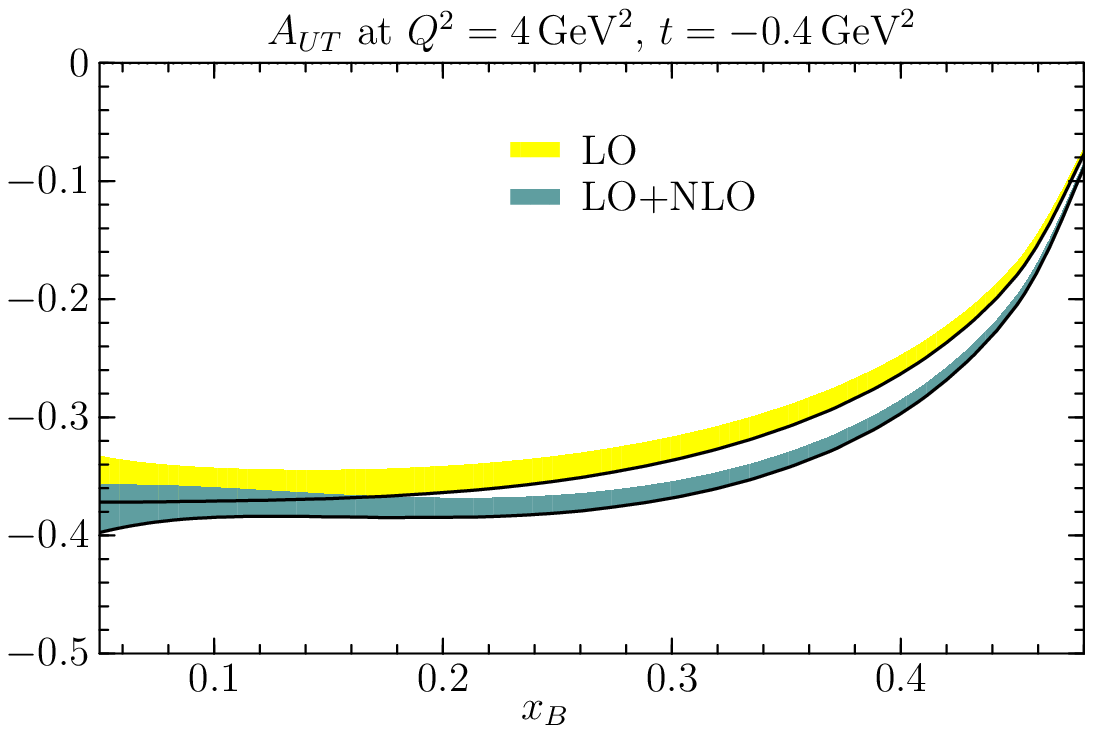}\hspace{1ex}
\includegraphics[width=\plotwidth]
{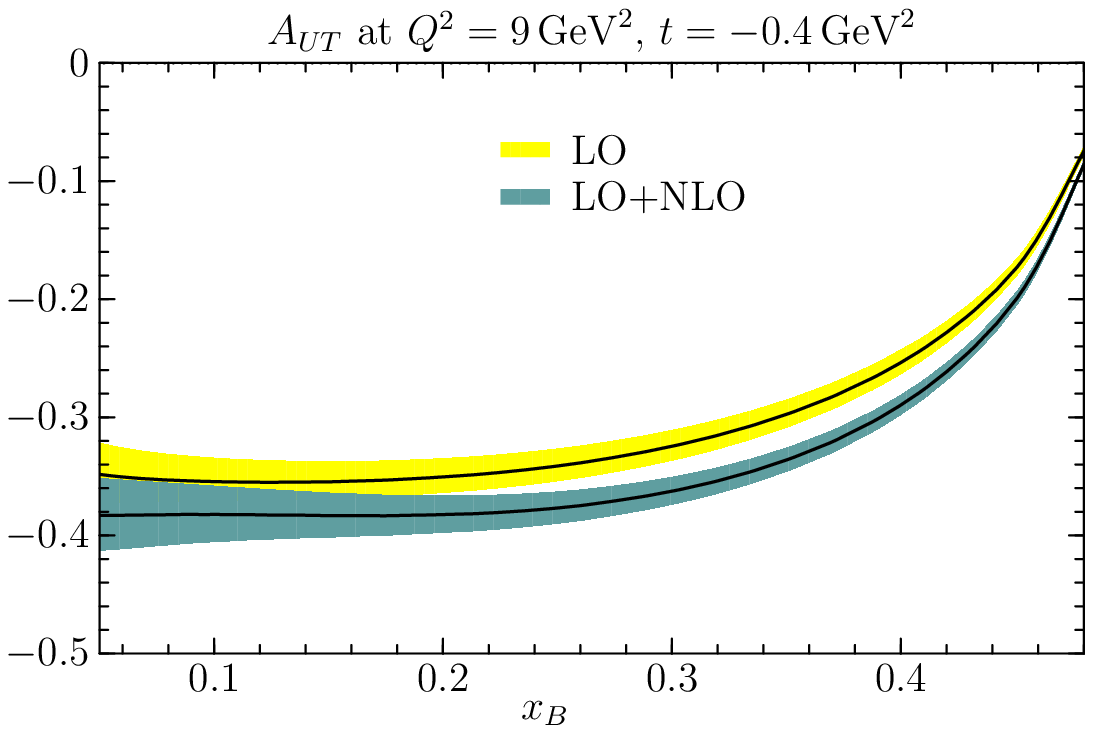}
\end{center}
\vspace{-1em}
\caption{\label{pi_asym} The transverse target spin asymmetry for
  $\pi^+$ production, as defined in \protect\eqref{AUT-def}.  The
  meaning of the bands and solid lines is as in
  Fig.~\protect\ref{cs_pi}.}
\end{figure}

In Fig.~\ref{cs_pi_HE} we show the separate contributions from the
terms with $|\widetilde{\mathcal{H}}_\pi|^2$ and with
$|\widetilde{\mathcal{E}}_\pi|^2$ in \protect\eqref{pi-cs}, as well as
the full result.  We see that at the value of $t$ chosen here, the
contribution from $|\widetilde{\mathcal{H}}_\pi|^2$ is more important,
mainly because of the suppression factor $(\Lambda^2-m_\pi^2)
/(\Lambda^2-t)$ in our model \eqref{Etilde-model} for $\widetilde{E}$.
The square of this factor is $0.36$ at $t=-0.4\gev^2$.

We compare the LO and NLO results for the cross section in
Fig.~\ref{cs_pi} and find that the NLO corrections are quite large,
even at $Q^2= 9\gev^2$.  In contrast, the corrections for the beam
spin asymmetry are very small as seen in Fig.~\ref{pi_asym}, in line
with the findings reported in \cite{Belitsky:2001nq}.  Note that with
our model $\widetilde{\mathcal{E}}_\pi$ is purely real, so that at
intermediate $x_B$ the large relative NLO corrections in
$\re\widetilde{\mathcal{H}}_\pi$ do not affect the numerator of
$A_{UT}$ in \eqref{pi-cs}.  Approximating the asymmetry as
\begin{equation}
A_{UT} \approx
  - \frac{\sqrt{t_0-t\rule{0pt}{1.8ex}}}{m_p} \;
    \frac{\xi \im\bigl( \widetilde{\mathcal{E}}_\pi^*\ms
          \widetilde{\mathcal{H}}_\pi^{} \bigr)}{\rule{0pt}{1.05em}
          | \widetilde{\mathcal{H}}_\pi^{} |^2}
= - \frac{\sqrt{t_0-t\rule{0pt}{1.8ex}}}{m_p} \;
    \biggl| \frac{\xi\ms \widetilde{\mathcal{E}}_\pi}{\rule{0pt}{1.05em}
    \widetilde{\mathcal{H}}_\pi}\, \biggr|
  \sin\delta_\pi
\end{equation}
with $\delta_\pi = \arg( \widetilde{\mathcal{H}}_\pi
/\widetilde{\mathcal{E}}_\pi)$, we can understand why only small
corrections are seen in this case: the relative phase $\delta_\pi$ is
well different from zero, and the NLO corrections increase both
$|\widetilde{\mathcal{H}}_\pi|$ and $|\widetilde{\mathcal{E}}_\pi|$.

\clearpage


\section{Summary}
\label{sec:sum}

In this work we have analyzed the NLO corrections for exclusive meson
production at large $Q^2$ in the collinear factorization approach.
Using the Gegenbauer expansion of meson distribution amplitudes, we
have rewritten the hard-scattering kernels of \cite{Ivanov:2004zv}
into functions depending on only one variable, and we have separated
the explicit logarithms in the factorization scale for the meson
distribution amplitude and the generalized parton distributions.

For vector meson production at small $x_B$ we find huge NLO
corrections even for $Q^2$ well above $10\gev^2$, in agreement with
the results obtained in \cite{Ivanov:2004zv}.  The corrections have
opposite sign compared to the Born term and can be traced back to BFKL
type logarithms in the hard-scattering kernels, which appear with
rather large numerical prefactors in this process.  We conclude at
this stage that a quantitative control of radiative corrections at
small $x_B$ will require resummation of these logarithms.  First steps
in this direction have been reported in \cite{Dima:2007}.  If
successful, such a resummation in combination with a dispersion
relation \cite{Anikin:2007yh} may also be useful for stabilizing the
real part of the amplitude, where we find very large NLO corrections
even at $x_B\sim 0.1$.

At intermediate to large $x_B$, typical of fixed-target experiments,
we have investigated the production of $\rho^0$, $\omega$, $\phi$ and
of $\pi^+$.  We find NLO corrections to the longitudinal cross
sections of up to $100\%$, which somewhat decrease in size when going
from $Q^2 = 4\gev^2$ to $9\gev^2$.  Note that the meson production
cross section depends \emph{quadratically} on generalized parton
distributions---the increased sensitivity to these basic quantities
comes with an increased sensitivity to higher-order corrections.  We
generally find that uncertainties on the cross section due to the
choice of renormalization and factorization scales are not too large
at LO and do not significantly decrease when going to NLO.  For scales
below $4\gev^2$, however, NLO corrections often grow out of control.
The cross section ratio for $\omega$ to $\rho$ production turns out to
be very stable under corrections, but less so the one for $\phi$ to
$\rho$.
For the transverse target polarization asymmetry $A_{UT}$ in $\pi^+$
production we find quite small NLO effects, confirming the results in
\cite{Belitsky:2001nq}.  For vector meson production this is however
not the case.  With the models we have used for the nucleon
helicity-flip distributions $E$, the numerator of the asymmetry in
this channel is dominated by the product $(\im\mathcal{E}_V) (\re
\mathcal{H}_V)$ in a wide range of kinematics and therefore suffers
from the perturbative instability we find for $\re\mathcal{H}_V$ at
small to intermediate $x_B$, even if the corrections to
$\im\mathcal{E}_V$ are not too large.  It is often assumed that
corrections tend to cancel in asymmetries.  The examples we have
studied show that this may hold in specific cases but not in others,
and that special care is needed for observables like $A_{UT}$ that
depend on the relative phase between amplitudes.

We should recall that in the kinematics we studied, one must expect
that our leading-twist results receive power corrections that cannot
be neglected when comparing with data.  They will certainly affect the
cross sections and will not always cancel in cross section ratios.  An
example is the transverse target polarization asymmetry in $\pi^+$
production.  The phenomenological estimates in
\cite{Vanderhaeghen:1999xj} found that the convolution
$\widetilde{\mathcal{H}}_\pi$ is decreased by effects of transverse
parton momentum in the hard scattering, whereas
$\widetilde{\mathcal{E}}_\pi$ is increased by the soft overlap
mechanism that has been extensively studied in the context of the pion
form factor.  Together, these corrections may significantly increase
leading-twist estimates for $A_{UT}$.

{}From our numerical studies we must conclude that a precise
quantitative interpretation of exclusive meson production requires
large $Q^2$, say above $10\gev^2$.  In addition it would be highly
valuable to have a consistent scheme for combining radiative with
power corrections, at least in parts.  Nevertheless, we find that
valuable information on generalized parton distributions can be
obtained also from data at lower $Q^2$.  In particular, a large
measured asymmetry $A_{UT}$ in vector meson production would give
valuable constraints on the size of the proton helicity-flip
distribution $E^g$ for gluons, which are most difficult to obtain in
deeply virtual Compton scattering or from lattice QCD calculations.


\section*{Acknowledgments} 

We gratefully acknowledge discussions with L.~Favart, H.~Fischer,
P.~Kroll, A.~Rostomyan and A.~Sch\"afer.  Special thanks are due to
D.~Yu.~Ivanov for numerous conversations and advice.
This work is supported by the Helmholtz Association, contract number
VH-NG-004.


\appendix

\section{Polylogarithms}
\label{app:polylog}

We collect here some properties of the polylogarithms that appear in
the hard-scattering kernels for meson production.  Their definitions
are
\begin{align}
\Li_2 z &= - \int_0^1 \frac{dt}{t}\, \ln(1- z t) \,,
&
\Li_3 z &= \int_0^1 \frac{dt}{t}\, \Li_2(z t) \,,
\end{align}
from which one readily obtains for the imaginary parts
\begin{align}
\im\ms\Bigl[ \Li_2(\yb + i\epsilon) \Bigr] &=
  \pi\ms \theta(-y)\ms \ln\yb \,,
& 
\im\ms\Bigl[ \Li_3(\yb + i\epsilon) \Bigr] &=
  \frac{\pi}{2}\, \theta(-y)\ms \ln^2\yb \,.
\end{align}
The limiting behavior for $y\to -\infty$ can be obtained from the
expansions
\begin{align}
  \label{poly-asy1}
\Li_2 y &= -\frac{\pi^2}{6} - \frac{1}{2}\ms \ln^2(-y)
   - \sum_{n=1}^\infty \frac{y^{-n}}{n^2} \,,
&
\Li_3 y &= -\frac{\pi^2}{6}\ms \ln(-y) - \frac{1}{6}\ms \ln^3(-y)
   + \sum_{n=1}^\infty \frac{y^{-n}}{n^3} \,,
\end{align}
which are valid for $y< -1$, and from
\begin{align}
  \label{poly-asy2}
\re\ms\Bigl[ \Li_2\yb \Bigr] &=
\frac{\pi^2}{3}
   - \frac{1}{2}\ms \ln^2\yb
   - \sum_{n=1}^\infty \frac{\yb^{-n}}{n^2} \,,
&
\re\ms\Bigl[ \Li_3\yb \Bigr] &=
\frac{\pi^2}{3}\ms \ln\yb
   - \frac{1}{6}\ms \ln^3\yb
   + \sum_{n=1}^\infty \frac{\yb^{-n}}{n^3} \,,
\end{align}
which holds for $y < 0$.  A useful relation finally is
\begin{equation}
\Li_2 y + \Li_2 \yb = \frac{\pi^2}{6} - (\ln y)\, (\ln\yb) \,.
\end{equation}
A wealth of further information can be found in \cite{Devoto:1983tc}.


\section{Hard-scattering kernels for higher Gegenbauer moments}
\label{app:kernels}

In this appendix we give the analogs of the hard-scattering kernels in
\eqref{kernels_asy} for Gegenbauer index $n=2$ and $n=4$.  For the
gluon kernel we find
\begin{align}
  \label{app:gluon}
t_{g,2}(y) &=
\biggl[ 2C_A\ms (y^2+\yb^2) - C_F\ms y \biggr] \frac{\ln y}{\yb}\,
  \ln\frac{Q^2}{\mu^2_{GPD}}
+ \frac{\beta_0}{2}\ms \ln\frac{\mu_R^2}{\mu^2_{GPD}}
- \frac{25}{12}\, C_F \ln\frac{Q^2}{\mu^2_{DA}}
\nonumber \\
&\quad
+ C_F \biggl[
  \frac{35}{36} (5-54y\yb)
- \frac{y}{2}\, \frac{\ln^2 y}{\yb}
- 7(\yb-y)\ms (1-30y\yb) \Li_2\yb
\nonumber \\
&\qquad\qquad
+ \left( \frac{1}{\yb} - \frac{3}{2}
  -\frac{392}{3}y+525y^2-420y^3 \right) \ln y
\biggr]
\nonumber \\
&\quad
+ C_A \biggl[
  - \frac{15}{4}\, (1-4y\yb)
  + \left( \frac1\yb-2y \right) \ln^2 y
  + (\yb-y)\ms (7-60y\yb) \Li_2\yb
\nonumber \\
&\qquad\qquad
- \left( \frac{23}{3\yb}+\frac{5}{6}-58y+150y^2-120y^3 \right) \ln y
\biggr]
\nonumber \\
&\quad
+ 6y\yb\ms \Bigl[ 5(1-4y\yb)\ms C_A - 14 (1-5y\yb)\ms C_F \Bigr]
  \left( 3\Li_3\yb - \ln y\, \Li_2 y - \frac{\pi^2}{6} \ln y \right)
+ \{y\to\yb\}\,,
\nonumber \\
t_{g,4}(y) &=
\biggl[ 2C_A\ms (y^2+\yb^2) - C_F\ms y \biggr] \frac{\ln y}{\yb}\,
  \ln\frac{Q^2}{\mu^2_{GPD}}
+ \frac{\beta_0}{2}\ms \ln\frac{\mu_R^2}{\mu^2_{GPD}}
- \frac{91}{30}\, C_F \ln\frac{Q^2}{\mu^2_{DA}}
\nonumber \\
&\quad
+ C_F \biggl[
  \frac{27287}{1800}-595y\yb+2520(y\yb)^2
- \frac{y}{2}\, \frac{\ln^2 y}{\yb}
+ 16(\yb-y) \Bigl( 1-105y\yb+630(y\yb)^2 \Bigr) \Li_2\yb
\nonumber \\
&\qquad\qquad
+ \left( \frac{1}{\yb} - \frac{5}{2}
  -\frac{11596}{15}y+9660y^2-34160y^3+45360y^4-20160y^5 \right) \ln y
\ms\biggr]
\nonumber \\
&\quad
+ C_A \biggl[
  - \frac{35}{16}\, (1-4y\yb) (5-72y\yb)
  + \left( \frac1\yb-2y \right) \ln^2 y
  + 2 (\yb-y) \Bigl( 8-315y\yb+1260(y\yb)^2 \Bigr) \Li_2\yb
\nonumber \\
&\qquad\qquad
- \left( \frac{257}{30\yb}+\frac{77}{60}-\frac{1741}{5}y
        +2940y^2-8960y^3+11340y^4-5040y^5 \right) \ln y
\biggr]
\nonumber \\
&\quad
+ 30y\yb\ms \Bigl[ 7(1-4y\yb)(1-6y\yb)\ms C_A 
\nonumber \\
&\qquad\qquad
  - 16\ms \Bigl( 1-14y\yb+42(y\yb)^2 \Bigr)\ms C_F \Bigr]
  \left( 3\Li_3\yb - \ln y\, \Li_2 y - \frac{\pi^2}{6} \ln y \right)
+ \{y\to\yb\} \,,  \phantom{\biggl[ \biggr]}
\end{align}
and for the pure singlet kernel
\begin{align}
  \label{app:singlet}
t_{b,2}(y) &=
2(\yb-y) \, \frac{\ln y}{\yb}\,
\biggl[ \ln\frac{Q^2}{\mu_{GPD}^2} - \frac{23}{6} \biggr]
+ (\yb-y)\, \frac{\ln^2 y}{\yb}
- \frac{15}{2} (\yb-y)
\nonumber \\
&\quad
+ 2(7-60y\yb) \Li_2\yb
- \left( \frac{5}{3}-90y+120y^2\right) \ln y
\nonumber \\
&\quad
+ 60 (\yb-y)\ms y\yb \left[ 3 \Li_3\yb 
  + \biggl( \Li_2\yb + \ln^2\yb - \frac{\pi^2}{3} \biggl)\ms \ln y
\right]
- \{y\to\yb\} \,,
\nonumber \\
t_{b,4}(y) &=
2(\yb-y) \, \frac{\ln y}{\yb}\,
\biggl[ \ln\frac{Q^2}{\mu_{GPD}^2} - \frac{257}{60} \biggr]
+ (\yb-y)\, \frac{\ln^2 y}{\yb}
- \frac{35}{8}\, (\yb-y) (5-72y\yb) 
\nonumber \\
&\quad
+ 4\ms \Bigl( 8-315y\yb+1260(y\yb)^2 \Bigr) \Li_2\yb
- \left( \frac{77}{30}-665y+4550y^2-8820y^3+5040y^4 \right) \ln y
\nonumber \\
&\quad
+ 420 (\yb-y)\ms y\yb\ms (1-6y\yb) \left[ 3 \Li_3\yb 
  + \biggl( \Li_2\yb + \ln^2\yb - \frac{\pi^2}{3} \biggl)\ms \ln y
\right]
- \{y\to\yb\} \,.
\end{align}
The quark non-singlet kernel reads
\begin{align}
  \label{app:non-sing}
t_{a,2}(y) &=
\beta_0 \left[ \frac{21}{4} - \ln y
  - \ln\frac{Q^2}{\mu^2_R} \right]
\nonumber \\
&\quad
+ C_F \biggl[
  \left( 3 + 2 \ln y \right)\ms \ln\frac{Q^2}{\mu^2_{GPD}}
- \frac{25}{6}\, \ln\frac{Q^2}{\mu^2_{DA}}
- \frac{1019}{72} - \left( \frac{1}{\yb} + \frac{7}{6} \right) \ln y
+ \ln^2 y
\biggr]
\nonumber \\
&\quad
+ \left( 2 C_F - C_A \right)
  \biggl\{ \frac{401}{12}-255y+270y^2
- \left( \frac{299}{3}-867y+1830y^2-1080y^3 \right) \ln\yb
\nonumber \\
&\qquad
+ \left( \frac{56}{3}-357y+1290y^2-1080y^3 \right) \ln y
+ 2\ms \bigl( 22-291y+780y^2-540y^3 \bigr)\,
       \bigl( \Li_2 y - \Li_2\yb \bigr)
\nonumber \\
&\qquad
+ 12\ms (1-21y+106y^2-175y^3+90y^4) \phantom{\biggl[ \biggr]}
\nonumber \\
&\qquad\quad
  \times \biggl[ 3 \bigl( \Li_3\yb + \Li_3 y \bigr)
  - \ln y\, \Li_2 y - \ln\yb\, \Li_2\yb
  - \frac{\pi^2}{6}\, \bigl( \ln y + \ln\yb \bigr) \biggr]
\biggr\} \,,
\nonumber \\
t_{a,4}(y) &=
\beta_0 \left[ \frac{31}{5} - \ln y
  - \ln\frac{Q^2}{\mu^2_R} \right]
\nonumber\\
&\quad
+ C_F \biggl[
  \left( 3 + 2 \ln y \right)\ms \ln\frac{Q^2}{\mu^2_{GPD}}
- \frac{91}{15}\, \ln\frac{Q^2}{\mu^2_{DA}}
- \frac{10213}{900} - \left( \frac{1}{\yb} + \frac{46}{15} \right) \ln y
+ \ln^2 y
\biggr]
\nonumber \\
&\quad
+ \left( 2 C_F - C_A \right)
  \biggl\{ \frac{4903}{40}-\frac{5775}{2}y+\frac{57085}{4}y^2
           -23310y^3+11970y^4
\nonumber \\
&\qquad
- \left( \frac{21109}{60}-\frac{41451}{5}y+\frac{103285}{2}y^2
         -125020y^3+129150y^4-47880y^5 \right) \ln\yb
\nonumber \\
&\qquad
+ \left( \frac{2899}{60}-\frac{11001}{5}y+\frac{45535}{2}y^2
         -78400y^3+105210y^4-47880y^5 \right) \ln y
\nonumber \\
&\qquad
+ \bigl( 137-4506y+35280y^2-100380y^3+117180y^4-47880y^5 \bigr)\,
  \bigl( \Li_2 y - \Li_2\yb \bigr) \phantom{\biggl[ \biggr]}
\nonumber \\
&\qquad
+ 30\ms \bigl( 1-48y+580y^2-2590y^3+5166y^4-4704y^5+1596y^6 \bigr)
  \phantom{\biggl[ \biggr]}
\nonumber \\
&\qquad\quad
  \times \biggl[ 3 \bigl( \Li_3\yb + \Li_3 y \bigr)
  - \ln y\, \Li_2 y - \ln\yb\, \Li_2\yb
  - \frac{\pi^2}{6}\, \bigl( \ln y + \ln\yb \bigr) \biggr]
\biggr\} \,.
\end{align}
Using \eqref{I_gluon}, \eqref{I_a} and the representation
\begin{equation}
\gamma_n = (-1)^{n+1}\,
  2 C_F \int_0^1 dz\, (1-z)\ms (3 + 2 \ln z)\, C_n^{3/2}(2z-1)
\end{equation}
of the anomalous dimensions, we can give a closed form for the scale
dependent terms for all even $n$,
\begin{align}
  \label{higher-kern}
t_{g,n}(y) &=
\biggl[ 2C_A\ms (y^2+\yb^2) - C_F\ms y \biggr] \frac{\ln y}{\yb}\,
  \ln\frac{Q^2}{\mu^2_{GPD}}
 + \frac{\beta_0}{2}\ms \ln\frac{\mu_R^2}{\mu^2_{GPD}}
 - \frac{\gamma_n}{2}\ms \ln\frac{Q^2}{\mu^2_{DA}} + \{y\to\yb\}
 + \ldots \,,
\nonumber \\
t_{b,n}(y) &=
2(\yb-y) \, \frac{\ln y}{\yb}\, \ln\frac{Q^2}{\mu_{GPD}^2} - \{y\to\yb\}
 + \ldots \,,
\nonumber \\
t_{a,n}(y) &= \beta_0 \biggl[ \frac{19}{6} + \frac{\gamma_n}{2 C_F}
   - \ln y - \ln\frac{Q^2}{\mu_{R}^2} \biggr]
 + C_F (3 + 2\ln y) \ln\frac{Q^2}{\mu_{GPD}^2}
 - \gamma_n \ln\frac{Q^2}{\mu_{DA}^2} + \ldots \,,
\end{align}
where the terms denoted by $\ldots$ are independent of $Q^2$ and the
scales and do not involve $\beta_0$.  {}From the scale dependence
\eqref{DA-evol} of the Gegenbauer coefficients of the meson
distribution amplitude we can readily reconstruct their evolution
equation
\begin{equation}
\mu^2 \frac{\dd}{\dd\mu^2}\, a_n(\mu)
  = - \frac{\alpha_s(\mu)}{4\pi}\, \gamma_n\ms a_n(\mu) +
  O(\alpha_s^2) \,.
\end{equation}
With \eqref{gegen-kernels} and \eqref{higher-kern} we see that the
$\mu_{DA}$ dependence of the process amplitude \eqref{gegen-amp}
cancels up to terms of order $\alpha_s^3$, as it must be.


\section{Evolution kernels}
\label{app:evolution}

For definiteness we give here the LO evolution kernels for GPDs, which
we have used to check the scale invariance of the NLO amplitude for
meson production as explained in Sect.~\ref{sec:kernels}.  The
non-singlet evolution equation reads
\begin{equation}
  \label{gpd_ns_evol}
\mu^2 \frac{\dd}{\dd\mu^2}\, F^{NS}(x,\xi,t) =
  \int_{-1}^1 \frac{dy}{|\xi|}\,
  V^{NS}\biggl( \frac{x}{\xi},\frac{y}{\xi} \biggr)
  F^{NS}(y,\xi,t) \,,
\end{equation}
where $F^{NS}$ can be a flavor non-singlet combination such as
$F^{u(+)} - F^{d(+)}$, or the charge-conjugation odd combination
$F^{q(-)}(x,\xi,t) = F^q(x,\xi,t) + F^q(-x,\xi,t)$ for a single quark
flavor.
In the gluon and quark singlet sector we have a matrix equation
\begin{equation}
  \label{gpd_s_evol}
\mu^2\frac{\dd}{\dd\mu^2}\,
\begin{pmatrix}
F^S(x,\xi,t)\; \\[1.8ex]
F^g(x,\xi,t)
\end{pmatrix}
= \int_{-1}^1 \frac{dy}{|\xi|}
\begin{pmatrix}
            V^{qq}\left( \frac{x}{\xi},\frac{y}{\xi} \right) &
\xi^{-1}\ms V^{qg}\left( \frac{x}{\xi},\frac{y}{\xi} \right)
\\[1.2ex]
\xi  V^{gq}\left( \frac{x}{\xi},\frac{y}{\xi} \right) \; &
     V^{gg}\left( \frac{x}{\xi},\frac{y}{\xi} \right)
\end{pmatrix}
\;
\begin{pmatrix}
F^S(y,\xi,t)\; \\[1.8ex]
F^g(y,\xi,t)
\end{pmatrix}
\end{equation}
with $F^{S}$ defined in \eqref{def-FS}.
At $O(\alpha_s)$ one has $V^{NS}(x,y) = V^{qq}(x,y)$ and
\begin{align}
  \label{evol-kernels}
V^{qq}(x,y)
&= \frac{\alpha_s}{4\pi}\, C_F
\left[ \rho(x,y)\, \frac{1+x}{1+y} \left(1+\frac{2}{y-x}\right)
     + \{x\rightarrow -x,y\rightarrow -y\} \right]_+ \,,
\nonumber \\
V^{qg}(x,y) &= -\frac{\alpha_s}{4\pi}\, 2 T_F\ms n_f
\left[ \rho(x,y)\, \frac{1+x}{(1+y)^2}\, (1-2x+y-xy)
- \{x\rightarrow -x,y\rightarrow -y\} \right] \,,
\nonumber \\
V^{gq}(x,y) &= \frac{\alpha_s}{4\pi}\, C_F
\left[ \rho(x,y)\, \left( (2-x) (1+x)^2
     - \frac{(1+x)^2}{1+y} \right)
- \{x\rightarrow -x,y\rightarrow -y\} \right] \,,
\nonumber \\
V^{gg}(x,y) &= \frac{\alpha_s}{4\pi}\, C_A
\left[ \rho(x,y)\, \frac{(1+x)^2}{(1+y)^2}
       \left(2+\frac{2}{y-x}\right)
     + \{x\rightarrow -x,y\rightarrow -y\} \right]_+
\nonumber \\
&+ \frac{\alpha_s}{4\pi}\, C_A
\left[ \rho(x,y)\, \frac{(1+x)^2}{(1+y)^2}\, (1-2x+2y-xy)
     + \{x\rightarrow -x,y\rightarrow -y\} \right]
\nonumber \\
&+ \frac{\alpha_s}{4\pi}\,
   \left( \beta_0 - \frac{14}{3}\, C_A \right) \delta(x-y)
\end{align}
with $T_F = 1/2$ and the remaining constants as given in
\eqref{color_factors}.  The plus-prescription appearing in $V^{qq}$
and $V^{gg}$ is defined by
\begin{equation}
  \label{plus_description}
\bigl[ f(x,y) \bigr]_+ = f(x,y) - \delta(x-y) \int dz\,f(z,y) \,,
\end{equation}
and the function $\rho(x,y)$ specifies the support as
\begin{equation}
  \label{rho_function}
\rho(x,y) = \theta\left(\frac{1+x}{1+y}\right)
            \theta\left(1-\frac{1+x}{1+y}\right) \sgn(1+y)
= \theta(y-x)\, \theta(x+1) - \theta(x-y)\, \theta(-x-1) \,.
\end{equation}

The evolution equations for polarized GPDs read as in
\eqref{gpd_ns_evol} and \eqref{gpd_s_evol}, with the unpolarized
matrix elements $F$ and kernels $V$ replaced by their polarized
counterparts $\widetilde{F}$ and $\widetilde{V}$.  With
$\widetilde{F}^{q(+)}$ and $\widetilde{F}^{q(-)}$ defined in
\eqref{def-Ftilde-pm} above, $\widetilde{F}^{NS}$ can be either a
flavor non-singlet combination like $\widetilde{F}^{u(+)} -
\widetilde{F}^{d(+)}$ or a charge-conjugation odd combination
$\widetilde{F}^{q(-)}$, whereas the flavor singlet combination is
given by
\begin{equation}
\widetilde{F}^{S} = \widetilde{F}^{u(+)} +
\widetilde{F}^{d(+)} + \widetilde{F}^{s(+)} \,.
\end{equation}
To $O(\alpha_s)$ the polarized evolution kernels are
\begin{equation}
\widetilde{V}^{NS}(x,y) = \widetilde{V}^{qq}(x,y) = V^{qq}(x,y)
\end{equation}
and
\begin{align}
  \label{pol-kernels}
\widetilde{V}^{qg}(x,y) &= -\frac{\alpha_s}{4\pi}\, 2 T_f\ms n_f
\left[ \rho(x,y)\, \frac{1+x}{(1+y)^2}
     - \{x\rightarrow -x,y\rightarrow -y\} \right] \,,
\nonumber \\
\widetilde{V}^{gq}(x,y) &= \frac{\alpha_s}{4\pi}\, C_F
\left[ \rho(x,y)\, \frac{(1+x)^2}{1+y}
     - \{x\rightarrow -x,y\rightarrow -y\} \right] \,,
\nonumber \\
\widetilde{V}^{gg}(x,y) &= \frac{\alpha_s}{4\pi}\, C_A
\left[ \rho(x,y)\, \frac{(1+x)^2}{(1+y)^2} 
       \left( 2+\frac{2}{y-x} \right)
     + \{x\rightarrow -x,y\rightarrow -y\} \right]_+
\nonumber \\
&+ \frac{\alpha_s}{4\pi}\,
   \left( \beta_0 - \frac{14}{3}\, C_A \right) \delta(x-y) \,.
\end{align}

The kernels given here agree with those in \cite{Blumlein:1999sc} if
one takes into account that any contribution to $V^{gq}(x,y)$ which is
even in $y$ at fixed $x$ will drop out in the convolution
\eqref{gpd_s_evol}.
Taking the limit $\xi\rightarrow 0$ as
\begin{equation}
  \label{kernel-forward}
\lim_{\xi\rightarrow 0^+}
\frac{1}{\xi}
\begin{pmatrix}
\phantom{\frac{1}{\xi}}\ms
   V^{qq}\left( \frac{z}{\xi},\frac{1}{\xi} \right) &
\frac{1}{\xi}\ms
   V^{qg}\left( \frac{z}{\xi},\frac{1}{\xi} \right) \\[1.2ex]
\frac{\xi}{z}\ms
   V^{gq}\left( \frac{z}{\xi},\frac{1}{\xi} \right) &
\frac{1}{z}\ms
   V^{gg}\left( \frac{z}{\xi},\frac{1}{\xi} \right) 
\end{pmatrix}
= \begin{pmatrix}
P^{qq}(z) & P^{qg}(z) \\[1.8ex]
P^{gq}(z) & P^{gg}(z)
\end{pmatrix}
\end{equation}
one obtains the usual DGLAP evolution kernels from
\eqref{evol-kernels}, and in analogy one recovers the polarized DGLAP
kernels from \eqref{pol-kernels}.  The factors $\frac{1}{z}$ in front of
$V^{gq}$ and $V^{gg}$ reflect the different forward limits of the
quark and gluon GPDs.


\end{document}